\documentclass[aps,twocolumn,showpacs,amsmath,amssymb,nofootinbib]{revtex4-1}

\usepackage[pdftex]{graphicx}
\usepackage[usenames,dvipsnames]{xcolor}
\usepackage{epstopdf}
\usepackage[hidelinks]{hyperref}

\usepackage[normalem]{ulem}

\graphicspath{{FIGs/}}

 \usepackage[utf8]{inputenc}
\usepackage[french]{babel}

\usepackage{beramono}
\usepackage{listings}

\usepackage{lmodern}

\lstdefinelanguage{Julia}%
  {morekeywords={abstract,break,case,catch,const,continue,do,else,elseif,%
      end,export,false,for,function,immutable,import,importall,if,in,%
      macro,module,otherwise,quote,return,switch,true,try,type,typealias,%
      using,while},%
   sensitive=true,%
   alsoother={\$},%
   morecomment=[l]\#,%
   morecomment=[n]{\#=}{=\#},%
   morestring=[s]{"}{"},%
   morestring=[m]{'}{'},%
     literate={é}{{\'e}}1
           {è}{{\`e}}1
           {ù}{{\`u}}1
}[keywords,comments,strings]%

\newcommand{\FRrefsec}[1]{sec.~\ref{#1}}
\newcommand{\FRreffig}[1]{fig.~\ref{#1}}
\newcommand{\FRrefeq}[1]{eq.~\ref{#1}}

\newcommand{\Ham}{\hat{\mathcal{H}}}
\newcommand{\Spin}{\hat{S}}
\newcommand{\A}{\hat{A}{}}
\newcommand{\B}{\hat{B}{}}
\newcommand{\M}{\hat{M}}
\newcommand{\densmat}{\hat{\rho}}
\newcommand{\U}{\hat{U}{}}
\newcommand{\D}{\hat{D}{}}
\newcommand{\V}{\hat{V}{}}
\newcommand{\X}{\hat{X}{}}
\newcommand{\W}{\hat{W}{}}
\newcommand{\bigOmega}{\hat{\Omega}{}}
\newcommand{\bigLambda}{\hat{\Lambda}{}}
\newcommand{\bigGamma}{\hat{\Gamma}{}}

\newcommand{\I}{\hat{I}}
\newcommand{\0}{\hat{0}}

\usepackage{chngcntr}

\begin{document}

\title{Méthodes de calcul avec réseaux de tenseurs en physique}
\author{Thomas E.~Baker}
\affiliation{Institut quantique \& Département de physique, Université de Sherbrooke, Sherbrooke, Québec J1K 2R1 Canada}
\author{Samuel Desrosiers}
\affiliation{Institut quantique \& Département de physique, Université de Sherbrooke, Sherbrooke, Québec J1K 2R1 Canada}
\author{Maxime Tremblay}
\affiliation{Institut quantique \& Département de physique, Université de Sherbrooke, Sherbrooke, Québec J1K 2R1 Canada}
\author{Martin P.~Thompson}
\affiliation{Institut quantique \& Département de physique, Université de Sherbrooke, Sherbrooke, Québec J1K 2R1 Canada}

\begin{abstract}
Cet article se veut un survol des réseaux de tenseurs et s'adresse aux débutants en la matière. Nous y mettons l’accent sur les outils nécessaires à l’implémentation concrète d’algorithmes. Quatre opérations de base (remodelage, permutation d’indices, contraction et décomposition) qui sont couramment utilisées dans les algorithmes de réseaux de tenseurs y sont décrites. Y seront aussi couverts la notation diagrammatique, intrication, les états en produit de matrices (MPS), les opérateurs en produit de matrices (MPO), état projeté de paires intriquées (PEPS), l'approche par renormalisation d’enchevêtrement multi-échelle (MERA), la décimation par bloc d’évolution temporelle (TEBD) et le groupe de renormalisation de tenseurs (TRG).
\end{abstract}
\maketitle

\vspace{0.25cm}

\section{Introduction}
\label{FRintro}

Les méthodes exactes de résolution de systèmes quantiques sont difficiles à appliquer aux problèmes de grande taille. Il est alors nécessaire d’utiliser des méthodes approximatives et les réseaux de tenseurs figurent parmi les méthodes les plus utilisées à cet effet.
Les méthodes des réseaux de tenseurs se basent sur la séparation d’une fonction d’onde quantique en facteurs matriciels (ou tensoriels), un pour chacun des sites du système étudié. Le premier exemple notable d’une telle séparation est l’approche de la matrice de transfert pour la résolution du modèle d’Ising~\cite{ising1925beitrag,onsager1944crystal,reif2009fundamentals}. 

Pour résoudre de grands systèmes à l’aide de réseaux de tenseurs, nous devons considérer des représentations efficaces de la fonction d’onde. Pour obtenir une telle représentation, nous utilisons la troncature des tenseurs pour réduire le nombre de degrés de liberté du réseau et ne conserver que les plus significatifs~\cite{schollwock2005density,schollwock2011density,orus2014practical,bridgeman2017hand}. Cet approche est en lien direct avec le groupe de renormalisation~\cite{wilson1975renormalization} et se base sur l’intrication entre les sites. Ces méthodes de réseaux de tenseurs sont applicables sur des systèmes classiques ou quantiques.

La formulation moderne de ces problèmes se base sur les états de produits de matrices (MPS, \textit{Matrix Product State}) \cite{schollwock2011density}. L'un des premiers exemples d’un tel objet fut l’état de Affleck-Kennedy-Lieb-Tasaki (AKLT), utilisé pour décrire un système de spins~\cite{affleck2004rigorous,schollwock2005density}. Depuis, plusieurs algorithmes ont été développés, notamment le groupe de renormalisation par matrice-densité (DMRG, \textit{Density Matrix Renormalization Group})~\cite{white1992density}.

Dans cette revue des réseaux de tenseurs, nous nous concentrons sur les opérations de base nécessaires à la manipulation des tenseurs. À la \FRrefsec{FRwhytensors}, nous commençons par une discussion de ce que sont les tenseurs. À la \FRrefsec{FRdiagrams}, nous introduisons une notation schématique qui permet de simplifier le traitement analytique des réseaux de tenseurs. À la \FRrefsec{FRbasicops}, nous présentons quatre opérations de base s'appliquant aux tenseurs. Dans la \FRrefsec{FRentangle}, nous discutons de la connexion entre la théorie de l'information et les propriétés d'intrication. À la \FRrefsec{FRsec:MPS}, nous résumons les formes de fonctions d'onde les plus communes. Finalement, à la \FRrefsec{FRalgorithms}, nous présentons deux exemples d'algorithmes qui utilisent les concepts précédents pour trouver l'état fondamental d'un système physique.

\section{Les réseaux de tenseurs}
\label{FRwhytensors}

Pour un problème de physique quantique donné, nous devons résoudre l’équation de Schrödinger:
\begin{equation}
  \Ham{}\psi = E\psi,
\end{equation}
pour un hamiltonien $\Ham{}$, une énergie, $E$, ainsi qu’une fonction d’onde $\psi$. L’objectif principal est de trouver l’état fondamental du système et son énergie. Également, nous pouvons chercher les propriétés des excitations des systèmes dépendent du temps et des systèmes hors équilibre.

\begin{figure}[b]
  \includegraphics[width=0.75\columnwidth]{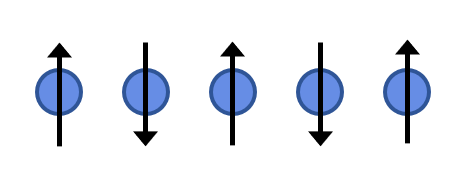}
  \caption{
    Une chaîne de spin est l’un des exemples les plus couramment utilisés pour illustrer les méthodes de réseaux de tenseurs. Les flèches représentent l’orientation des spins qui, dans un système réel, peuvent être en superposition.
  }
  \label{FRspinsys}
\end{figure}

Trouver la solution exacte d’un tel système est extrêmement difficile. Les méthodes standards qui sont présentées dans un cours d’introduction à la physique quantique deviennent rapidement impraticables pour des systèmes plus complexes. Par exemple, la façon la plus directe de trouver l’état fondamental d’un hamiltonien est par diagonalisation exacte. Pour ce faire, nous représentons le hamiltonien comme une matrice et nous utilisons des méthodes numériques pour trouver les valeurs propres. Cette méthode est très coûteuse en termes de mémoire et d’espace de calcul. C’est pourquoi nous ne pouvons l’utiliser que pour de petits systèmes.

Pour illustrer l’espace nécessaire à la représentation d’un hamiltonien sur un ordinateur, nous considérons une chaîne de spins (\FRreffig{FRspinsys}). Ces systèmes sont utilisés pour modéliser le magnétisme dans les matériaux. Les électrons peuvent subir des interactions d’échange, même s’ils sont localisés sur chaque atome. Un modèle commun est celui d’Ising~\cite{reif2009fundamentals}, représentant un système de $N$ spins
\begin{equation}
  \label{FRHamIsing}
  \Ham{} = -J \sum_{i = 1}^{N - 1} \Spin{}^z_i \cdot \Spin{}^z_{i + 1}.
\end{equation}
Les spins adjacents interagissent selon leur composante en $z$ et, de façon générale, il est possible d’avoir une réalisation de toutes les amplitudes de spins $(s=\frac12, 1, \frac32,\ldots)$. La constante $J$ représente l’interaction d’échange entre les spins. Un autre exemple commun de hamiltonien est le modèle de Heisenberg~\cite{townsend2000modern},
\begin{equation}
  \Ham{}=-J\sum_{i=1}^{N-1}\mathbf{S}_i\cdot \mathbf{S}_{i+1},
\end{equation}
où $\mathbf{S}= (\Spin{}^x,\Spin{}^y,\Spin{}^z)$, un vecteur qui contient les matrices de spin~\cite{townsend2000modern}.  Ainsi, il y a $2s+1$  états quantiques disponibles pour ce spin et $(2s+1)^N$ pour un réseau de $N$ sites. Comme le hamiltonien représente l'interaction entre ces spins, nous pouvons le représenter comme une matrice carrée de dimension $(2s+1)^{N} \times (2s+1)^{N}$. Le nombre de composantes du hamiltonien croît exponentiellement avec la taille du système. De nos jours, la taille maximale d'un système soluble par diagonalisation exacte est d'environ 50 sites \cite{wietek2018sublattice}.

Il existe des méthodes pour trouver seulement les plus petites valeurs propres de $\Ham{}$, tel l’algorithme de Lanczos~\cite{press1989numerical}. Cependant, ces méthodes sont limitées par la représentation en mémoire de la fonction d'onde aussi.

Nous voulons décomposer un système quantique en facteurs afin de réduire la complexité de la résolution du problème. Chacun de ces facteurs est représenté par un tenseur. Pour cette revue, nous considérons un tenseur comme la généralisation d’une matrice, c’est-à-dire comme un tableau multidimensionnel de nombres complexes\footnote{Pour une définition formelle des tenseurs, consulter ref.~\onlinecite{boas2006mathematical}.}. Nous allons ainsi utiliser des réseaux de tenseurs qui ont une croissance linéaire avec le nombre de sites du modèle. Nous serons alors en mesure de résoudre des systèmes de plusieurs milliers de sites. Une fois que le système quantique est décomposé convenablement, nous pouvons concevoir des algorithmes qui agissent sur une partie du système à chaque étape. Il est alors possible d’obtenir la solution globale après plusieurs itérations. 
En plus d’une croissance en complexité raisonnable avec la taille du système, les méthodes de réseaux de tenseurs ont également l’avantage de ne pas présenter de problèmes de signe~\cite{troyer2005computational} qui sont communs avec les algorithmes de Monte Carlo quantiques. Il y a cependant quelques limitations à l’utilisation des réseaux de tenseurs. Notamment, ces derniers performent beaucoup mieux lorsque les interactions et les corrélations sont à courte portée.

\section{Représentation graphique}
\label{FRdiagrams}

\begin{figure}[t]
  \includegraphics[width=\columnwidth]{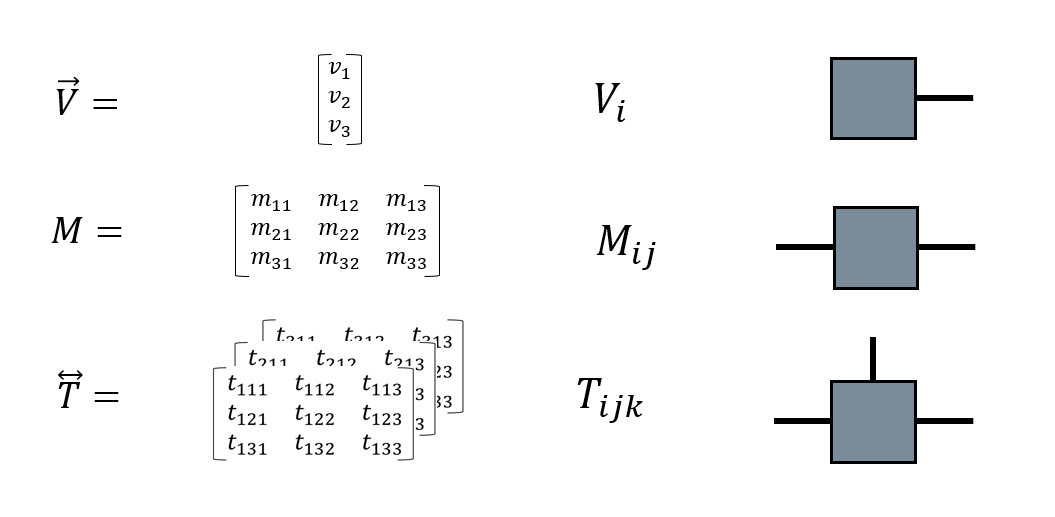}
\caption{Un tenseur de rang un (vecteur, haut), deux (matrice, milieu) et trois (bas), sont représentés respectivement par tous leurs éléments, par la notation indicielle et par la représentation en diagrammes.}
  \label{FRrepresent}
\end{figure}

Un tenseur peut être classifié par son rang. Le rang d’un tenseur correspond au nombre d’indices qu’il possède. Par exemple, un scalaire est un tenseur de rang 0, un vecteur est un tenseur de rang 1 et une matrice est un tenseur de rang 2. Nous dénotons un tenseur $T$ de rang 3 avec les indices $\mu$, $\nu$, et $\eta$ comme suit:
\begin{equation}
  \label{FRtensor}
  T_{\mu\nu\eta}.
\end{equation}
Chaque indice correspond à une dimension du tenseur et peut prendre différentes valeurs\footnote{Des indices en positions supérieures et inférieures sont parfois utilisés pour différencier des tenseurs covariants et contravariants. Cependant, cette distinction n’est généralement pas nécessaire pour les réseaux de tenseurs.}.

Écrire les tenseurs de cette manière avec tous leurs indices peut devenir très compliqué dans le cadre d’un réseau de tenseurs complet. Pour éviter cela, nous allons introduire une représentation par diagrammes comme dans ref.~\onlinecite{penrose1971angular}. Dans la \FRreffig{FRrepresent}, des tenseurs de rang un, deux et trois sont représentés avec les diagrammes correspondants. Nous y remarquons que les différents tenseurs sont représentés par un carré duquel sortent des lignes, chacune représentant un des indices. En général, dans un diagramme, les lignes ne sont pas explicitement identifiées par les indices qu’elles représentent. Cette information peut être fournie arbitrairement. 

\begin{figure}[t]
  \includegraphics[width=0.5\columnwidth]{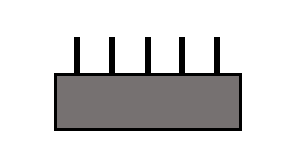}
  \caption{La représentation d’un tenseur de rang 5.}
  \label{FRwavefunctionDiagram}
\end{figure}

Quand nous décrivons une fonction d’onde, $\psi_{\sigma_1\sigma_2\ldots}$, par un tenseur, les lignes verticales du diagramme prennent un sens précis selon la convention. Chaque ligne représente un indice ($\sigma_i$) associé au spin de chacun des sites du système. Par exemple, un tenseur de rang cinq équivalent à la fonction d’onde d’un système de cinq sites est représenté avec cinq lignes verticales comme dans la \FRreffig{FRwavefunctionDiagram}. 

\begin{figure}[b]
  \includegraphics[width=0.9\columnwidth]{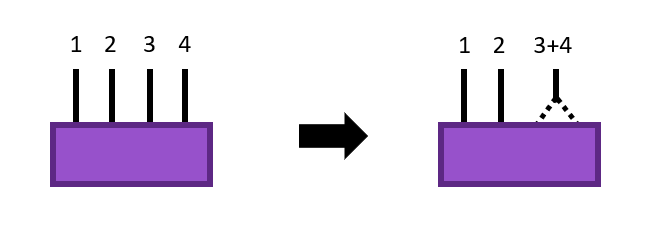}
  \caption{Un remodelage sur un tenseur de rang quatre où les troisième et quatrième indices sont fusionnés.}
  \label{FRreshape}
\end{figure}

Dans les sections suivantes, nous présentons les éléments de base des réseaux de tenseurs. Ces éléments se retrouvent dans les principales bibliothèques numériques~\cite{DMRjulia,tensorsnet,pfeifer2014ncon,stoudenmireitensor,Xtensor,tenpy,alvarez0209,al2017tensor,kao2015uni10,chanblock,chanPySCF,evoMPS,simpleDMRG,bauer2011alps,de2008density} et sont utiles pour la majorité des algorithmes.

\section{Opérations de base}\label{FRbasicops}

Tout comme pour les scalaires auxquels nous pouvons appliquer certaines opérations (addition, multiplication, etc.), nous pouvons appliquer certaines opérations aux tenseurs. Les opérations introduites ici sont à considérer comme la base minimale indispensable pour comprendre les réseaux de tenseurs et les méthodes algorithmiques les utilisant. Certaines combinaisons de ces opérations apparaissent dans chaque algorithme. Toutes ces opérations sont généralement disponibles et faciles à utiliser dans des langages de programmation de haut niveau tels Python et Julia.

Nous présentons quatre opérations. Les deux premières (remodelage et permutation) s’appliquent sur un seul tenseur. Les deux autres opérations (contraction et décomposition) impliquent plus d’un tenseur.

\subsection{Remodelage}

\begin{table}
\begin{tabular}{| c || c | c | c |}
\hline
Indice& $x$ & $y$ & $z$\\
\hline
1 & 1 & 1 & 1\\
2 & 2 & 1 & 1\\
3 & 3 & 1 & 1\\
\vdots & \vdots & \vdots & \vdots\\
$w_x$ & $w_x$ & 1 & 1\\
$1+w_x$ & 1 & 2 & 1\\
$2+w_x$ & 2 & 2 & 1\\
$3+w_x$ & 3 & 2 & 1\\
\vdots & \vdots & \vdots & \vdots\\
$2w_x$ & $w_x$ & 2 & 1\\
$1+2w_x$ &1 & 3 & 1\\
\vdots & \vdots & \vdots & \vdots\\
$w_x+w_x(w_y-1)$ & $w_x$ & $w_y$ & 1\\
$1+w_xw_y(w_z-1)$ & 1 & 1 & $w_z$\\
\vdots & \vdots & \vdots & \vdots\\
$w_x+w_x\Big((w_y-1)+w_y(w_z-1)\Big)$ & $w_x$ & $w_y$ & $w_z$\\
\hline
\end{tabular}
\caption{Tableau représentant le remodelage d’un tenseur de rang 3 de format prédéterminé  $(w_x,w_y,w_z)$ en un vecteur (un tenseur de rang 1). Les valeurs à gauche sont les valeurs de l’indice du vecteur permettant de se référer aux éléments. Les différents triplets de valeurs à droite sont les valeurs des indices du tenseur de rang trois se référant aux mêmes éléments. Nous nous référons d’abord aux éléments du tenseur initial suivant une certaine règle d’itération sur ses indices (ici, en partant du premier, $w_x$). Le remodelage est toujours inversible. La valeur d’indice du vecteur pour un certain élément selon ses coordonnées $(x,y,z)$ est donnée par $x+w_x((y-1)+w_y(z-1))$.}
\label{FRtabreshape}
\end{table}

Nous pouvons changer le rang d’un tenseur en regroupant certains de ses indices ou en subdivisant un indice. Par exemple, le tenseur de rang trois de l’ \FRrefeq{FRtensor} peut être remodelé en regroupant les indices $\mu$ et $\nu$ pour obtenir le tenseur de rang 2,
\begin{equation}\label{FRreshapeTensor}
T_{\gamma\eta}
\end{equation}
où $\gamma=(\mu\nu)$. Un tenseur peut être remodelé de façon à joindre n’importe quelle combinaison d’indices. Dans la \FRreffig{FRreshape}, nous illustrons un tenseur de rang quatre remodelé en un tenseur de rang trois.

Pour remodeler un tenseur, une règle doit d’abord être définie. Un exemple de règle est montré dans le tableau \ref{FRtabreshape} pour un tenseur de rang trois, de dimensions $(w_x,w_y,w_z)$ remodelé en un vecteur de taille $w_x\times w_y\times w_z$. Le premier indice est choisi pour être le premier incrémenté. Au fur et à mesure que nous incrémentons, un compteur (colonne de gauche) indique la position dans le vecteur résultant.

En général, nous pouvons remodeler un tenseur en un autre tenseur de n’importe quel rang. La seule condition nécessaire est que ces deux tenseurs doivent contenir le même nombre d’éléments. Par exemple, nous pouvons remodeler un tenseur de rang 3 de dimensions $(10, 5, 20)$ en un tenseur de rang 5 de dimensions $(2, 5, 5, 10, 2)$. Nous pouvons également remodeler ce tenseur sous forme d’un tenseur de rang 100, avec les dimensions ($1000, 1,1,1,1,1,1,1,\ldots$), comme celui-ci contient toujours le même nombre d’éléments.

Le remodelage est une opération mathématique reliant des tenseurs de rangs différents. Cependant, les indices du tenseur résultant un sens physique lors de la représentation d’un système quantique. L’opération de remodelage permet de distribuer la base d’états sur différents sites en un groupe à gauche et un autre à droite. Un point sur lequel nous reviendrons aux sections suivantes.

Le remodelage est une opération au coût de calcul négligeable. Cela s’explique par le fait qu’un tenseur est représenté dans la mémoire d’un ordinateur par un vecteur avec la dimension de chacun des indices. Seules les valeurs des dimensions sont modifiées lors du remodelage. Il existe d’autres façons de remodeler un vecteur, mais cet exemple appuie le fait que le remodelage est une opération très peu coûteuse.

\subsection{Permutation d'indices}

\begin{figure}[t]
  \includegraphics[width=0.9\columnwidth]{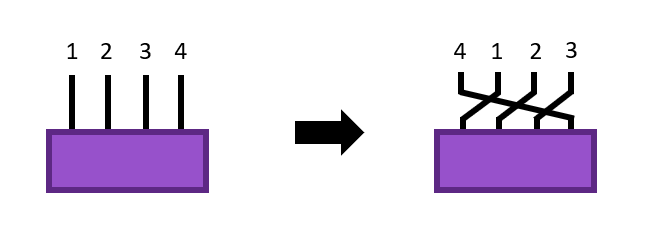}
  \caption{Une permutation d'indices pour un tenseur de rang 4 où le quatrième indice devient le premier.}
  \label{FRpermute}
\end{figure}  

Notez que dans le remodelage de l’\FRrefeq{FRreshapeTensor}, nous ne pouvions pas remodeler les indices $\mu$ et $\eta$ ensembles puisque l’indice $\nu$ se situe entre les deux. Selon la règle définie au tableau \ref{FRtabreshape}, lors d’un remodelage, les éléments d’un tenseur sont référés selon l’ordre des indices (nous considérons toutes les valeurs de $x$ avant de faire varier $y$ pour réordonner les éléments). De façon générale, nous ne pouvons remodeler un tenseur qu’en opérant sur des indices voisins, à moins de permuter les différents indices ({\it i.e.}, $T_{\mu\nu\eta}\rightarrow T_{\mu\eta\nu}$).   

Analytiquement, la permutation d’indices ne demande aucun effort de calcul. Nous pouvons simplement retracer les pattes comme dans la \FRreffig{FRpermute}. Cependant, lors d’applications pratiques dans un algorithme, échanger l’allocation de la mémoire pour changer l’ordre des indices nécessite une opération spéciale. La permutation d'indices consiste en la création d’un nouveau tenseur dans lequel l’information du premier tenseur sera copiée, en suivant le nouvel ordre d’indices. Généralement, il est recommandé d’éviter de faire cette opération autant que possible, vu son coût élevé.

\subsection{Contraction}\label{FRcontraction}

En plus de manipuler les indices d’un tenseur, nous pouvons combiner et diviser les tenseurs. Afin de joindre deux tenseurs, nous appliquons l’opération de contraction sur certains indices de ces deux tenseurs. Par exemple,
\begin{equation}\label{FRcontract}
C_{ik}=\sum_jA_{ij}B_{jk}
\end{equation}
représente une contraction des tenseurs de rang 2 (matrices) $\A$ et $\B$ sur l'indice $j$ commun aux deux. L’opération de contraction requiert une sommation sur chacun des indices contractés. Graphiquement, nous représentons cette opération par une ligne qui relie deux sommets.

\begin{figure}[t]
  \includegraphics[width=0.6\columnwidth]{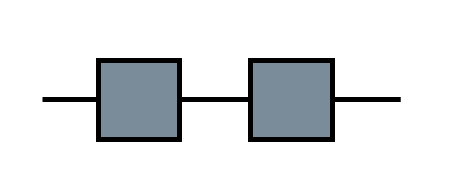}
\caption{Une représentation graphique du produit matriciel.}
  \label{FRmatrix}
\end{figure}

\begin{figure}[b]
  \includegraphics[width=\columnwidth]{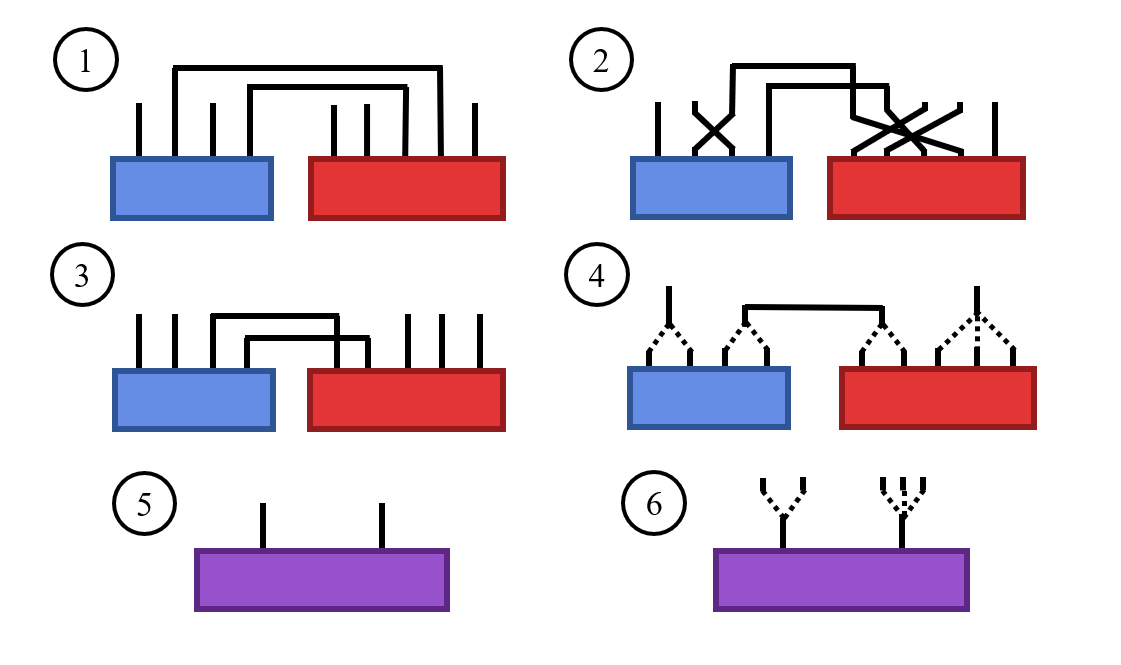}
\caption{En choisissant une certaine combinaison de permutations et de remodelages, il est possible de convertir toute contraction en produit matriciel. \textbf{1)} Une contraction de deux tenseurs sur deux indices. \textbf{2)} Permutation des indices des deux tenseurs afin que les indices non contractés soient au début du tenseur de droite et à la fin de celui de gauche. \textbf{3)} Les tenseurs sont représentés avec le nouvel ordre de leurs indices. Les indices contractés doivent suivre le même ordre dans chacun des tenseurs. \textbf{4)} Un remodelage des tenseurs est appliqué pour obtenir des tenseurs de rang 2. \textbf{5)} Le produit matriciel est appliqué. \textbf{6)} Remodelage de la matrice résultante afin de ramener les indices non contractés des tenseurs initiaux. }
  \label{FRcontraprod}
\end{figure}

Numériquement, nous pouvons effectuer la contraction de tenseurs par une série de sommes sur chacun des indices contractés. Cependant, il existe une approche alternative qu'est plus vite à convertir le problème en un produit matriciel, représenté schématiquement à la \FRreffig{FRmatrix}. Les étapes générales permettant de transformer un tenseur arbitraire en une matrice sont d’appliquer une permutation d’indices suivie d’un remodelage.  Ce processus est illustré plus en détail à la figure  \ref{FRcontraprod}, où des tenseurs de rang quatre et cinq sont contractés sur certains indices communs à l’étape 1. L’étape 2 de la \FRreffig{FRcontraprod} montre que le premier tenseur subit une permutation plaçant les indices contractés à droite des autres. Le second tenseur subit aussi une permutation plaçant les indices contractés cette fois-ci à gauche des autres. Les indices contractés des deux tenseurs doivent suivre le même ordre entre eux. Les tenseurs peuvent être représentés selon le nouvel ordre de leurs indices, comme à l’étape 3. À l’étape 4, les tenseurs sont alors remodelés de façon à combiner les indices non contractés ensemble, puis les indices contractés.

Les deux tenseurs sont alors sous forme de matrices. Sur le tenseur de gauche, les indices non contractés combinés représentent maintenant les lignes de la première matrice. Les indices connectés représentent alors les colonnes de la matrice. C’est le cas inverse pour la seconde matrice. Après le produit matriciel, nous obtenons une matrice résultante (étape 5). Finalement, à l’étape 6, cette matrice est remodelée de façon à retrouver les indices non contractés des tenseurs initiaux.

Le coût de calcul pour appliquer une contraction est déterminé par la taille de chaque indice des tenseurs contractés. Dans notre cas, en supposant que chaque indice est de taille $\chi$, le nombre total d’indices ouverts (5) et reliés (2) entraîne que le coût de calcul est de $\mathcal{O}(\chi^7)$. Le même résultat s’obtient en débutant par la définition de la contraction où l’on fait la somme sur chacun des indices contractés; écrit de manière mathématique:
\begin{equation}
C_{\alpha\beta\gamma\delta\epsilon}=\sum_{\zeta\eta}A_{\alpha\zeta\beta\eta}B_{\gamma\delta \eta\zeta\epsilon}.
\end{equation}
Cela montre que deux sommes (sur $\eta$ et $\zeta$) contribuent au coût de l’opération. Cependant, pour assigner des valeurs à l’ensemble des éléments du tenseur de rang 5, $C_{\alpha\beta\gamma\delta\epsilon}$, nous itérons sur chacun des indices non contractés. Au total, l’opération numérique de contraction nécessite 7 boucles (une pour chaque indice sur lequel nous itérons), chacune contribuant un facteur  $\chi$ au coût de calcul.

\subsection{Décomposition de tenseurs et de matrices}

En plus de combiner des tenseurs, il est souvent nécessaire de décomposer un tenseur. De la même manière que la contraction de tenseurs peut être convertie en un produit de matrices, la décomposition d’un tenseur peut être convertie en une décomposition de matrices.

\subsubsection{Décomposition en valeurs propres}

La décomposition en valeurs propres est bien connue en physique. Celle-ci correspond à la diagonalisation d’une matrice. La matrice (tenseur de rang 2) se décompose alors en trois matrices:
\begin{equation}\label{FREigdecomp}
\M = \U \bigOmega \U^\dagger
\end{equation}
Deux de ces matrices ($\U$ et $\U^\dagger$) sont liées par l’opérateur adjoint ($\dagger$) \cite{townsend2000modern}. Ces deux matrices sont unitaires et satisfont $\U\U^\dagger=\U^\dagger \U=\I$, où $\I$ est l’opérateur identité. La matrice $\bigOmega{}$ est diagonale et contient les valeurs propres de $m$. Les matrices $\bigOmega{}$ et $\U$ ont les mêmes dimensions. Nous pouvons appliquer la décomposition en valeurs propres seulement sur des matrices carrées. Cependant, nous ne pouvons pas généraliser cette décomposition à tous les tenseurs.

\subsubsection{Décomposition en valeurs singulières}\label{FRsec:svd}

\begin{figure}[t]
\includegraphics[width=0.8\columnwidth]{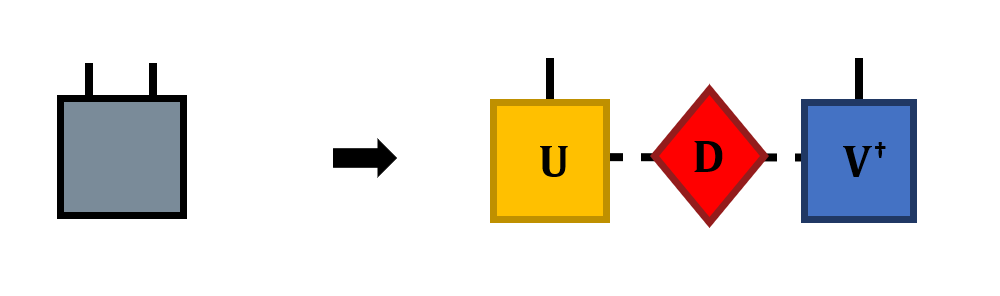}
\caption{Une représentation graphique de la décomposition en valeurs singulières (SVD).}
\label{FRSVD}
\end{figure}

Nous pouvons séparer un tenseur en trois autres tenseurs avec la décomposition en valeurs singulières (SVD  \textit{Singular Value Decomposition}). La SVD décompose une matrice rectangulaire M en trois matrices:
\begin{equation}\label{FRSVDdecomp}
\M{} = \U{}\D{}\V^\dagger.
\end{equation}
Cette opération est représentée graphiquement à la \FRreffig{FRSVD}.

\begin{figure}[b]
  \includegraphics[width=0.9\columnwidth]{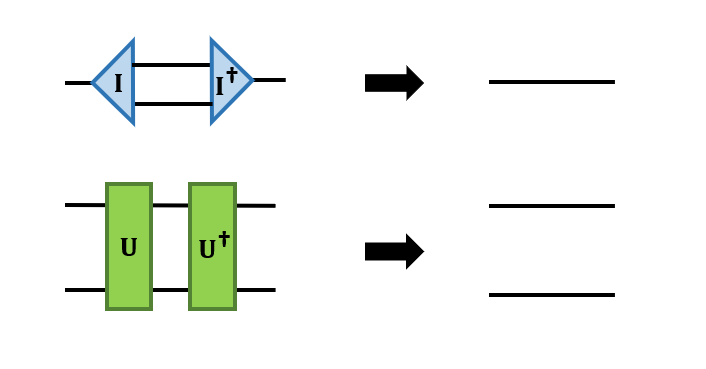}
  \caption{Les opérateurs unitaires et les isométries se contractent à l'identité. Les branches résultantes représentent l'identité (haut: $\I$, bas: $\I\otimes \I$).}
  \label{FRuniso}
\end{figure}

Les matrices $\U$ (rectangle jaune) et $\V^\dagger$ (rectangle bleu) sont des isométries et leur produit est l'identité dans un sens seulement. Si $\M{}$ est de dimensions $(a, b)$, $\U^\dagger \U=\I_{a\times a}$ et $\V^\dagger \V=\I_{b\times b}$ (voir \FRreffig{FRuniso}), mais $\U\U^\dagger$ et $\V\V^\dagger$ ne correspondent pas à une matrice identité complète généralement.

Les dimensions des matrices $\U$ et $\V^\dagger$ sont liées aux dimensions de $\M{}$. Le format de $\U$ sera de dimensions $(a, m)$ où $m\leq \mathrm{min}(a,b)$ \cite{schollwock2005density}. Le format de $\V^\dagger$ est alors $(m, b)$. Il existe d’autres formulations de la SVD où la dimension intérieure de la décomposition est choisie différemment mais, par convention ici, cette dimension maximale correspond à la plus petite des dimensions extérieures.

Finalement, la matrice $\D$ (losange rouge) est une matrice diagonale assignant un poids aux différents vecteurs dans les matrices $\U$ et $\V^\dagger$: 
\begin{equation}\label{FRlambdaN}
\D=\left(\begin{array}{ccccc}
\lambda_1&0&0&\cdots&0\\
0&\lambda_2&0&\cdots&0\\
0&0&\lambda_3&\ddots&0\\
\vdots&\vdots&\ddots&\ddots&\vdots\\
0&0&0&\cdots&\lambda_m
\end{array}\right).
\end{equation}

La matrice $\D$ exprime les plus importants états de base contenus dans $\U$ et $\V$. Il est fréquent que seules quelques valeurs de $\D$ soient importantes. Les valeurs les plus faibles de $\D$ peuvent tronquer \cite{schollwock2005density}. Cependant, la troncature des petites valeurs de $\D$ ne changera pas la norme de la matrice beaucoup. 

Pour effectuer une SVD, nous notons que $\M{}\M{}^\dagger=\U\D^2\U^\dagger$ et $\M{}^\dagger \M{}=\V\D^2\V^\dagger$. Toute matrice rectangulaire peut ainsi être multipliée par elle-même pour donner une matrice carrée. Cette matrice peut alors être décomposée par décomposition en valeurs propres pour obtenir les différents éléments de la SVD. Ce n’est pas la seule façon d’obtenir une SVD. En pratique, les outils standards n’utilisent pas la décomposition de $\M{}\M{}^\dagger$ ou de $\M{}^\dagger\M{}$ afin d’éviter une perte de précision venant de la mise au carré des valeurs de $\D$. Cependant, cette méthode met en évidence le lien étroit entre la SVD et la décomposition en valeurs propres.

\begin{figure}[t]
  \includegraphics[width=0.9\columnwidth]{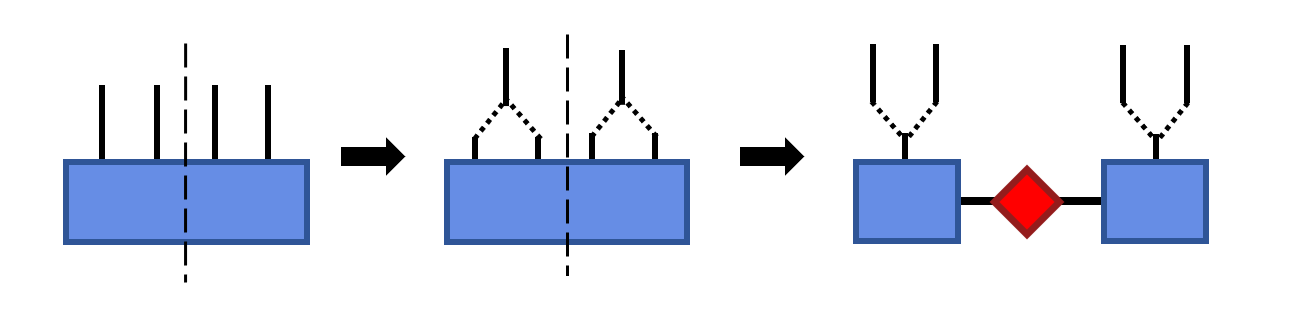}
\caption{Une décomposition d'un tenseur de rang quatre peut être effectuée en remodelant le tenseur et en appliquant une SVD sur la matrice résultante. La procédure ressemble à celle effectuée pour préparer un tenseur à la contraction dans la \FRreffig{FRcontraprod}.}
  \label{FRSVDFIG}
\end{figure}

Nous pouvons utiliser la SVD pour décomposer tout type de tenseur. La \FRreffig{FRSVDFIG} illustre la procédure à effectuer. Nous y remodelons un tenseur de rang 4 en un tenseur de rang 2 avant d’appliquer la SVD et de remodeler les tenseurs résultants afin de retrouver les indices appropriés. Cette approche est similaire à la procédure de la \FRrefsec{FRcontraction} pour la contraction.

\section{{Intrication}}\label{FRentangle}

La SVD n'est pas l’unique choix pour la décomposition de tenseurs. Il existe en effet plusieurs méthodes de décompositions matricielles pour les matrices rectangulaires: la décomposition $QR$ \cite{schollwock2005density,press1989numerical}, par exemple. Comme les choix sont multiples, nous allons démontrer l’importance de la méthode SVD dans le contexte des réseaux de tenseurs. L’explication est liée à la théorie de l’information qui étudie la manière dont deux objets communiquent entre eux~\cite{ash1965information}. Dans le cas d’un réseau de tenseurs, l’idée est la même: nous nous intéressons à la manière dont deux tenseurs interagissent. Une discussion exhaustive de cette théorie n’est pas nécessaire pour comprendre les bases des réseaux de tenseurs, mais il doit tout de même être mentionné qu’elle joue un rôle sous-jacent dans le développement d’algorithmes.

La compréhension de la SVD passe par la définition de la matrice $\D$.  Tel qu'introduit dans la section précédente, la matrice $\M{}$ été décomposé.  Dans un système physique, $\hat M$ égale la fonction d’onde $\psi$.  Notons que les valeurs propres $\rho_i$ de la matrice de densité de $\psi$, $\densmat{}=\psi\psi^\dagger$, sont indépendantes du choix de base de calcul. La matrice de densité est donnée par $\densmat{}=\U{}\bigOmega \U{}^\dagger$, où $\bigOmega$ est une matrice diagonale contenant tous les éléments $\rho_i$, et $\U{}$ ($\V$) contient les fonctions de base à gauche (droite) du site de $\psi$. Quand nous obtenons $\psi$ de $\densmat{}(=\U\sqrt{\bigOmega}\V^\dagger\V\sqrt{\bigOmega}\U^\dagger)$, nous trouvons $\psi=\U{}\D\V^\dagger$ où $\bigOmega=\D^2$.  La SVD contient donc toutes les composantes nécessaires de la matrice de densité.

Une quantité intéressante reliée à cette séparation est l’entropie d’intrication, aussi appelée entropie de von Neumann \cite{reif2009fundamentals},
\begin{equation}\label{FRentropy}
S=-\sum_i\rho_i\log_2\rho_i.
\end{equation}

\begin{figure}[b]
  \includegraphics[height=0.3\columnwidth]{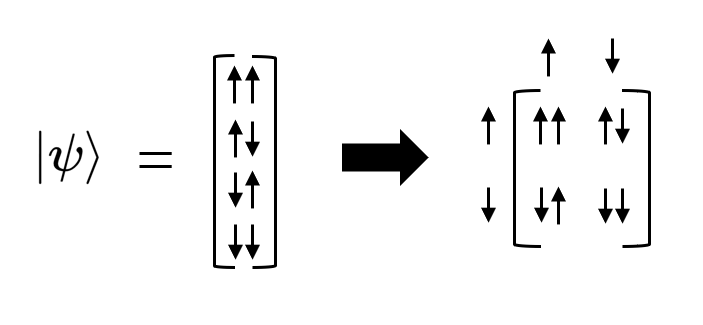}
  \caption{Vecteur décrivant un état de deux particules de spin $\frac12$. Il peut être remodelé sous forme de matrice. À présent, chaque indice de la matrice représente le spin de chacun des sites.}
  \label{FRvectmat}
\end{figure}

Certains cas spécifiques d’états quantiques peuvent être analysés en termes d’intrication. Nous présentons ici deux exemples illustrant les extremums d’intrication. Un premier exemple est l’état
\begin{equation}
|\psi\rangle=(|\uparrow\uparrow\rangle+|\downarrow\uparrow\rangle)/\sqrt2=\left(\frac{|\uparrow\rangle+|\downarrow\rangle}{\sqrt2}\right)\otimes|\uparrow\rangle
\end{equation}
qui a la SVD
\begin{equation}
\U{}\D\V^\dagger=\left(\begin{array}{cc}
0 & 1\\
1 & 0
\end{array}\right)\left(\begin{array}{cc}
1 & 0\\
0 & 0
\end{array}\right)\left(\begin{array}{cc}
1 & 1\\
1 & -1
\end{array}\right)\Big/\sqrt2
\end{equation}
où $|\uparrow\rangle=(1\;0)^T$, $|\downarrow\rangle=(0\;1)^T$ et $|\psi\rangle$ ont été remodelés comme dans la \FRreffig{FRvectmat}. Le symbole $T$ représente la transposée. L'intrication entre ces deux spins est $S=0$ (nous considérons seulement les valeurs non nulles de $\D$) puisque nous n’avons qu’une valeur propre dans \FRrefeq{FRentropy}. Ces spins sont qualifiés de {\it non intriqués}.

En second exemple, considérons l’état ;
\begin{equation}
|\psi\rangle=(|\uparrow\uparrow\rangle+|\downarrow\downarrow\rangle)/\sqrt2.
\end{equation}
Celui-ci est dit {\it maximalement intriqué} puisque les valeurs propres de $\D$ sont toutes deux $1/\sqrt2$. L'entropie est maximale puisque toutes les valeurs de $\D$ sont égales. Notons que l'état singulet $|\psi\rangle=(|\uparrow\downarrow\rangle-|\downarrow\uparrow\rangle)/\sqrt2$  est aussi maximalement intriqué avec $S=1$.

Dans un système physique typique, la structure de $\D$ donnera une intrication quelque part entre ces deux exemples. Un point supplémentaire est mentionné dans l’Appendice~\ref{FRsec:correlation}: la structure de $\D$ peut être classée globalement en deux catégories pour un problème physique donné.

\begin{figure}[t]
  \includegraphics[width=\columnwidth]{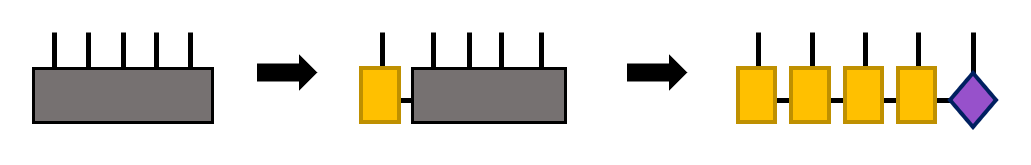}
\caption{Un vecteur à plusieurs sites peut être remodelé et divisé par SVD plusieurs fois de manière à ce que chaque site soit représenté par un tenseur. Le dernier tenseur est le seul pouvant être non unitaire, puisqu'il contient toute l'information de la matrice $\D$ de la SVD subséquente.}
  \label{FRMPSFIG}
\end{figure}

\section{État de produit de matrices}\label{FRsec:MPS}

\begin{figure}[b]
  \includegraphics[width=0.9\columnwidth]{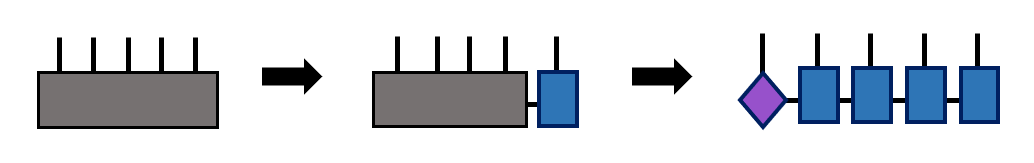}
  \caption{Le MPS peut aussi être construit de droite à gauche.}
  \label{FRleftMPS}
\end{figure}

Nous sommes enfin prêts à dériver un réseau de tenseurs à partir d’un système quantique selon les méthodes de la \FRrefsec{FRbasicops}. Pour cela, nous partirons d’une fonction d’onde complète puis nous appliquerons une série de remodelages et de SVD pour obtenir la représentation site par site de la fonction d’onde, illustrée à la \FRreffig{FRMPSFIG}. Il est aussi possible d’aller de droite à gauche tel qu’à la \FRreffig{FRleftMPS}. Cela permet d’obtenir une représentation équivalente d’une même fonction d’onde sous forme d’État de produit de matrices (MPS, \textit{Matrix Product State}). Lors de cette procédure, l’étape cruciale est le remodelage des indices physique des sites comme à la \FRreffig{FRMPSFIG}. Notons que nous commençons à partir d’une fonction d’onde prérésolue à des fins de démonstration uniquement. En général, la fonction d’onde sera dérivée d’un hamiltonien décomposé en réseau de tenseurs.
Ces deux algorithmes sont discutés après l’introduction d’un premier réseau de tenseurs représentant une fonction d’onde: l’État de produit de matrices (MPS, \textit{Matrix Product State}).

Même si les opérations peuvent être déduites à partir de la \FRreffig{ENMPSFIG}, nous allons les présenter en détail dans cette section. Tout en montrant les étapes de la dérivation du MPS, il est naturel de se demander ce que signifie le nouvel indice horizontal.  Pour bien comprendre le résultat, nous montrons ci-après la signification des étapes de la génération du MPS à deux reprises.  La première fois, nous montrons explicitement comment chaque étape est remodelée et où les indices physiques sont placés dans la nouvelle forme.  Ensuite, nous revoyons chaque étape, mais en gardant à l'esprit les fonctions de base de chaque indice. Cela nous permettra de comprendre pleinement la signification d'un indice horizontal qui est introduit.  Sur chaque diagramme, nous indiquerons les fonctions de base, même si cette information n'est généralement pas fournie.  L'index - également appelé index des liens - représente les fonctions de base de tous les autres sites du système.

\begin{figure}[t]
  \includegraphics[width=0.9\columnwidth]{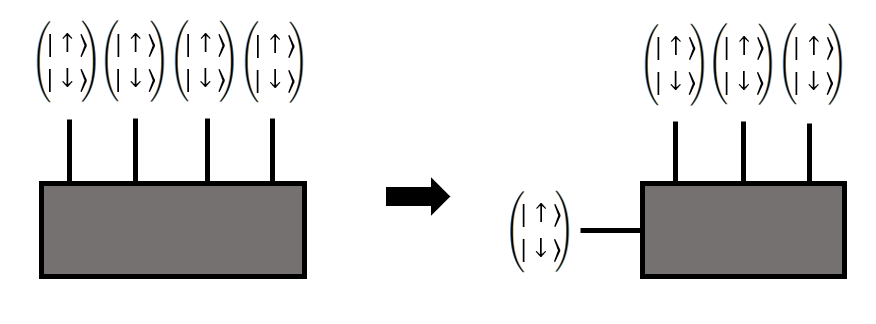}
  \caption{Remodelage d'un index en préparation du premier site à séparer.}
  \label{FRMPSfirst}
\end{figure}

\begin{figure}[b]
  \includegraphics[width=0.7\columnwidth]{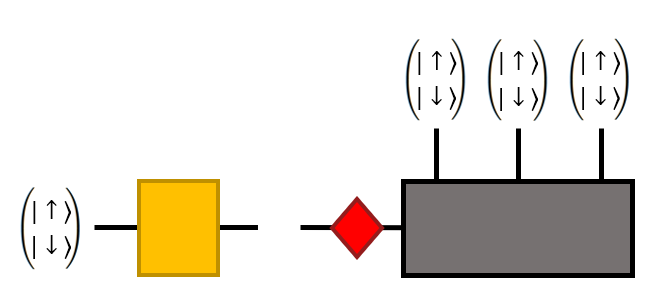}
  \caption{Il est possible d'effectuer une SVD sur la matrice précédente de manière à ce que le tenseur $\U{}$ généré représente localement notre premier site. La matrice $\D$ décrivant le poids de toutes les combinaisons de vecteurs peut être contractée au reste de la fonction d'onde.}
  \label{FRMPSsecond}
\end{figure}

\subsection{Séparer une fonction d'onde en un état de produit de matrices}

Nous commençons par une fonction d’onde provenant d’une diagonalisation exacte. La fonction d’onde complète peut être remodelée de manière à isoler du reste les degrés de liberté d’un site, comme illustré à la \FRreffig{FRMPSfirst}. Pour effectuer cette tâche, un programme est disponible dans l'annexe~\ref{FRMPScode} en guise d’exemple. Les indices verticaux d’une fonction d’onde sont appelés indices physiques, puisqu’ils expriment les degrés de liberté physique de chaque site ({\it i.e.}. Pour un système de spin il s’agit de les états sont $|\uparrow\rangle,|\downarrow\rangle$ tandis que pour le modèle de Hubbard ce serait $|0\rangle,|\uparrow\rangle,|\downarrow\rangle,|\uparrow\downarrow\rangle$). Une SVD peut ensuite être appliquée telle que représentée à la \FRreffig{FRMPSsecond}. Il est important de mettre un indice à gauche et le reste à droite: c’est ce qui permet la séparation de l’indice physique du premier site. La matrice $\U{}$, provenant de la SVD, peut être remodelée en tenseur représentant le premier site du MPS. Un indice additionnel est apparu au cours de la procédure, il connecte les tenseurs gauche et droit. Ce type d’indice, dessiné à l’horizontale sur la figure, est qualifié d'\textit{indice de lien}.

\begin{figure}[t]
  \includegraphics[width=0.8\columnwidth]{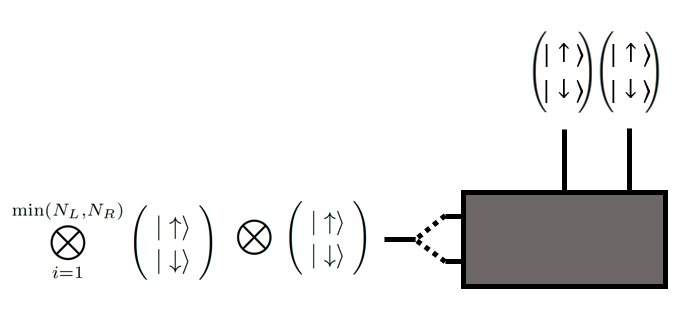}
  \caption{Le reste de la fonction d’onde doit être remodelé afin que l’indice physique suivant soit combiné à l’indice de lien.
  }
  \label{FRrestreshape}
\end{figure}

\begin{figure}[b]
  \includegraphics[width=1\columnwidth]{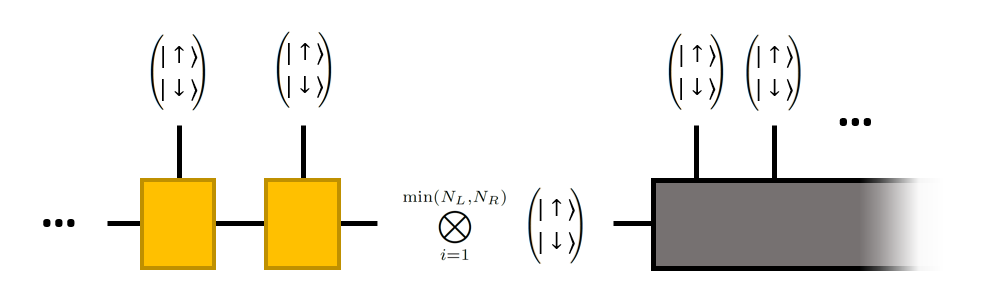}
  \caption{Au centre d’un MPS, les fonctions de base sont les produits de Kronecker de tous les états de base à gauche ou à droite (le moins élevé des deux).}
  \label{FRMPSfourth}
\end{figure}

La même idée peut ensuite être appliquée aux sites restants de la fonction d’onde. Le tenseur restant qui contient les autres degrés de liberté doit être remodelé afin que l’indice physique suivant soit combiné à l’indice de lien précédent. C’est ce qui est montré à la \FRreffig{FRrestreshape}. La matrice $\U{}$ peut encore une fois être remodelée pour donner un tenseur de la forme requise de la \FRreffig{FRMPSfourth} (la base de l’indice de lien y est indiquée). La procédure peut être répétée jusqu’à ce que chaque site soit représenté par un tenseur individuel. À présent, nous avons obtenu une représentation locale de chaque site du système ainsi qu'un premier exemple de fonction d’onde exprimée comme un réseau de tenseurs, un MPS.

Le coût associé à la séparation d’un tenseur est l’ajout d’un nouvel indice, le même qui apparaît dans \FRrefsec{FRsec:svd} lors de l’exécution de la SVD (voir \FRreffig{FRSVD}). Ce nouvel indice contient l’information pour chaque site différent du site actuel. La section suivante vise à comprendre la signification de cet indice de lien et comment il apparaît.

\subsection{Signification des indices horizontaux}

Tout au long des diagrammes de la section précédente, nous avons pris soin d’exprimer les fonctions de base de chaque indice, même si elles ne sont normalement pas affichées. Nous fournissons une analyse plus minutieuse de ces états dans cette section afin de pouvoir comprendre ce que représentent les nouveaux indices de lien.

Pour comprendre la signification de l'indice horizontal, nous devons comprendre ce que chaque valeur de l'indice représente, et nous le faisons en termes de base. Tout d'abord, nous commençons par examiner comment les fonctions de base sont jointes et séparées sur les indices physiques avant de discuter des indices de liaison. Comme nous l'avons déjà montré, les fonctions de base de l'indice physique sont $|\uparrow\rangle,|\downarrow\rangle$. Si nous voulons écrire la base de deux indices physiques, nous utilisons  un produit de Kronecker ($\otimes$, voir l'annexe~\ref{FRkronecker}). Cette opération prend tous les états à droite de l’opérateur et les joint aux états à gauche. La fonction d’onde à deux sites peut être écrite dans la base
\begin{equation}\label{FRjoinbasis}
\left(\begin{array}{c}
|\uparrow\rangle\\
|\downarrow\rangle
\end{array}\right)\otimes\left(\begin{array}{c}
|\uparrow\rangle\\
|\downarrow\rangle
\end{array}\right)\equiv\left(\begin{array}{c}
|\uparrow\uparrow\rangle\\
|\uparrow\downarrow\rangle\\
|\downarrow\uparrow\rangle\\
|\downarrow\downarrow\rangle
\end{array}\right),
\end{equation}
où le premier et le second vecteur représentent respectivement les états possibles du premier et du second site.

Pour plusieurs sites, nous pouvons écrire
\begin{equation}
\left(\begin{array}{ccccc}
|\uparrow\uparrow\uparrow\uparrow\uparrow\uparrow\uparrow\dots\rangle\\
|\downarrow\uparrow\uparrow\uparrow\uparrow\uparrow\uparrow\dots\rangle\\
|\uparrow\downarrow\uparrow\uparrow\uparrow\uparrow\uparrow\dots\rangle\\
\vdots
\end{array}\right)\equiv\bigotimes_{i=1}^N\left(\begin{array}{c}
|\uparrow\rangle\\
|\downarrow\rangle
\end{array}\right).
\end{equation}

Lorsque la base est remodelée en un nouveau groupe d’indices, les fonctions de base sont également séparées selon ce regroupement. Par exemple, dans le cas de deux sites, la \FRreffig{FRvectmat} montre une fonction d’onde à deux sites remodelée. 

Si nous prenons une fonction d'onde et appliquons une SVD, le nouvel indice peut être compris en termes de base des indices physiques. Considérons la SVD de la fonction d'onde à deux sites de l'\FRrefeq{FRMPSfirst} remodelée comme dans la \FRreffig{FRvectmat}. Nous ne considérerons que les fonctions de base impliquées dans ce qui suit et non la manière dont elles sont pondérées.  En examinant $\hat U$ ($\hat V$), une colonne (ligne) est une combinaison linéaire des fonctions de base des indices extérieurs, pondérée par les valeurs de $\hat U$ ($\hat V$). Chaque colonne (ligne) représente une autre valeur de l'indice horizontal et une combinaison linéaire différente des fonctions de base.

La plus grande dimension de l'indice horizontal est le minimum entre les deux côtés, $\min(N_L,N_R)$ basé sur la définition de la SVD qui est utilisée ici. Nous aurions pu choisir une définition différente, mais c'est le choix standard ici, à la fois pour la vitesse de calcul et le fait que les informations physiques doivent être indépendantes de la base définie, de sorte que la taille du nouvel indice peut être représentée à gauche ou à droite. Pour décrire les fonctions de base de l'index horizontal, nous pouvons écrire
\begin{equation}
\bigotimes_{i=1}^{\mathrm{min}(N_L,N_R)}\left(\begin{array}{c}
|\uparrow\rangle\\
|\downarrow\rangle
\end{array}\right).
\end{equation}
C'est la base de l'indice horizontal. Cette expression signifie que les indices de liaison sont des combinaisons linéaires de toutes les autres fonctions de base sur tous les autres indices physiques.

\subsection{Troncature}

Lors de la décomposition d’une fonction d’onde complète par utilisation de SVD successives (voir la \FRrefsec{FRsec:MPS}), nous pouvons déterminer la taille de l’indice de lien entre chaque site en connaissant les dimensions de la matrice sur laquelle nous applique la SVD. Pour une chaîne de spins  $\frac12$, l’indice physique est toujours de taille deux. Le premier indice de lien, reliant le premier site au reste de la fonction d’onde, sera de taille deux, car la matrice sur laquelle la SVD est appliquée est de format $2\times 2^{N-1}$. La taille de la matrice $\D$ correspond alors au minimum des deux dimensions. Suivant la même logique, l’indice de lien suivant sera de taille quatre, puis huit, seize, trente-deux, etc. À un certain point, les indices de lien rapetisseront, limités par la taille du second indice de la matrice à décomposer, jusqu’à atteindre une taille de deux, à l'autre extrémité du MPS. L'indice augmente ainsi de façon exponentielle jusqu'à l'indice situé au milieu de la chaine qui sera de dimension $2^{N/2}$. Ensuite, la dimension de l'indice diminue de façon symétrique.

Cependant, cette croissance exponentielle de l’indice de lien nous ramène précisément au problème lié à la résolution exacte de la fonction d’onde à la \FRrefsec{FRwhytensors}, qui justifiait l’utilisation des réseaux de tenseurs. Heureusement, il est parfois possible de tronquer l’indice de lien sans trop affecter la précision de l'état en utilisant la SVD introduite à la \FRrefsec{FRsec:svd}.

\begin{figure}[t]
  \includegraphics[width=\columnwidth]{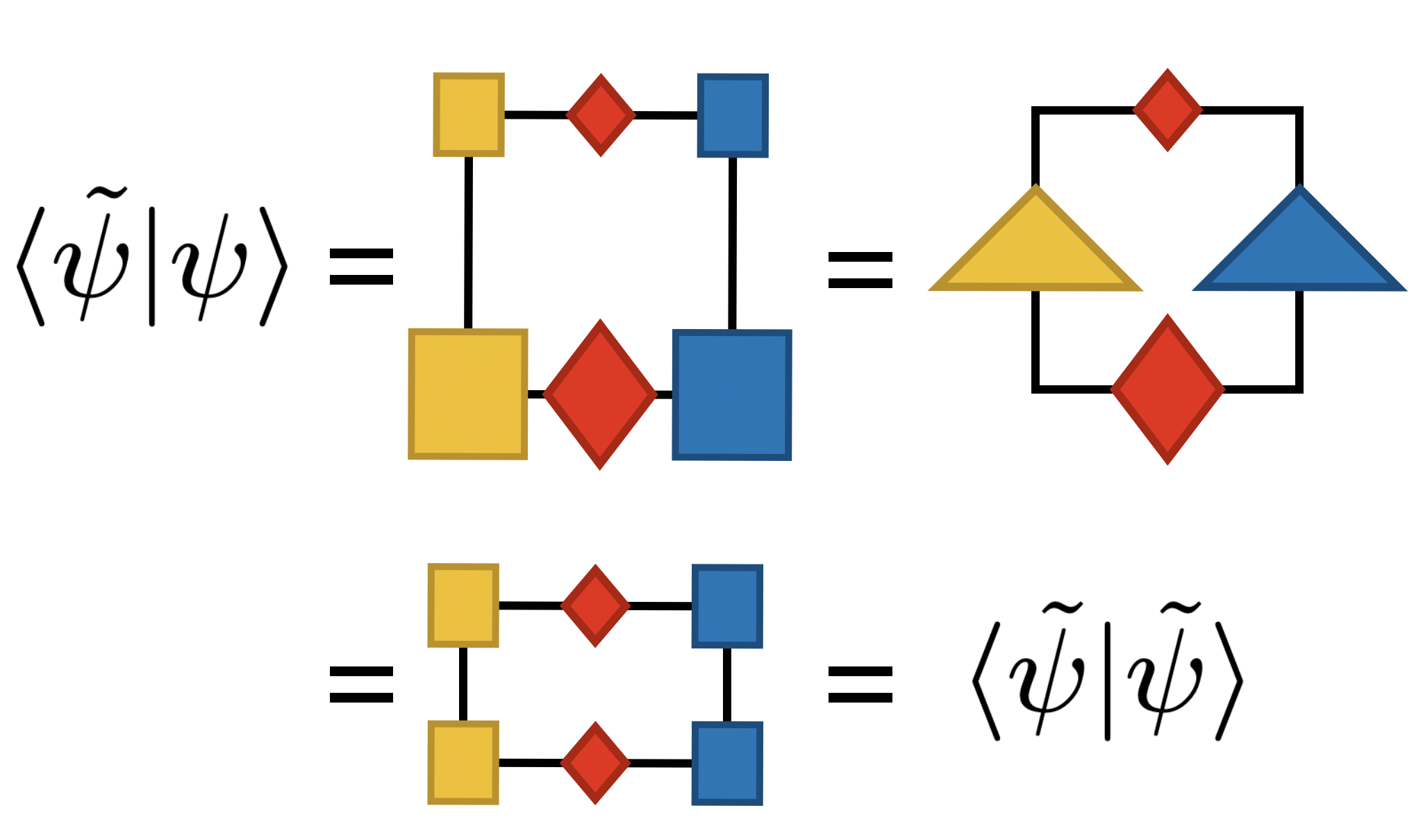}
\caption{L'égalité $\langle\tilde{\psi}|\psi\rangle=\langle\tilde{\psi}|\tilde{\psi}\rangle$ (\FRrefeq{FRtruncerr}) est ici justifiée. La fonction d'onde tronquée par la SVD est représentée par des tenseurs plus petits au haut du diagramme. $\tilde \V^\dagger \V$ et $\tilde \U{}^\dagger \U{}$ se contractent alors en isométries qui ne lieront linéairement que les valeurs de $\D$ conservées dans la fonction d'onde tronquée. Cette contraction est donc équivalente à $\langle\tilde{\psi}|\tilde{\psi}\rangle$.}
  \label{FRerrorjust}
\end{figure}

La matrice $\D$ est typiquement aménagée en ordonnant les valeurs singulières en ordre décroissant sur la diagonale en partant du coin supérieur gauche de la matrice. Nous pouvons se demander à quel point la précision de notre représentation du système est affectée en tronquant les valeurs les plus faibles. Une bonne façon d’évaluer cette perte de précision est d’évaluer la norme de la fonction d’onde approximative résultante (notée avec un tilde), $|\tilde\psi\rangle$. La différence entre la fonction d’onde tronquée (ou approximée) et la fonction d’onde complète est
\begin{equation}\label{FRtruncerr}
\left\||\psi\rangle-|\tilde{\psi}\rangle\right\|^2=1+\delta-2\delta=1-\delta
\end{equation}
où $\delta=\langle\tilde{\psi}|\psi\rangle=\langle\tilde{\psi}|\tilde{\psi}\rangle = \sum_{i=1}^{m}\lambda_i^2$ s’appelle l’\textit{erreur de troncature}. Cette égalité ($\langle\tilde{\psi}|\psi\rangle=\langle\tilde{\psi}|\tilde{\psi}\rangle$) s’observe en schématisant la SVD appliquée sur la fonction d’onde complète et sa version tronquée, comme illustré à la \FRreffig{FRerrorjust} voir \FRrefsec{FRmpslimits} ou annexe~\ref{FRsec:correlation} pour plus d'informations). Nous remarquons alors que $\tilde \V^\dagger \V$ et $\tilde \U{}^\dagger \U$ se multiplient en isométries qui viendront tronquer les valeurs de $\D$ non présentes dans  $|\tilde\psi\rangle$.

Afin d'illustrer l’impact de cette procédure sur les observables du problème, rappelons que l’énergie, $E$, est liée à la matrice densité selon $\mathrm{Tr}(\densmat{}\Ham{})\equiv E$ \cite{reif2009fundamentals}. En ne tronquant que de petits éléments de $\densmat{}$, l’énergie globale du système ne changera que très peu. Heureusement, pour plusieurs cas d’intérêt, une grande partie des valeurs singulières sont faibles (voir annexe \ref{FRsec:correlation}). La taille de la matrice $\D$ peut donc être grandement réduite pour plusieurs systèmes physiques.

Troncature élimine la base d’états associés à un poids faible dans la matrice densité. Il s’agit là d’un concept fondamental aux réseaux de tenseurs. 

\subsection{Mesures et jauges}\label{FRgauge}

Les mesures liées aux opérateurs locaux sont parfaitement adaptées à un MPS. Rappelons que la SVD produit deux tenseurs isométriques ($\U{}$ et $\V^\dagger$) ainsi que la matrice $\D$, seul objet qui associe un quelconque poids à la décomposition. Les tenseurs à gauche du centre d’orthogonalité ($\U{}$) sont dits  \guillemotleft{} normalisés à gauche \guillemotright{} alors que ceux  de droite ($\V^\dagger$) sont  \guillemotleft{} normalisés à droite \guillemotright{}. Si deux tenseurs normalisés à gauche sont contractés, le résultat est équivalent à l’identité (voir \FRreffig{FRuniso}). L’équivalent est aussi vrai pour les tenseurs normalisés à droite. La propriété générale selon laquelle toute l’information non unitaire d’un MPS est contenue dans un seul tenseur est appelée jauge~\cite{vidal2003efficient}. Dans le cas d’une mesure locale, les tenseurs n’ont pas tous besoin d’être évalués. Une application pratique de cette propriété est montrée ici. 

\begin{figure}[b]
\includegraphics[width=0.8\columnwidth]{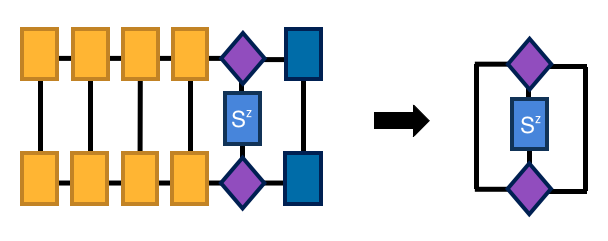}
\caption{Les mesures locales peuvent être faites de manière bien plus simple en utilisant le MPS. Ici l’opérateur est équivalent à une mesure locale de spin sur un seul site. L’application de cet opérateur n’implique que trois tenseurs.}
\label{FRMPOFIG}
\end{figure}

Un exemple de mesure locale de l’opérateur $\Spin{}^z$ avec $\hbar=1$,
\begin{equation}\label{FRSzDef}
\Spin{}^z=\frac12\left(\begin{array}{cc}
1 & 0\\
0 & -1
\end{array}\right),
\end{equation}
sur un seul site est illustré à la \FRreffig{FRMPOFIG}. En appliquant la même mesure sur la fonction d’onde complète, il aurait fallu écrire $\Spin_5^z=\I\otimes \I\otimes \I\otimes \I\otimes \Spin{}^z\otimes \I$, qui est une matrice bien plus grande que celle de l’équation \eqref{FRSzDef}.

\begin{figure}[t]
  \includegraphics[width=0.8\columnwidth]{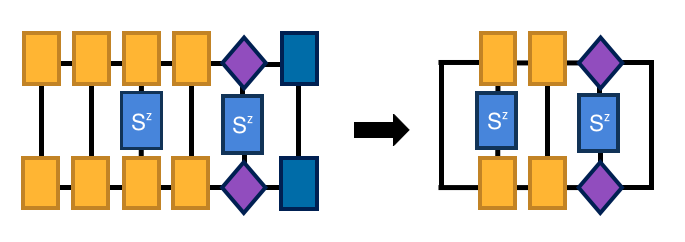}
\caption{Pour une mesure s’appliquant à deux sites, le problème se réduit aux tenseurs des sites d’intérêts et aux tenseurs situés entre ceux-ci. Ceci est une conséquence de la brisure d’orthogonalité lors de la contraction par la gauche.}
  \label{FRMPO2site}
\end{figure}

Pour une mesure appliquée à deux sites ou plus, comme à la \FRreffig{FRMPO2site} ($\langle\psi|\Spin_3^z\Spin_5^z|\psi\rangle$), l’orthogonalité n’est pas conservée pour tous les sites. En contractant les tenseurs normalisés à gauche, l’apparition d’un opérateur brise la relation d’orthogonalité. Les tenseurs se trouvant entre les différents opérateurs ne se contractent donc pas à l’identité. Comme le centre d’orthogonalité a été placé sur le même site que celui sur lequel s’applique l’autre opérateur, le reste des tenseurs, tous normalisés à droite, se contractent à l’identité.

\begin{figure}[b]
  \includegraphics[width=\columnwidth]{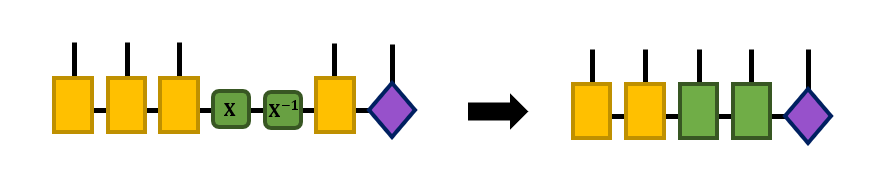}
\caption{Le MPS s’avère non unique, car il représente toujours une fonction d’onde complète après l’introduction d’une matrice et de son opposée entre deux tenseurs.}
  \label{FRnounique}
\end{figure}

Il n’existe pas un MPS unique représentant une fonction d’onde. En effet, entre deux indices de lien, une identité peut être introduite sous forme d’un terme $\X\X^{-1}$. En contractant $\X$ avec le tenseur du MPS à sa gauche et  $\X^{-1}$ avec le tenseur à sa droite, les deux tenseurs du MPS en sont affectés. Comme cette opération est équivalente à l’application d’une identité, le MPS reste équivalent à l’état original, comme à la \FRreffig{FRnounique}. Par conséquent, deux MPS peuvent avoir une structure différente tout en menant aux mêmes résultats.

\subsection{{Opérateurs en produit de matrices}}\label{FRMPO}

Le MPS est une représentation locale en termes de tenseurs d’une fonction d’onde. Pour avoir une représentation complète d’un problème de mécanique quantique en termes de réseaux de tenseurs locaux, il nous faut une construction équivalente pour les opérateurs: l’opérateur en produit de matrices (MPO).

\begin{figure}[t]
  \includegraphics[width=0.45\columnwidth]{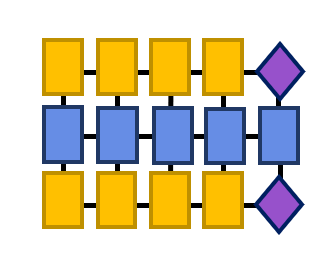}
\caption{Ce diagramme représente la valeur moyenne d’un opérateur MPO dans un état MPS, $\langle\psi|\Ham{}|\psi\rangle$.}
  \label{FRUnitMPO}
\end{figure}

Par exemple, le hamiltonien du modèle d'Ising peut être construit avec le MPO suivant:
\begin{equation}\label{FRbulkMPO}
\Ham{}_{a,a'}^\mathrm{(Ising)}=\left(\begin{array}{ccc}
\I & \0 & \0\\
\Spin{}^z & \0 & \0\\
\0 & -J\Spin{}^z & \I\\
\end{array}\right).
\end{equation}
Comme les éléments de cette \guillemotleft{} matrice \guillemotright{} sont eux-mêmes des opérateurs, il s’agit en fait d’un tenseur de rang 4 (voir \FRreffig{FRUnitMPO}). Les lignes et les colonnes de la matrice représentent les indices horizontaux du MPO, et les opérateurs individuels ont un indice physique pour se connecter à la fonction d'onde et un second pour se connecter à son dual de la fonction d'onde. Lorsque l’on contracte une chaîne de ces matrices d’opérateurs le long des indices de liaison pour former le hamiltonien d’un système, celui-ci se trouve dans le coin inférieur gauche de la matrice finale. Donc, pour un système aux conditions aux limites ouvertes, il faut terminer les extrémités de la chaîne avec deux vecteurs: $(0, 0, 1)$ à gauche et $(1,0,0)^T$ à droite.  La détermination de la forme du MPO suit quelques règles simples, mais il existe également des méthodes permettant de déterminer automatiquement la forme des opérateurs \cite{hubig2017generic}.

Tout comme les MPS, la représentation en MPO n’est pas unique. Trivialement, des rangées et des colonnes de zéro pourraient être ajoutées aux matrices sans changer l’opérateur représenté.

\begin{figure}[b]
  \includegraphics[width=0.9\columnwidth]{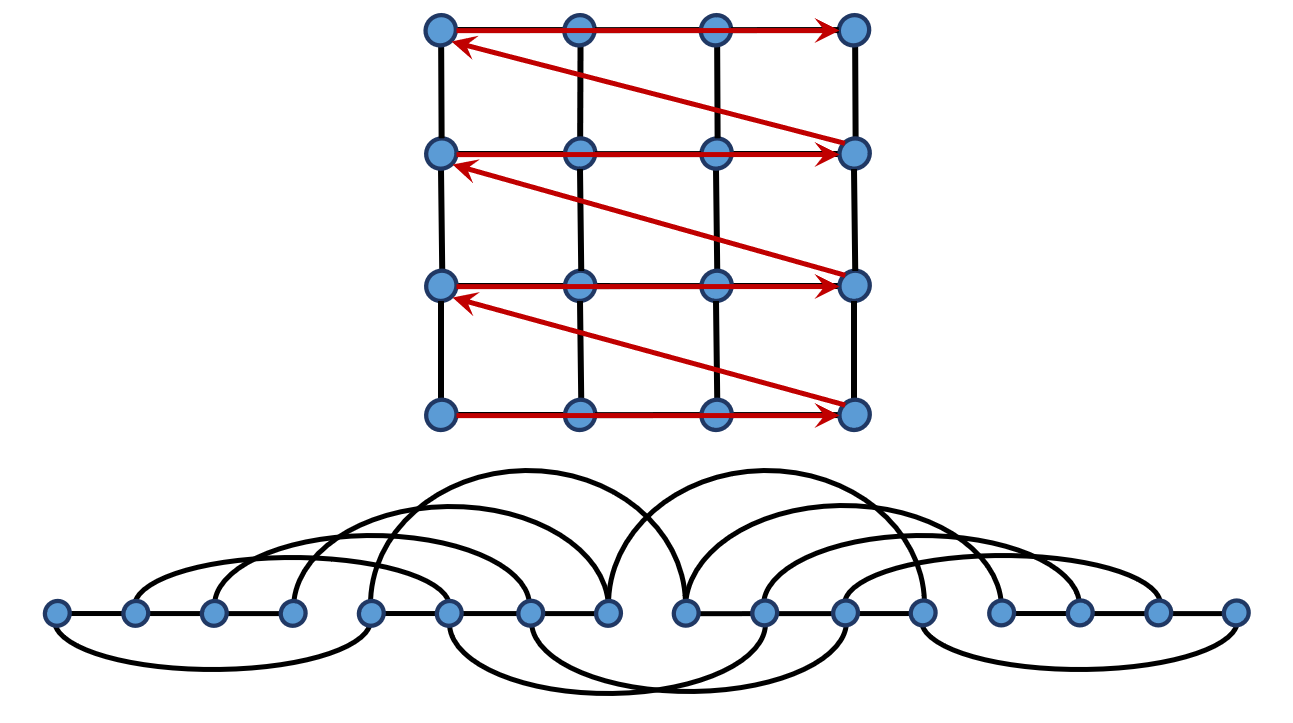}
\caption{Un réseau en deux dimensions peut être développé en une ligne. Le chemin utilisé peut varier selon la situation.}
  \label{FRsnake}
\end{figure}

Pour ajouter des termes couplant des sites plus distants, il faut augmenter la dimension des matrices d’opérateurs. Par exemple, le modèle d’Ising avec couplage au premier et second voisin
\begin{equation}
\Ham{}=\sum_i \left(-J_1 \Spin{}^z_{i}\Spin{}^z_{i+1}-J_2 \Spin{}^z_i\Spin{}^z_{i+2}\right)
\end{equation}
nécessite le MPO
\begin{equation}\label{FRLRI}
\Ham{}_{a,a'}^\mathrm{(Ising)}=\left(\begin{array}{cccc}
\I & \0 & \0 & \0\\
\Spin{}^z & \0 & \0 & \0\\
\0 & \I & \0 & \0\\
\0 & -J_1 \Spin{}^z & -J_2 \Spin{}^z & \I\\
\end{array}\right).
\end{equation}
Typiquement, pour chaque terme agissent à un site donné du système, une colonne et une rangée supplémentaire doivent être ajoutées. Pour construire un MPO pour un réseau bidimensionnel, il faut choisir un chemin traversant tous les sites du système une seule fois. Ce chemin définit un problème effectif en une dimension; un exemple est illustré à la \FRreffig{FRsnake}. La portée effective du couplage pour le MPO dépend de la largeur du système, et donc la taille des matrices d’opérateurs est proportionnelle à la largeur du système. Ceci compromet l’applicabilité des MPO et MPS aux systèmes de dimension supérieure à un.

Certains couplages à longue portée peuvent être écrits avec un MPO de petite dimension. Par exemple, un couplage en exponentielle de la distance entre deux sites
\begin{equation}
\Ham{}=\sum_{i\neq j}e^{-\frac{|i-j|}\xi}\Spin{}^z_{i}\Spin{}^z_{j},
\end{equation}
 couple tous les sites d'un système entre eux. La MPO générant ce hamiltonien s’écrit comme suit
\begin{equation}
\Ham{}_{a,a'}=\left(\begin{array}{cccc}
\I & \0 & \0\\
\Spin{}^z & \kappa\I & \0\\
\0 & \Spin{}^z & \I
\end{array}\right)
\end{equation}
où $\kappa\equiv\exp(-1/\xi)$. Écrire le MPO de ce système avec la méthode décrite pour l’\FRrefeq{FRLRI} aurait requis des matrices de taille proportionnelle au nombre de sites dans le système. Trouver la plus petit expression en MPO pour le hamiltonien d’un système est critique à la bonne performance des algorithmes de MPS.

\subsection{Limites de l’état de produit de matrices}\label{FRmpslimits}

Quelle taille doit avoir une fonction d’onde pour décrire un système donné ? Quelle quantité d’information pouvons-nous tronquer et encore conserver les propriétés du système que l’on veut résoudre? Cela dépend des systèmes, mais il y a quelques propriétés simples qui s’appliquent à tout système physique d’intérêt.

Grossièrement, il y a deux cas qui vont différencier la performance d'un réseau de tenseurs et la troncature maximale nécessaire pour obtenir une solution acceptable. Une bonne façon de penser à ces deux cas est de se situer dans la limite thermodynamique (durant les calculations, nous utilisons un système finis mais c'est approximatif de la limite). Dans le premier cas, il y un \textit{gap} (ou espacement) entre les deux énergies les plus faibles (voir annexe~\ref{FRsec:correlation}).  Dans l’autre cas, il n’y a pas de gap entre les valeurs propres et elles forment un spectre continu. Le MPS est bien adapté aux états avec gap. De plus, si l’interaction est locale, c'est-à-dire si l’interaction s’étend seulement à quelques sites dans un chemin unidimensionnel, alors le MPS peut être tronqué sans sacrifier trop de précision~\cite{verstraete2006matrix}.  Dans un exemple simple, comme le modèle de Heisenberg de 10 sites, 50 états peuvent être conservés sur le lien et nous obtiendrons encore une bonne précision en comparaison avec la diagonalisation exacte. Si nous avons des interactions à longue portée, alors l’indice de lien doit garder plus d’information. Notez que sur un système de taille finie, il y a toujours au moins un petit gap et que dans de nombreux cas, le MPS peut se rapprocher d'un système d'intérêt et ne nécessitera une dimension plus importante que si le système n'est pas idéalement adapté au MPS \cite{schollwock2005density}.

Nous pouvons toujours développer un système de dimension supérieure en un système à une dimension. Cela fonctionne avec une efficacité variable, comme dans la \FRreffig{FRsnake}.  L’augmentation de la taille est le résultat de l’augmentation de la portée de l’interaction, qui est introduit dans la \FRrefsec{FRMPO}. La croissance du MPO avec la dimension du modèle se rapporte à la loi de domaine de l’intrication et prend son nom de l’étude des trous noirs~\cite{eisert2010colloquium}. La loi de domaine de l’intrication stipule que la quantité d’informations transmises entre deux domaines d’un système peut être exprimée en fonction de la frontière entre ces régions. La loi est valide pour les états aux limites du spectre des valeurs propres, c'est-à-dire pour l'état fondamental. Pour les états du centre du spectre, il y a une loi de volume qui impose un comportement différent pour l'intrication. 

Pour un problème unidimensionnel le domaine entre le côté droit et gauche du système est zéro dimensionnel. Le domaine entre les côtés droit et gauche d’un problème en deux dimensions est unidimensionnel. En général, le domaine entre les deux côtés est $d-1$ dimensionnel pour un problème à $d$ dimensions. Alors, augmenter la dimension du problème va augmenter la quantité d’information d’un côté à l’autre du système. 

\begin{figure}[b]
  \includegraphics[width=0.9\columnwidth]{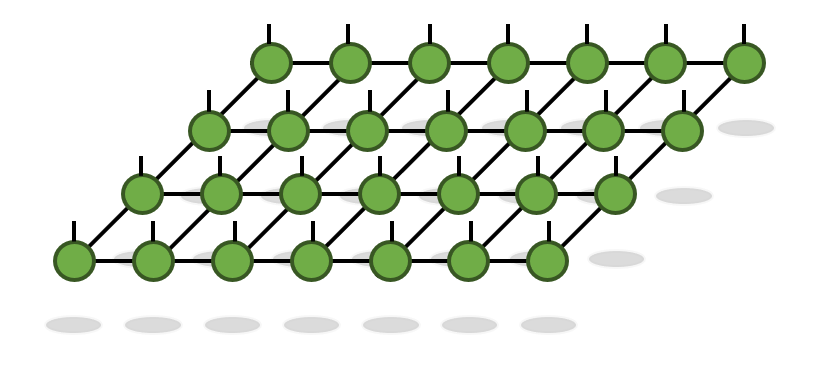}
\caption{Le PEPS est une représentation bidimensionnelle de la fonction d’onde. Cela permet parfois une meilleure représentation des systèmes bidimensionnels que le MPS.}
  \label{FRPEPS}
\end{figure}

\begin{figure}[t]
  \includegraphics[width=0.8\columnwidth]{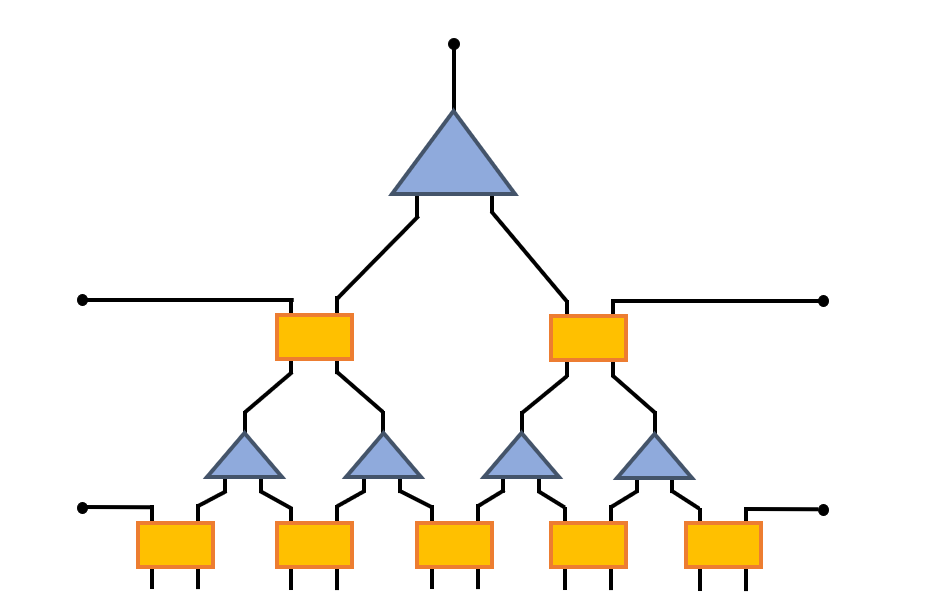}
\caption{Représentation d’un MERA. }
  \label{FRMERA}
\end{figure}

\subsection{Au-delà de l’état de produit de matrices}

Le MPS n’est qu’une forme particulière de réseau de tenseurs pour représenter une fonction d’onde. Dans cette section, nous présentons ici deux formes additionnelles.

Toutes nos discussions jusqu’à présent portaient sur des systèmes où les côtés droit et gauche sont aisément identifiables ({\it i.e.}, système à une dimension), et les MPS fonctionnent mieux lorsque les corrélations diminuent exponentiellement comme dans un système avec gap avec interactions locales. Pour un système bidimensionnel, il peut être avantageux d’utiliser la version bidimensionnelle du MPS appelé \textit{état projeté de paires intriquées} (PEPS, \textit{Projected Entangled Pair State})~\cite{verstraete2004renormalization,verstraete2008matrix}.  Les PEPS (\FRreffig{FRPEPS}) utilisent des tenseurs de rang 5 (avec un indice physique et 4 indices de liens) comme éléments du réseau pour un réseau carré (et un autre réseau aurait un rang différent; par exemple, le réseau ruche d’abeilles aurait un tenseur de rang 4). La taille de la dimension de lien peut être plus petite que dans le cas du calcul d’un MPS sur deux dimensions en choisissant cette forme, mais le rang du tenseur est plus élevé. 

Contracter complètement le réseau de tenseurs dans le PEPS est extrêmement difficile~\cite{valiant1979complexity}.  Il n’y a pas d’algorithme connu qui permet de contracter un tenseur arbitraire efficacement, cependant il existe des algorithmes qui tentent de le faire de la meilleure façon~\cite{pfeifer2014faster,kourtis2019fast}.

Pour une autre approche, rappelons qu’un MPS peut aussi gérer les corrélations sans gap, mais cela requiert de conserver plus de valeurs singulières de la SVD. Nous avons aussi la possibilité d’utiliser un autre réseau de tenseur nommé \textit{approche de renormalisation de l’intrication multi-échelles} (MERA, \textit{multi-scale entanglement renormalization ansatz}) qui est spécifiquement conçu pour représenter les corrélations avec ou sans gap \cite{vidal2007entanglement,vidal2008class}. Un MERA est présenté dans \FRreffig{FRMERA}. Pour voir pourquoi le MERA est meilleur pour les systèmes sans gap, nous référerons à l’appendice~\ref{FRsec:correlation}.

\begin{figure}[b]
  \includegraphics[width=0.9\columnwidth]{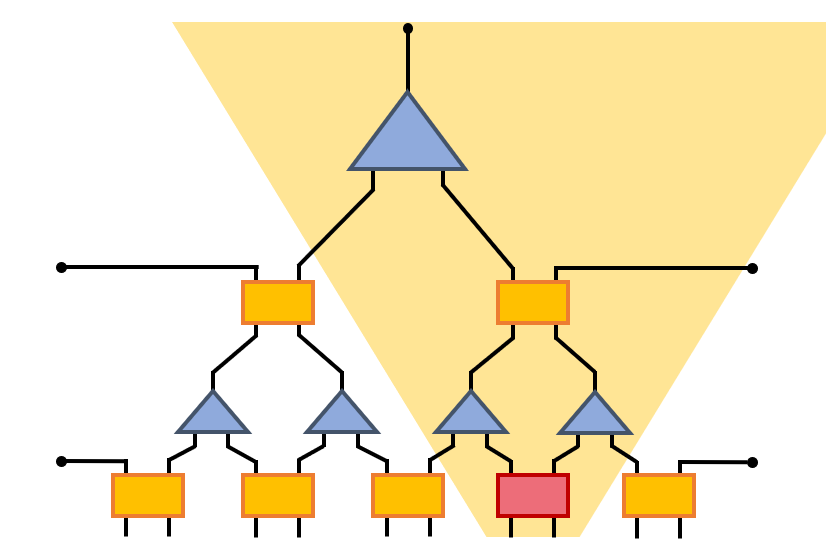}
\caption{Un changement appliqué à un tenseur du MERA affectera seulement les tenseurs avec lesquelles il partage un indice. Donc, tous les tenseurs dans le cône jaune seront affectés.}
  \label{FRlightcone}
\end{figure}

\begin{figure}[t]
  \includegraphics[width=0.8\columnwidth]{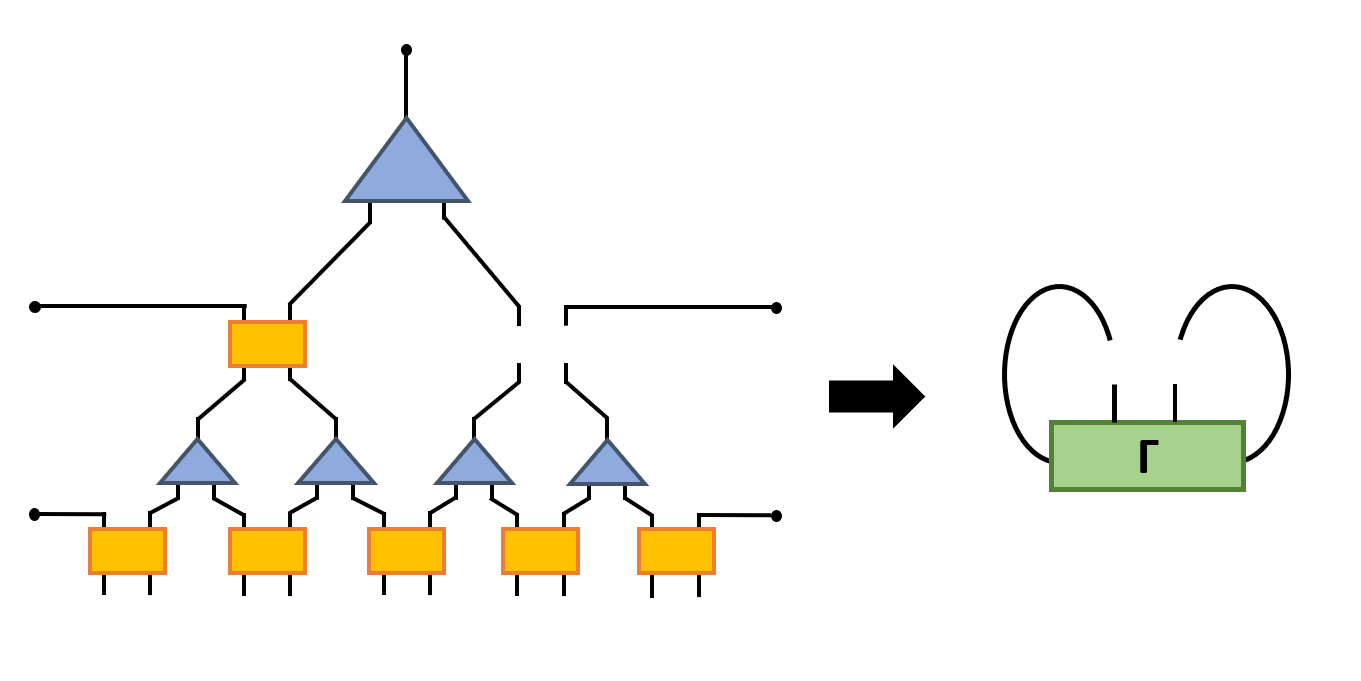}
\caption{Pour entraîner un tenseur du MERA, nous créons un environnement fait de la contraction de tous les autres tenseurs du réseau. Le tenseur résultant sera celui minimisant la trace (voir texte).}
  \label{FRMERAsolve}
\end{figure}

Une propriété du MERA est qu’il possède un cône causal: seuls les tenseurs à l’intérieur de ce cône seront affectés lorsque l’on altère le tenseur situé au sommet. Par exemple, modifier le tenseur unitaire dans \FRreffig{FRlightcone} sur la dernière ligne ne va pas affecter les unitaires à gauche ou à droite. Tous les tenseurs plus hauts dans le réseau dont les indices sont reliés à ceux-ci, seront affectés par le changement. Puisque seulement quelques tenseurs sont altérés par l’inclusion d’un opérateur, cette propriété peut être comparée à la jauge pour le MPS; mais il s’agit d’une propriété différente.

Notons qu’il existe aussi une version bidimensionnelle du MERA~\cite{evenbly2010frustrated}. Des MERA ont aussi été utilisés pour simuler des théories des champs conformes qui peuvent être utiles dans l’étude de l’holographie, par exemple~\cite{swingle2012entanglement}.

Une méthode commune pour déterminer les tenseurs unitaires ou isométriques dans un MERA est de maximiser la trace de l’opérateur quand il est appliqué sur le réseau. Par exemple, nous pouvons enlever l’un des tenseurs du réseau (voir \FRreffig{FRMERAsolve}) et contracter tous les autres indices. Cela produit un tenseur $\bigGamma$. Pour générer le bon tenseur unitaire ($\W$), la condition $\max\mathrm{Tr}\left(\W\bigGamma\right)$ doit être satisfaite. Pour déterminer comment maximiser cette expression exactement, notons que $\bigGamma$ peut être décomposé avec une SVD, $\bigGamma=\U{}\D\V^\dagger$. Si nous choisissons $\W=\V\U{}^\dagger$, alors sous propriété cyclique de la trace, la condition se réduit à la trace sur $\D$. Cela est alors relié à la maximisation de l’intrication du système.

Plus d’information à propos de ces réseaux de tenseurs se trouve dans la réf.~\onlinecite{evenbly2009algorithms,evenbly2016entanglement}.

\section{{Algorithmes}}\label{FRalgorithms}

Pour résoudre un problème de mécanique quantique, les tenseurs d’un réseau sont transformés à l’aide des quatre opérations de base de la \FRrefsec{FRbasicops}. Les différents algorithmes de réseaux de tenseurs sont des utilisations différentes de ses opérations.

\subsection{Évolution temporelle}
Il y a deux types d’algorithmes d’évolution temporelle. La première utilise l’opérateur d’évolution du système $e^{-i\Ham{}t}$, où $t$ est l’intervalle de temps, pour simuler la dynamique d’un état. L’autre utilise l’opérateur d’évolution en temps imaginaire $e^{-\beta\Ham{}}$, où $\beta$ est le temps imaginaire, pour obtenir l’état fondamental. Par application répétée de $e^{-\beta\Ham{}}$ avec une valeur finie de $\beta$ sur un état quelconque, l’état fondamental du système peut être approchée.

Pour appliquer ces opérateurs exponentiels sur un MPS, il faut procéder à une factorisation qui nous permet d’agir localement sur le MPS. Il n’existe malheureusement pas d’identité exacte permettant d’accomplir ceci pour les exponentielles de matrice. En pratique, nous utilisons l’identité de Trotter-Suzuki \cite{suzuki1985decomposition} pour effectuer la décomposition de façon approximative:
\begin{equation}\label{FRBCH}
e^{-i(\Ham{}_A+\Ham{}_B)t}\approx e^{-i\Ham{}_At}e^{-i\Ham{}_Bt}.
\end{equation}
Cette expression commet une erreur de l’ordre de $t^2[\Ham{}_A,\Ham{}_B]$. L’approximation est donc bonne pour des temps suffisamment courts. Une solution pour un temps $Nt$ peut être obtenue en appliquant l’expression $N$ fois sur un état.

L’évolution temporelle compressée (TEBD) est une méthode pour simuler l’évolution temporelle d’un MPS \cite{vidal2003efficient,vidal2004efficient}. L’algorithme procède ainsi:
\begin{enumerate}
  \item Nous appliquons au centre d’orthogonalité ainsi qu’à ses voisins pertinents tous les termes de l’exponentielle qui agissent sur le centre d’orthogonalité. Suite à cette étape, le centre et les voisins sur lesquels nous venons d’agir sont combinés en un seul tenseur de rang supérieur.

 \includegraphics[width=0.9\columnwidth]{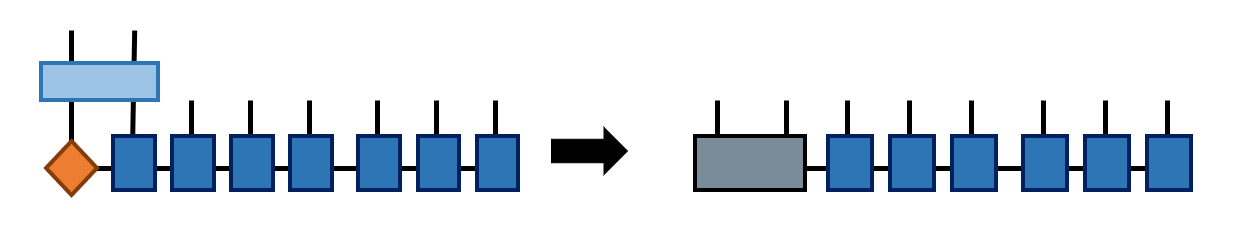}
  \item Nous effectuons une SVD pour extraire le tenseur du site qui était le centre, et nous déplaçons le centre d’orthogonalité au premier site voisin.
  \item Nous recommençons l’étape 1 jusqu’à ce qu’on atteigne l'extrémité du MPS, en s’assurant d’appliquer une seule fois tous les termes de l’opérateur d’évolution.
\end{enumerate}
  \includegraphics[width=\columnwidth]{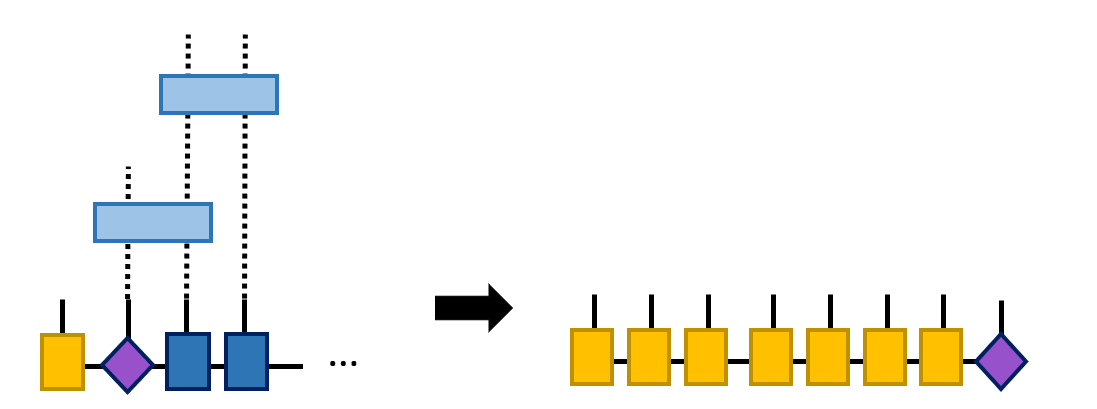}
Une fois l'extrémité du MPS atteinte, un pas temporel est effectué. Pour faire un autre pas, nous recommençons dans la direction inverse.

\subsection{Groupe de renormalisation de tenseurs }

\paragraph{Système classique.--} Les réseaux de tenseurs peuvent aussi être utilisés pour résoudre des problèmes de physique statistique classique. Ces systèmes sont doublement intéressants à cause du principe de correspondance\cite{bohr1920serienspektra} qui nous dit qu’un système quantique en $d$ dimensions a un analogue classique en $d+1$ dimensions.

\paragraph{Groupe de renormalisation.--} Kadanoff a introduit en 1966 une méthode de renormalisation pour les systèmes de spins \cite{kadanoff1966scaling}. Le principe de cette méthode est de regrouper les spins en domaines. Ces domaines forment alors les sites d’un réseau avec un pas plus grand. En effectuant cette opération de renormalisation à répétition sur un système, nous atteignons éventuellement un point fixe qui révèle des caractéristiques importantes du système. Les algorithmes de renormalisation de tenseur ont la même propriété de créer une séquence de réseaux avec des pas de plus en plus grands, et l’état des domaines est filtré avec une décomposition aux valeurs singulières.

\paragraph{Mise en place du problème.--}
Le point de départ des algorithmes de réseaux de tenseurs pour les systèmes classiques est la fonction de partition $\mathcal{Z}$ sous la forme d’un réseau de tenseurs locaux à l’échelle microscopique:
  \includegraphics[width=\columnwidth]{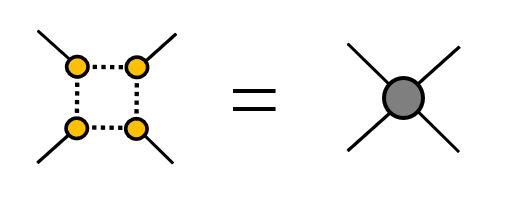}

Par exemple, la contribution d’un groupe de quatre sites formant une plaquette à la fonction de partition du modèle d'Ising sur un réseau carré est:
\begin{equation}
\mathcal{Z}_{ijk\ell}=e^{-\beta\Ham{}_{ijk\ell}}=\exp\Big(\beta J(\sigma_i\sigma_j+\sigma_j\sigma_k+\sigma_k\sigma_\ell+\sigma_\ell\sigma_i)\Big)
\end{equation}
où $\sigma_i=\pm1$ est le spin au site $i$. Chacun des éléments de ce tenseur correspond à une configuration différente de la plaquette. La fonction de partition est alors donnée par la somme sur les configurations du produit de chacun de ces tenseurs locaux: c’est la contraction d’un réseau de tenseurs. Effectuer la contraction de ce réseau exactement n’est possible que pour de très petits systèmes. Pour les autres, il faut faire appel aux idées du groupe de renormalisation pour extraire l’information la plus pertinente à chaque échelle.

\paragraph{Algorithme.--}
Le groupe de renormalisation de tenseurs (TRG) \cite{levin2007tensor} est un exemple de mise en application des idées du groupe de normalisation aux réseaux de tenseurs. Les tenseurs locaux sont contractés, formant un réseau avec un pas plus grand, et une décomposition aux valeurs singulières est utilisée pour identifier les configurations les plus pertinentes et rejeter les autres. Nous répétons ensuite ces étapes.

La première étape est d’effectuer une séquence de SVD pour séparer le tenseur initial:

\includegraphics[width=\columnwidth]{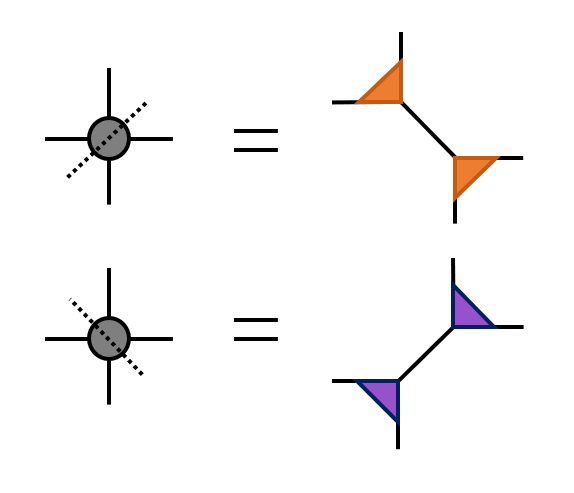}

La matrice $\D$ de chacun des liens est factorisée $\sqrt{\D}\cdot\sqrt{\D}$ et une racine est contractée avec les tenseurs $\U{}$ et $\V^\dagger$ respectivement.

\includegraphics[width=\columnwidth]{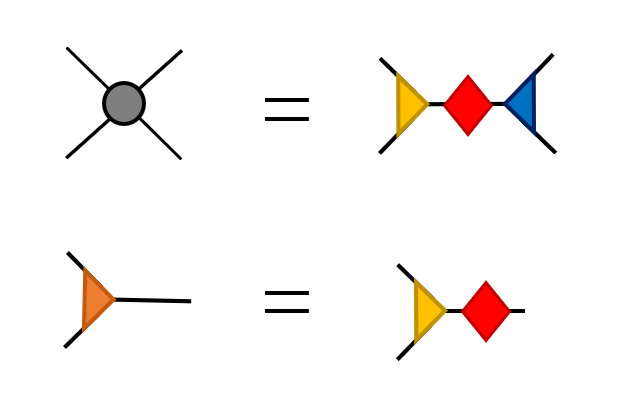}

Après avoir décomposé un site, les tenseurs sont contractés de façon a former un réseau à plus grande échelle.

\includegraphics[width=\columnwidth]{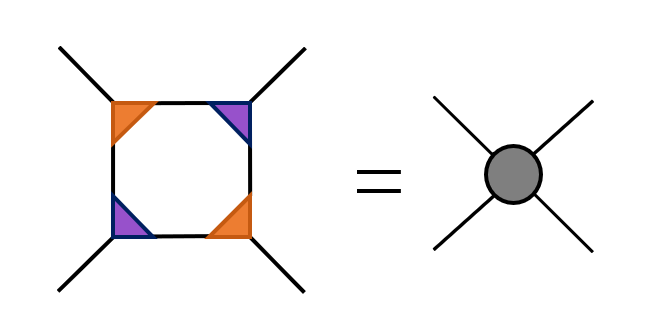}

On contracte les tenseurs avec le patron précédent parce que l’on exploite l’invariance par translation du système. La figure ci-dessous montre les tenseurs du réseau et que leurs décompositions mènent bien au patron de contraction montré précédemment:

\includegraphics[width=0.95\columnwidth]{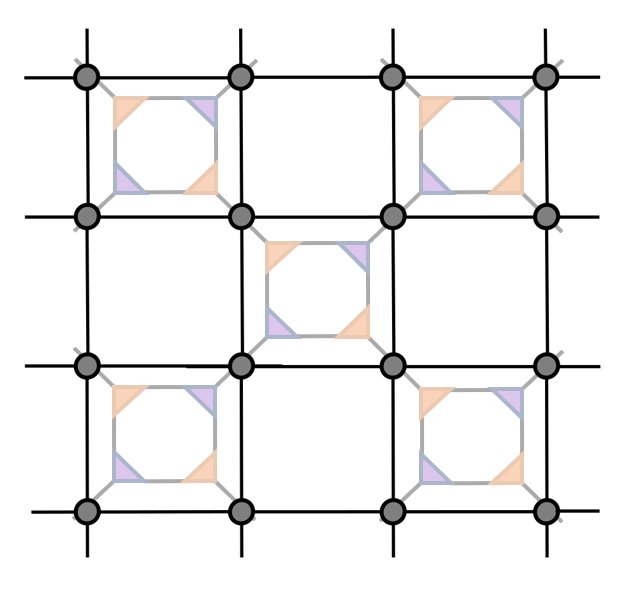}

Le site à l’échelle supérieure est formé d’une plaquette de quatre tenseurs. À chaque étape de renormalisation, nous divisons le nombre de tenseurs présent dans le réseau par quatre. Cet algorithme peut être utilisé pour des systèmes finis ou infinis, et pour d’autres problèmes que les spins d’Ising sur un réseau carré.
D’autres algorithmes, telle la méthode de renormalisation de réseau de tenseurs (TNR) \cite{evenbly2015tensor,evenbly2015tensorB,evenbly2017algorithms}, sont en mesure de résoudre ces systèmes bidimensionnels classiques. La TNR est un raffinement de la TRG qui parvient à résoudre la fonction de partition avec une meilleure précision.

\section{Autres applications des réseaux de tenseurs}

Nous avons porté une attention particulière aux méthodes pour lesquelles il est nécessaire de considérer l’intrication et les interactions entre les sites. Il existe plusieurs autres applications de ces méthodes.

Par exemple, en informatique quantique, il est possible de représenter les portes logiques agissant sur les qubits à l’aide de diagrammes~\cite{nielsen2002quantum,bravyi2014efficient,darmawan2017tensor,tremblay2018depth}. Ces portes peuvent générer de l’intrication entre les qubits. Notons cependant qu’il n’est pas d’usage courant de faire la renormalisation en tronquant certaines valeurs dans la matrice densité. Les opérations présentées à la \FRrefsec{FRsec:svd} ne sont pas nécessaires puisque l’ensemble des portes logiques sont unitaires.

Pour les problèmes où il n’est pas nécessaire de considérer l’intrication pour effectuer la renormalisation, il est parfois possible d’utiliser la notation graphique présentée ci-haut. Entre autres, il est possible d’utiliser les réseaux de tenseurs en apprentissage automatique (réseaux de neurones)\cite{novikov2016exponential,stoudenmire2016supervised,evenbly2019number} ou en compression d'image \cite{evenbly2018representation}. La représentation graphique est également utile pour illustrer rapidement un problème. Notons que pour les méthodes que nous avons présentées dans cette revue, il est important de considérer l’intrication entre les site, alors que ce n’est pas toujours le cas pour d’autres contextes.

\section{Conclusion}

Les méthodes de réseaux de tenseurs sont applicables dans plusieurs situations et convergent rapidement pour des problèmes. Bien comprendre comment utiliser ces outils permet de déterminer de manière efficace les propriétés d’un système quantique ou classique. Une représentation graphique est utilisée et cette dernière permet d’appliquer les réseaux de tenseurs à d’autres domaines.

Nous avons effectué un survol des opérations de base des réseaux de tenseurs: le remodelage, la permutation d’indices, la contraction et la décomposition. De plus, nous avons montré le lien entre la décomposition en valeurs singulières et l’intrication entre deux parties d’un réseau. Pour modéliser une fonction d’onde, nous avons introduit l’état de produit de matrices, qui représente efficacement un système avec un gap. La représentation en réseau de tenseurs d’un hamiltonien, l'opérateur de produit de matrices, peut être utilisée de pair avec cette représentation de la fonction d’onde. Finalement, nous avons présenté des algorithmes, dont l’évolution temporelle compressée et le groupe de renormalisation des tenseurs, qui utilisent ces outils pour trouver l’état fondamental d’un système quantique et il est également possible de le faire pour des systèmes classiques.

\section{{Remerciements}}

S.D. tient à remercier l’Institut Transdisciplinaire d’Information Quantique (INTRIQ) pour son soutien financier. T.E.B. est récipiendaire d'une bourse postdoctorale de l’Institut quantique et de l'INTRIQ.  Cette recherche a été entreprise en partie grâce au financement du Fond d’excellence en recherche Apogée Canada.

Les auteurs remercient pour leurs commentaires Sarah Labbé, Sara Turcotte, Agustin Di Paolo, Sarah E.~Grefe, Martin Schnee, Maxime Dion, Colin Trout, Jessica Lemieux, Alexandre Prémont-Foley, Prosper Reulet, Yunlong Lian, Catherine Leroux, Thomas Gobeil, David Poulin, David Sénéchal, Andreas Bill, Yves Bérubé-Lauzière, Yves Grosdidier et Glen Evenbly.

\begin{appendix}

\section{Plus de détails sur la différence entre un MPS et un MERA}\label{FRsec:correlation}

Le concept d'état fondamental quantique avec intrication peut être plus généralement reformulé en termes de corrélations. Le concept de corrélation est très important en physique. Puisque la physique est l’étude des conséquences de l’interaction entre différents corps, étudier les corrélations (ou ce qui arrive aux autres parties d'un système lorsqu’une partie est perturbée) est essentiellement l’étude de la physique elle-même. Les corrélations peuvent apparaître dans plusieurs contextes différents, mais nous allons étendre ce concept ici pour illustrer autrement les différences entre le MPS et le MERA.

\begin{figure}[b]
  \includegraphics[width=\columnwidth]{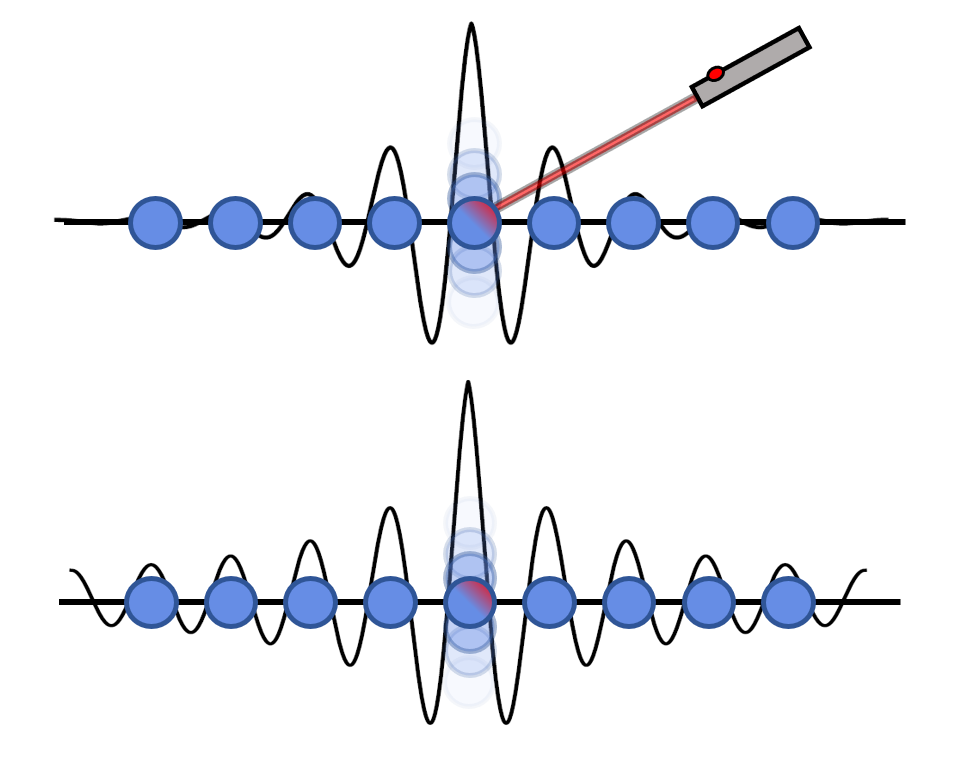}
\caption{Représentation de la décroissance des corrélations après une perturbation pour un système gap (haut) et sans gap (bas). Les corrélations sans gap s’étendent plus loin que la décroissance exponentielle des interactions avec gap.}
  \label{FRperturbe}
\end{figure}

Lorsqu’un système est perturbé, la perturbation peut affecter d’autres sites éloignés (voir \FRreffig{FRperturbe}). L'effet dépend en général de la distance et il n’y a que deux types de corrélations qui peuvent être présentes dans un système (voir \cite{hastings2004locality} pour une preuve mathématique). Si le système a un gap entre la première et la seconde valeur propre, alors les corrélations vont décroître exponentiellement.  Si les valeurs propres n’ont pas de gap entre les deux premiers états (ou si les valeurs propres sont vraiment très proches les unes des autres), alors les corrélations décroissent comme une loi de puissance. En résumé, une corrélation $\mathcal{C}$ peut prendre l’une des formes suivantes:
\begin{equation}\label{FRcorrelations}
\mathcal{C}\sim\begin{cases}
\exp(-x/\xi)&\mathrm{avec\;gap}\\
(x/\xi)^\gamma&\mathrm{sans\;gap.}\\
\end{cases}
\end{equation}

\begin{figure}[t]
  \includegraphics[width=\columnwidth]{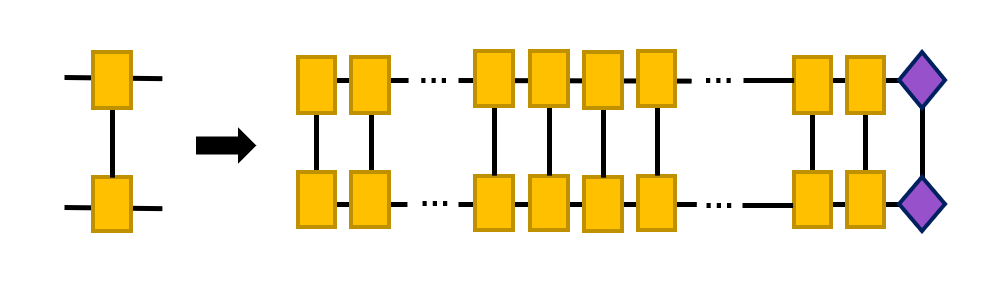}
\caption{Une série de matrices de transfert.}
  \label{FRtransfermatrix}
\end{figure}

En formant la matrice de transfert d’un MPS avec son conjugué hermitien comme dans la \FRreffig{FRtransfermatrix}, nous pouvons contracter une chaîne de ces opérateurs, en supposant une chaîne suffisamment longue. Il est bien connu que multiplier des matrices en succession peut révéler la plus petite valeur propre. Par exemple, chaque matrice de transfert $T$ peut être décomposée en une décomposition en valeurs propres $\U\bigLambda \U^\dagger$. Un produit de celles-ci peut donner 
\begin{equation}\label{FRtransfers}
\prod_{i=1}^N T_i=\left(\U{}\bigLambda \U{}^\dagger\right)\left(\U{}\bigLambda \U{}^\dagger\right)\left(\U{}\bigLambda \U{}^\dagger\right)\ldots=\U{}\bigLambda^N\U{}^\dagger.
\end{equation}
La matrice $\bigLambda$ est diagonale et contient les valeurs propres où
\begin{equation}\label{FRlambdaN}
\bigLambda^N=\left(\begin{array}{cccc}
\varepsilon_1^N&0&0&\cdots\\
0&\varepsilon_2^N&0&\cdots\\
0&0&\varepsilon_3^N&\ddots\\
\vdots&\vdots&\ddots&\ddots
\end{array}\right).
\end{equation}
S’il y a un gap entre les valeurs propres, la plus grande valeur propre, en valeur absolue, ($\varepsilon_1$) va apparaître beaucoup plus grande que toutes les autres lorsqu’elle est élevée à la puissance $N$.  Par conséquent, la seule valeur dans \FRrefeq{FRlambdaN} qui est non-négligeable est $\varepsilon_1$.

La série de matrices de transfert définie dans \FRrefeq{FRtransfers} peut transporter l’information d’une perturbation d’un site à un autre situé $N$ sites plus loin. Il est alors raisonnable de se demander comment les deux sites sont corrélés. Puisque nous savons que la structure de la corrélation est donnée par \FRrefeq{FRcorrelations}, nous voyons qu’une corrélation avec un gap peut être réécrite
\begin{equation}
\exp\left(-\frac{N}{{\ln(\varepsilon_1)^{-1}}}\right)
\end{equation}
d’où l’on peut clairement identifier la longueur de corrélation $\xi^{-1}=\ln(\varepsilon_1)$.  Cela montre comment la plus petite valeur propre est caractérisée par un MPS. Un système sans gap donnerait une somme de termes exponentiels qui deviendrait une loi de puissance.

Si nous formons une matrice de transfert pour un MERA, nous pouvons le faire à n’importe quel niveau. Si nous choisissons un bon niveau, alors nous obtenons le même résultat que pour un MPS. Cependant, la matrice de transfert équivalente formée au premier niveau du MERA s’étend sur 4 sites plutôt que 2, le prochain niveau s’étend sur 8 sites, etc. De cette manière, le MERA permet d’exposer des corrélations sur différentes échelles de longueur. Ainsi, grâce à cette approche, nous pouvons encoder des corrélations avec une portée beaucoup plus grande, incluant celle d’un système sans gap que le MERA peut très bien gérer.

\section{Produit de Kronecker et somme directe}\label{FRkronecker}
\hspace{-1cm}

Le produit de Kronecker entre deux tenseurs, $A \otimes B$, est défini comme la multiplication de tous les éléments du tenseur $B$ avec tous les éléments du tenseur $A$. Par exemple,
\begin{eqnarray}
\A\otimes \B&=&\left(\begin{array}{cc}
a_{11}&a_{12}\\
a_{21}&a_{22}
\end{array}\right)\otimes\left(\begin{array}{cc}
b_{11}&b_{12}\\
b_{21}&b_{22}
\end{array}\right)\\
&=&\left(\begin{array}{cc}
a_{11}\B&a_{12}\B\\
a_{21}\B&a_{22}\B
\end{array}\right)\\
&=&\left(\begin{array}{cccc}
a_{11}b_{11} & a_{11}b_{12} & a_{12}b_{11} & a_{12}b_{12}\\
a_{11}b_{21} & a_{11}b_{22} & a_{12}b_{21} & a_{12}b_{22}\\
a_{21}b_{11} & a_{21}b_{12} & a_{22}b_{11} & a_{22}b_{12}\\
a_{21}b_{21} & a_{21}b_{22} & a_{22}b_{21} & a_{22}b_{22}\\
\end{array}\right)
\end{eqnarray}

Le produit de Kronecker peut être considéré comme la cinquième opération sur les tenseurs. Cependant, ce dernier est rarement utilisé dans les algorithmes et n’a pas été inclus dans la liste de la section \FRrefsec{FRbasicops}. Un raisonnement similaire peut être utilisé pour d’autres opérations comme la somme directe $\A \oplus \B$, définie pour deux matrices $\A$ et $\B$ comme
\begin{equation}
  \A\oplus \B=\left(\begin{array}{cc}
  \A&\0\\
  \0&\B
  \end{array}\right).
\end{equation}

\section{Exemple de code pour la construction d'un état de produit matrix}\label{FRMPScode}

Ici, nous fournissons un code simple permettant d’obtenir sous forme de MPS une fonction d’onde quelconque. Ce code est rédigé dans le langage de programmation Julia.

\begin{widetext}
\begin{lstlisting}
#############################################################
#
#  Conversion de fonctions d'onde en états de produit matriciel
#
#############################################################
# Fait par T.E. Baker, S. Desrosiers, M. Tremblay et M.P. Thompson (2019)
# Voir la licence accompagnant ce programme
# Ce code est natif du langage de programmation Julia (v1.1.1)
#

using LinearAlgebra

# Création d'un vecteur d'état normalisé aléatoire de taille N
physInd = 2 # Taille de l'indice physique
vect = rand(ComplexF64, physInd^N, 1) # Initialisation d'un état aléatoire
vect /= norm(vect) # Normalisation de la fonction d'onde

function makeMPS(vect,physInd,N)
  mps = []
  Lindsize = 1 # Taille actuelle de l'indice de gauche
  
  # Vecteur remodelé isolant le premier indice de taille 2
  M = reshape(vect, physInd, physInd^(N-1))

  # Boucle de construction du MPS 
  for i=1:N-1
    U, D, V = svd(M) # Application de la SVD
    temp = reshape(U,Lindsize,physInd,size(D,1))
    push!(mps, temp)
    Lindsize = size(D,1)
    DV = Diagonal(D)*V'
    if i == N-1
      temp = reshape(DV,Lindsize[1],physInd,1)
      push!(mps,temp)
    else
      Rsize = cld(size(M,2),physInd) #Division entière, arrondit
      M = reshape(DV,size(D,1)*physInd,Rsize)
    end
  end
  return mps
end

mps = makeMPS(vect,physInd,N)
\end{lstlisting}
\end{widetext}

\end{appendix}
  
\end{document}


\title{Basic tensor network computations in physics}
\author{Thomas E.~Baker}
\affiliation{Institut quantique \& Département de physique, Université de Sherbrooke, Sherbrooke, Québec J1K 2R1 Canada}
\author{Samuel Desrosiers}
\affiliation{Institut quantique \& Département de physique, Université de Sherbrooke, Sherbrooke, Québec J1K 2R1 Canada}
\author{Maxime Tremblay}
\affiliation{Institut quantique \& Département de physique, Université de Sherbrooke, Sherbrooke, Québec J1K 2R1 Canada}
\author{Martin P.~Thompson}
\affiliation{Institut quantique \& Département de physique, Université de Sherbrooke, Sherbrooke, Québec J1K 2R1 Canada}

\begin{abstract}
This article is intended as an overview of tensor networks for beginners looking to perform computations. We focus on the tools necessary for the concrete implementation of algorithms. Four basic operations (reshaping, permutation, contraction, and decomposition) that are commonly used in tensor network algorithms are covered. The article also briefly reviews diagrammatic notation, entanglement, matrix product states (MPS), matrix product operators (MPO), projected entangled pair states (PEPS), multi-scale entanglement renormalization ansatz (MERA), time evolution block decimation (TEBD), and tensor renormalization group (TRG).
\end{abstract}
\maketitle

\vspace{0.25cm}

\section{Introduction}
\label{ENintro}

Exact methods of solving quantum systems are difficult to apply to large scale problems. It is necessary to use approximate methods, and tensor networks are among the most widely used methods for this purpose.  Tensor network methods are based on the separation of a quantum wave function into a series of sub-parts, one for each site of the studied system. The first notable example of such a separation is the transfer matrix approach which can be used to solve the Ising  model~\cite{ising1925beitrag,onsager1944crystal,reif2009fundamentals}. 

To solve large systems using tensor networks, we must consider effective representations of a wave function. To obtain such a representation, we truncate the tensors of the tensor networks to reduce the number of degrees of freedom and keep only the most relevant ones~\cite{schollwock2005density,schollwock2011density,orus2014practical,bridgeman2017hand}. This approach is directly linked to the renormalization group~\cite{wilson1975renormalization} and is based on the entanglement between the sites. These tensor network methods are applicable to classical or quantum systems.

The modern formulation of these problems is based on Matrix Product States (MPS) \cite{schollwock2011density}. One of the first examples of such an object was the Affleck-Kennedy-Lieb-Tasaki (AKLT) state, used to describe a spin system ~\cite{affleck2004rigorous,schollwock2005density}. Since then, several algorithms have been developed, including the Density Matrix Renormalization Group (DMRG) ~\cite{white1992density}.

In this introduction to tensor networks, we focus on the basic operations required to manipulate tensors. In \ENrefsec{ENwhytensors}, we start with a discussion of what tensors are. In \ENrefsec{ENdiagrams}, we introduce a diagrammatic notation that simplifies the description of the tensor networks. In \ENrefsec{ENbasicops}, we present four basic operations that apply to tensors. In \ENrefsec{ENentangle}, we discuss the connection between information theory and entanglement properties. In \ENrefsec{ENsec:MPS}, we summarize the most common wave function ansatzs. Finally, in the \ENrefsec{ENalgorithms}, we present two examples of algorithms that use previous concepts to find the ground state of a system.

\section{Why Tensors?}
\label{ENwhytensors}

For a given quantum physics problem, we must solve the Schrödinger equation,
\begin{equation}
  \Ham{}\psi = E\psi,
\end{equation}
for a Hamiltonian $\Ham{}$, an energy, $E$, as well as a wave function $\psi$.  The main objective is to find the fundamental energy and the ground state of the system, although we can also find excitations, time dependence, and non-equilibrium properties.

\begin{figure}[b]
  \includegraphics[width=0.75\columnwidth]{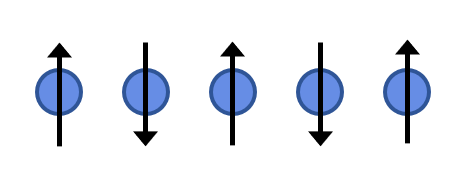}
  \caption{
    A spin chain is one of the most commonly used examples to illustrate tensor network methods. The arrows represent the orientation of the spin vectors that, in a real system, can be superimposed.
  }
  \label{ENspinsys}
\end{figure}

Finding the exact solution for such a system is extremely difficult. Standard methods that are presented in an introductory course in quantum physics quickly become impractical for more complex systems. For example, the most direct way to find the fundamental state of a Hamiltonian is by exact diagonalization. To do this, we represent the Hamiltonian as a matrix and use numerical methods to find the eigenvalues. This method is very expensive in terms of memory and computing space, so it is most commonly used for small systems.

To illustrate the space needed to represent a Hamiltonian on a computer, we consider a chain of spins (\ENreffig{ENspinsys}). These systems are used to model atoms whose nuclei are fixed in space. Electrons can experience exchange interactions, even if they are localized and the atoms are distant. A common model is the Ising model~\cite{reif2009fundamentals}, representing a system of $N$ spins
\begin{equation}
  \label{ENHamIsing}
  \Ham{} = -J \sum_{i = 1}^{N - 1} \Spin{}^z_i \cdot \Spin{}^z_{i + 1}.
\end{equation}
Adjacent spins interact according to their $z$-component and, in general, it is possible to have any amplitude for the spins $(s=\frac12,1,\frac32,\ldots)$. The constant $J$ represents the exchange interaction between the spins. Another common example Hamiltonian is the Heisenberg model~\cite{townsend2000modern},
\begin{equation}
  \Ham{}=-J\sum_{i=1}^{N-1}\mathbf{S}_i\cdot \mathbf{S}_{i+1},
\end{equation}
where $\mathbf{S}=( \Spin{}^x,\Spin{}^y,\Spin{}^z)$, a vector containing spin matrices~\cite{townsend2000modern}. Thus, there are $2s+1$ quantum states available for this spin and $(2s+1)^N$ for a network of $N$ sites. Since the Hamiltonian represents the interaction between these spins, we can represent it as a square matrix of dimension $(2s+1)^{N} \times (2s+1)^{N}$. The number of Hamiltonian components increases exponentially with the size of the system.  The upper limit for exact diagonalization is approximately 50 sites at present \cite{wietek2018sublattice}.

There are algorithms to find only the smallest eigenvalues of $\Ham{}$, such as the Lanczos algorithm~\cite{press1989numerical}. However, these methods are again limited by the representation in memory of the wave function.  

We want to break down a quantum system into sub-parts in order to reduce the complexity of solving the problem. Each of these sub-parts is represented by a tensor. For this review, we consider a tensor as the generalization of a matrix, ({\it i.e.}, as a multi-dimensional array of complex numbers\footnote{For a formal definition of tensors, see Ref.~\onlinecite{boas2006mathematical}.}).  We will then use tensor networks that have a linear growth with the number of sites in the model and solve systems with several thousand sites. Once the quantum system is properly decomposed, we can design algorithms that act on a sub-part of the system at each step. It is then possible to obtain the global solution after several iterations. 
In addition to growing in reasonable complexity with the size of the system, tensor network methods also have the advantage of not having sign problems~\cite{troyer2005computational} that are common with quantum Monte Carlo algorithms. However, there are some limitations to the use of tensor networks. In particular, the latter perform much better when interactions and correlations are short.

\section{Diagram Representation}
\label{ENdiagrams}

\begin{figure}[t]
  \includegraphics[width=\columnwidth]{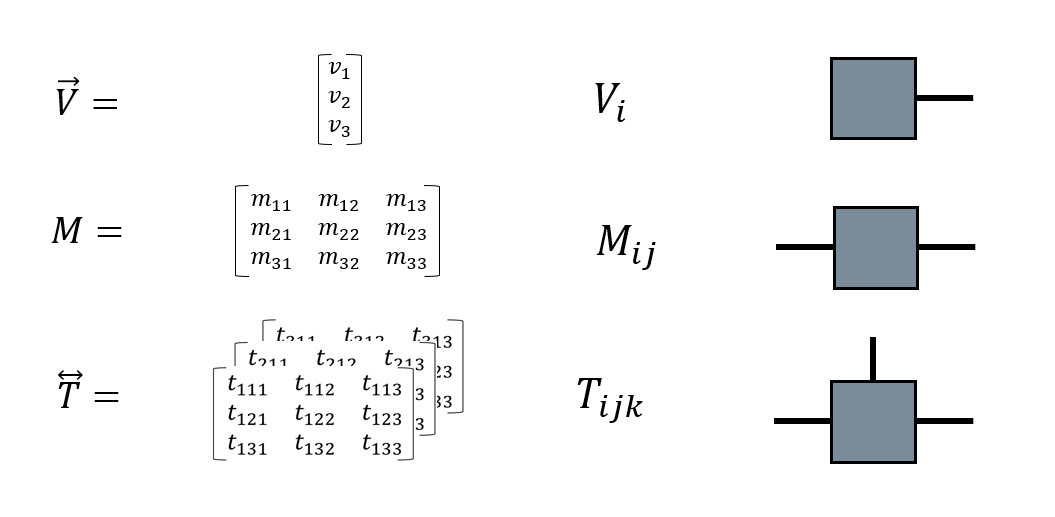}
\caption{A tensor of rank one (vector, top), two (matrix, middle) and three (bottom), are represented respectively by all their elements, by the Einstein index notation and by the diagram representation.}
  \label{ENrepresent}
\end{figure}

A tensor can be classified by its rank. The rank of a tensor corresponds to the number of indices it has. For example, a scalar is a tensor of rank 0, a vector is a tensor of rank 1 and a matrix is a tensor of rank 2. We denote a tensor $T$ of rank 3 as with the indices $\mu$, $\nu$, and $\eta$:
\begin{equation}
  \label{ENtensor}
T_{\mu\nu\eta}.
\end{equation}
Each index corresponds to a dimension of the tensor and can take different values\footnote{Indices in upper and lower positions are sometimes used to differentiate covariant and contravariant tensors. However, this distinction is generally not necessary for tensor networks.}.

Writing tensors in this way with all their indices can become very complicated as part of a complete tensor network. To avoid this, a diagram representation is introduced as in Ref.~\onlinecite{penrose1971angular}. In the \ENreffig{ENrepresent}, tensors of rank one, two, and three are represented with the corresponding diagrams. It can be seen that the different tensors are represented by a shape from which lines emerge, each representing one of the indices. In general, lines are not explicitly identified by the indices they represent in a diagram. This information may be provided arbitrarily. 

\begin{figure}[t]
  \includegraphics[width=0.5\columnwidth]{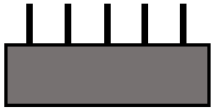}
  \caption{Diagrammatic representation of a tensor of rank 5.}
  \label{ENwave functionDiagram}
\end{figure}

When describing a wave function, $\psi_{\sigma_1\sigma_2\ldots}$, by a tensor, the vertical lines of the diagram take on a precise meaning according to the convention. Each line represents an index ($\sigma_i$) associated with the spin of each site in the system. For example, a tensor of rank five, equivalent to the wave function of a system of five sites, is represented with five vertical lines as in the \ENreffig{ENwave functionDiagram}.

In the following sections, we present the basic elements of tensor networks. These elements are used in the libraries for tensor network computation~\cite{DMRjulia,tensorsnet,pfeifer2014ncon,stoudenmireitensor,Xtensor,tenpy,alvarez0209,al2017tensor,kao2015uni10,chanblock,chanPySCF,evoMPS,simpleDMRG,bauer2011alps,de2008density} and are useful for most algorithms.

\section{Basic operations}\label{ENbasicops}

Just as we can apply certain operations (additions, multiplication, etc.) for scalars, we can apply certain operations to tensors. The operations introduced here are to be considered as the minimum essential operations for understanding tensor networks and the algorithmic methods using them. Some combinations of these operations appear in each tensor network algorithm. All these operations are generally available and easy to use in high-level programming languages such as Python and Julia.

We present four operations. The first two (reshaping and permutation) apply to a single tensor. The other two operations (contraction and decomposition) involve more than one tensor.

\subsection{Reshaping}

\begin{figure}[b]
  \includegraphics[width=0.9\columnwidth]{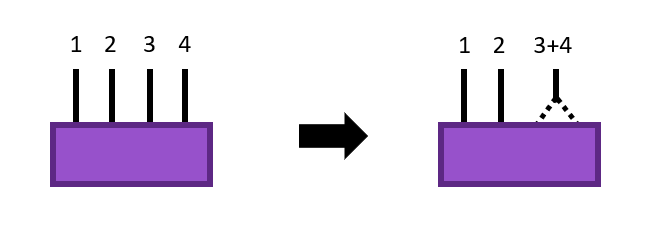}
  \caption{Reshaping a tensor of rank four where the third and fourth indices are merged.}
  \label{ENreshape}
\end{figure}

We can change the rank of a tensor by grouping or ungrouping some of its indices. For example, the rank-3 tensor of the \ENrefeq{ENtensor} can be reshaped by combining the $\mu$ and $\nu$ indices to obtain the rank 2 tensor,
\begin{equation}\label{ENreshapeTensor}
T_{\gamma\eta}
\end{equation}
where $\gamma=(\mu\nu)$. A tensor can be reshaped to join any combination of indices. In \ENreffig{ENreshape}, we illustrate a tensor of rank four reshaped into a tensor of rank three.

To reshape a tensor, a rule must first be defined. An example of a rule is shown in the Table~\ref{ENtabreshape} for a tensor of rank three, of dimensions $(w_x,w_y,w_z)$ reshaped into a vector of size $w_x\times w_y\times w_z$. The first index is chosen to be the first incremented index. As we increment, a counter (left column) identifies the position in the resulting vector.

In general, we can reshape a tensor into another tensor of any rank. The only necessary condition is that these two tensors must contain the same number of elements. For example, we can reshape a tensor of rank 3 with dimensions $(10, 5, 20)$ into a tensor of rank 5 with dimensions $(2, 5, 5, 10, 2)$. We can also reshape this tensor as a tensor of rank 100, with the dimensions ($1000, 1,1,1,1,1,1,1,1,1,1,\ldots$), as it always contains the same number of elements.

Reshaping is a mathematical operation that links tensors of different ranks. However, the resulting tensor indices will have a physical meaning when representing a quantum system. The reshape operation allows us to group basis states on different sites into a group on the left and another group on the right. A point we will return to in later sections.

\begin{table}
\begin{tabular}{| c || c | c | c |}
\hline
Index & $x$ & $y$ & $z$\\
\hline
1 & 1 & 1 & 1\\
2 & 2 & 1 & 1\\
3 & 3 & 1 & 1\\
\vdots & \vdots & \vdots & \vdots\\
$w_x$ & $w_x$ & 1 & 1\\
$1+w_x$ & 1 & 2 & 1\\
$2+w_x$ & 2 & 2 & 1\\
$3+w_x$ & 3 & 2 & 1\\
\vdots & \vdots & \vdots & \vdots\\
$2w_x$ & $w_x$ & 2 & 1\\
$1+2w_x$ &1 & 3 & 1\\
\vdots & \vdots & \vdots & \vdots\\
$w_x+w_x(w_y-1)$ & $w_x$ & $w_y$ & 1\\
$1+w_xw_y(w_z-1)$ & 1 & 1 & $w_z$\\
\vdots & \vdots & \vdots & \vdots\\
$w_x+w_x\Big((w_y-1)+w_y(w_z-1)\Big)$ & $w_x$ & $w_y$ & $w_z$\\
\hline
\end{tabular}
\caption{Table representing the reshape of a tensor of rank 3 of each value with dimensions $(w_x,w_y,w_z)$ into a vector (a tensor of rank 1). The values on the left refer to the index in the vector. The different triplets of values on the right are the values of the indices of the tensor of rank three referring to the same elements. We first refer to the elements of the initial tensor according to a certain iteration rule on its indices (here, starting from the first, $w_x$). The reshape is always reversible. The index value of the vector for a certain element according to its coordinates $(x,y,z)$ is given by $x+w_x((y-1)+w_y(z-1))$.}
\label{ENtabreshape}
\end{table}

The cost to reshape a tensor should be considered as a free operation. The reason for this is that a computer will store the tensor as a vector instead of a more complicated data structure. Stored with the elements of the tensor contained is another meta-data vector with the dimensions of each index of the tensor. Only the accompanying meta-data is changed on reshape. In general, there are many ways to reshape a vector, but should reinforce that reshaping is a very inexpensive operation.

\subsection{Permute Dimensions}

Note that in the reshape of the \ENrefeq{ENreshapeTensor}, we could not reshape the $\mu$ and $\eta$ indices together since the $\nu$ index is inbetween them. According to the rule defined in the Table \ref{ENtabreshape}, during a reshape, the elements of a tensor are referred according to the order of the indices (we consider all values of $x$ before varying $y$ to re-order the elements). Generally speaking, we can only reshape a tensor by operating on neighboring indices unless we change the order of indices ({\it i.e. }, $T_{\mu\nu\eta}\rightarrow T_{\mu\eta\nu}$)  first.   

\begin{figure}[t]
  \includegraphics[width=0.9\columnwidth]{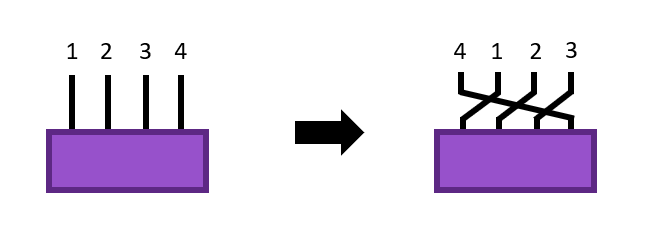}
  \caption{A permutation of indices for a tensor of rank 4 where the fourth index becomes the first.}
  \label{ENpermute}
\end{figure}

Analytically, the permutation of indices does not require any calculational effort. We can simply interchange the indices as in the \ENreffig{ENpermute}. However, in practical applications and algorithms, exchanging the memory allocation to change the order of the indices requires a special operation. Permuting dimensions consists of creating a new tensor of the required format in which the information from the first tensor will be copied, following the new order of indices. Generally, it is recommended to avoid doing this operation as much as possible, given its high cost.

\subsection{Contraction}\label{ENcontraction}

In addition to manipulating the indices of a tensor, we can join and separate the tensors. In order to join two tensors together, we apply the contraction operation on certain indices of these two tensors. For example,
\begin{equation}\label{ENcontract}
C_{ik}=\sum_jA_{ij}B_{jk}
\end{equation}
represents a contraction of the tensors of rank 2 (matrices) $\A$ and $\B$ on the $j$ index common to both. The contraction operation requires a summation on each of the contracted indices. Graphically, we represent this operation by an index that connects two vertices.

\begin{figure}[t]
  \includegraphics[width=0.6\columnwidth]{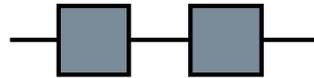}
\caption{A graphical representation of the matrix product.}
  \label{ENmatrix}
\end{figure}

\begin{figure}[b]
  \includegraphics[width=\columnwidth]{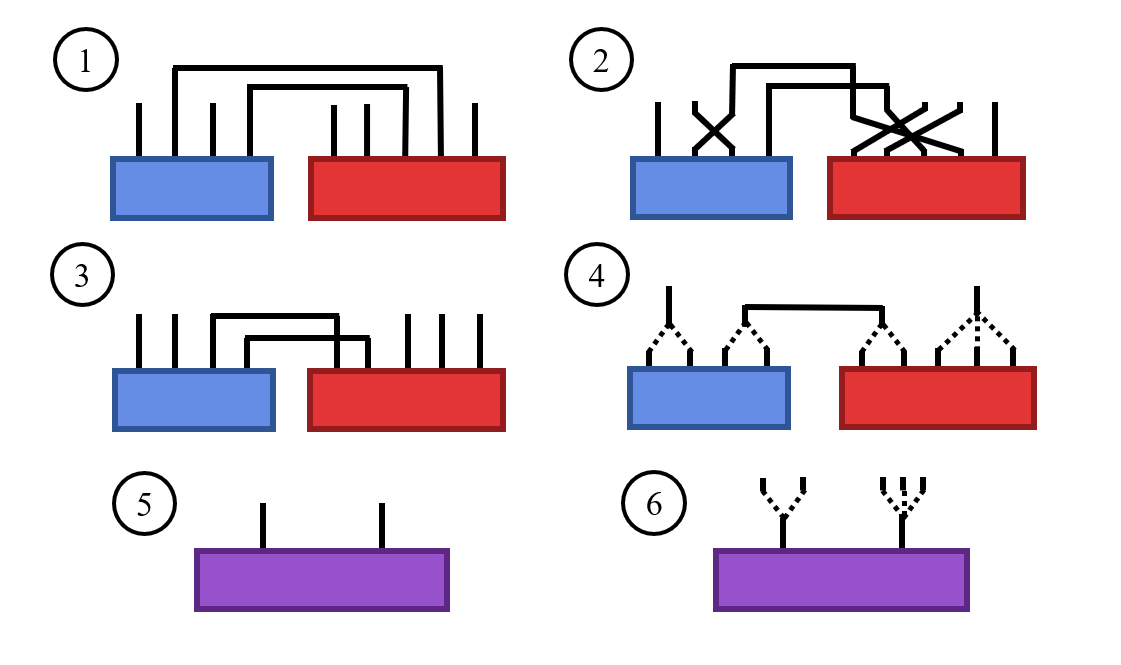}
\caption{By choosing a certain combination of permutations and reshapping, it is possible to convert any contraction into a matrix product. \textbf{1)} A contraction of two tensors on two indices. \textbf{2)} Switch the indices of the two tensors so that the non-contracted indices are at the beginning of the right tensor and at the end for the left one. Tensors are represented with the new order of their indices. The contracted indices must follow the same order in each of the tensors. \textbf{4)} A tensor reshape is applied to obtain tensors of rank 2. \textbf{5)} The matrix product is applied. \textbf{6)} Reshaping the resulting matrix to bring back the uncontracted indices of the initial tensors. }
  \label{ENcontraprod}
\end{figure}

Numerically, we can contract tensors by a series of sums on each of the contracted indices. However, there is an alternative that is faster consisting of converting the problem into a product of matrices, represented schematically in \ENreffig{ENmatrix}. The general steps to transform an arbitrary tensor into a matrix are to apply a permutation of indices followed by a reshape.  This process is illustrated in more detail in \ENreffig{ENcontraprod}, where tensors of rank four and five are contracted on some common indices in step 1. Step 2 of the \ENreffig{ENcontraprod} shows that the first tensor undergoes a permutation placing the contracted indices to the right of the others. The second tensor is also permuted, placing the indices contracted this time to the left of the others. The contracted indices of the two tensors must follow the same order between them. The tensors can be represented in the new order of their indices, as in step 3. In step 4, the tensors are then reshaped to combine the non-contracted indices together and then the contracted indices.

The two tensors are then in the form of matrices. On the left tensor, the combined non-contracted indices now represent the rows of the first matrix. The connected indices then represent the columns of the matrix. The opposite is true for the second matrix. After the matrix product, we obtain a resulting matrix (step 5). Finally, this matrix is reshaped to find the uncontracted indices of the initial tensors in step 6.

The cost of a contraction is determined by the size of each index of contracted tensors. In our case, assuming that each index is $\chi$, the total number of open (5) and common (2) indices implies that the computational cost is $\mathcal{O}(\chi^7)$. The same result is obtained by starting from the definition of contraction where the sum is calculated on each of the contracted indices. Written mathematically,
\begin{equation}
C_{\alpha\beta\gamma\delta\epsilon}=\sum_{\zeta\eta}A_{\alpha\zeta\beta}B_{\gamma\delta \eta\zeta\epsilon}.
\end{equation}
This shows that two indices (on $\eta$ and $\zeta$) contribute to the cost of the operation. However, to assign values to all the elements of the rank-5 tensor, $C_{\alpha\beta\gamma\delta\epsilon}$, we iterate on each of the non-contracted indices. In total, the numerical contraction operation requires 7 loops (one for each index on which it is iterated), each contributing a factor of $\chi$ to the computational cost.

\subsection{Decomposition of tensors and matrices}

In addition to combining tensors, it is often necessary to decompose a tensor. In the same way that tensor contraction can be converted into a product of matrices, the decomposition of a tensor can be converted into a matrix decomposition.

\subsubsection{Decomposition into eigenvalues}

The decomposition into eigenvalues is well known in physics. This corresponds to the diagonalization of a matrix. The matrix (tensor of rank 2) is then broken down into three matrices:
\begin{equation}\label{ENEigdecomp}
\M = \U{} \bigOmega{} \U{}^\dagger
\end{equation}
Two of these matrices ($\U{}$ and $\U{}^\dagger$) are linked by the dagger operator ($\dagger$) \cite{townsend2000modern}. These two matrices are unitary and satisfy $\U{}\U{}^\dagger=\U{}^\dagger \U{}=\I$, where $\I$ is the identity operator. The $\bigOmega{}$ matrix is diagonal and contains $m$ eigenvalues. The $\bigOmega{}$ and $\U$ matrix have the same dimensions. We can apply the eigenvalue decomposition only to square matrices. However, we cannot generalize this decomposition to all tensors and we must use another decomposition.

\subsubsection{Singular Value Decomposition}\label{ENsec:svd}

\begin{figure}[t]
\includegraphics[width=0.8\columnwidth]{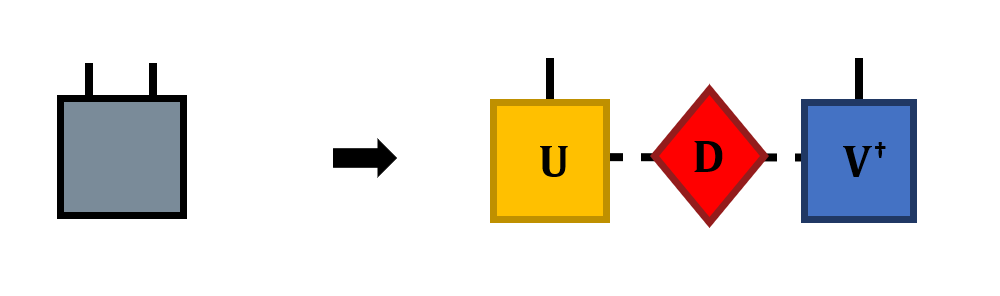}
\caption{A graphical representation of the decomposition into singular values (SVD). }
\label{ENSVD}
\end{figure}

We can separate a tensor into three other tensors with the singular values decomposition (SVD). The SVD breaks down a rectangular matrix M into three matrices:
\begin{equation}\label{ENSVDdecomp}
\M{} = \U{}\D{}\V^\dagger.
\end{equation}
This operation is graphically represented in the \ENreffig{ENSVD}. 

The dimensions of the $\U$ and $\V^\dagger$ matrices are related to the dimensions of $\M{}$. The dimensions of $\U$ will be $(a, m)$ where $m\leq \mathrm{min}(a,b)$ \cite{schollwock2005density}. The dimensions of $\V^\dagger$ are then $(m, b)$. There are other SVD formulations where the inner dimension of decomposition is chosen differently, but by convention here, the maximal dimension corresponds to the smallest of the outer dimensions.

\begin{figure}[b]
  \includegraphics[width=0.9\columnwidth]{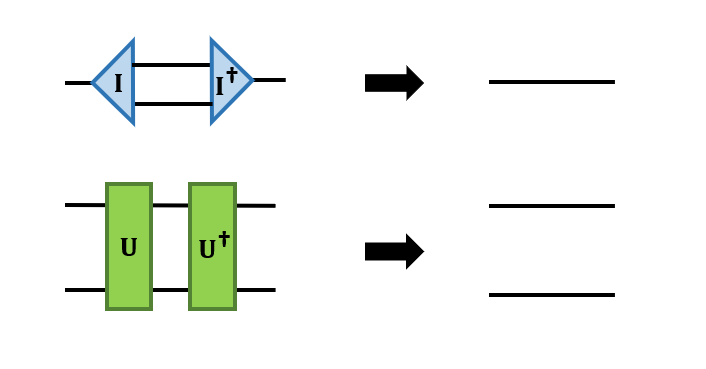}
\caption{  Unitaries and isometries contract to the identity. The resulting branches on the right represent identity (top: $\I$, bottom: $\I\otimes \I$).}
  \label{ENuniso}
\end{figure}

Finally, the $\D$ matrix (red diamond) is a diagonal matrix assigning a weight to the different vectors in the $\U$ and $\V^\dagger$ matrices:
\begin{equation}\label{ENlambdaN}
\D=\left(\begin{array}{ccccccc}
\lambda_1&0&0&\cdots&0\\
0&\lambda_2&0&\cdots&0\\\
0&0&\lambda_3&\ddots&0\\\
\vdots&\vdots&\ddots&\ddots&\vdots\\
0&0&0&\cdots&\lambda_m
\end{array}\right).
\end{equation}
The $\U$ (yellow rectangle) and $\V^\dagger$ (blue rectangle) matrices are isometries and multiply into an identity on only one side.  If the size of $\M{}$ is $(a, b)$, $\U^\dagger \U=\I_{a\times a}$ and $\V^\dagger \V=\I_{b\times b}$ (see \ENreffig{ENuniso}), but $\U\U^\dagger$ and $\V\V^\dagger$ are not the identity matrix generally. 

\begin{figure}[t]
  \includegraphics[width=0.9\columnwidth]{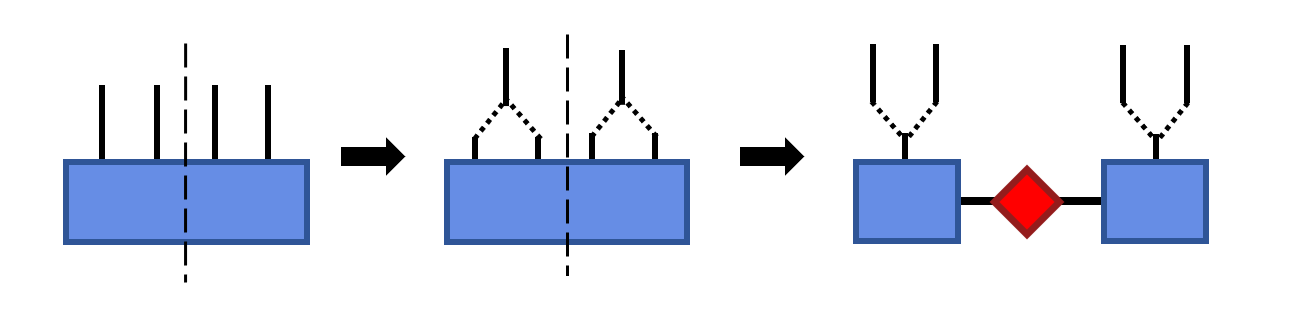}
\caption{A decomposition of a tensor of rank four can be performed by reshaping the tensor and applying a SVD to the resulting matrix. The procedure is similar to the one performed to prepare a tensor for contraction in the \ENreffig{ENcontraprod}.}
  \label{ENSVDFIG}
\end{figure}

The $\D$ matrix expresses the most important basis states contained in $\U$ and $\V$. It is common for only a few $\D$ values to be large. The lowest values of $\D$ can be truncated \cite{schollwock2005density}. However, truncation of the small $\D$ values will not change the norm of the original matrix by much. 

To perform a SVD, we note that $\M{}\M{}^\dagger=\U\D^2\U^\dagger$ and $\M{}^\dagger \M{}=\V\D^2\V^\dagger$. Any rectangular matrix can thus be multiplied by itself to obtain a square matrix. This matrix can then be decomposed by the eigenvalue decomposition to obtain the different elements of the SVD. This is not the only way to obtain a SVD, in fact, computer programs generally do not use the decomposition of $\M{}\M{}^\dagger$ or $\M{}^\dagger \M$ to avoid a loss of accuracy due to squaring the $\D$ values. However, this method highlights the close link between the SVD and the decomposition into eigenvalues.

We can use the SVD to break down any type of tensor. Figure~\ref{ENSVDFIG} shows the procedure to be performed. We reshape a tensor of rank 4 into a tensor of rank 2 before applying the SVD and reshaping the resulting tensors to find the appropriate indices. This approach is similar to the procedure in \ENrefsec{ENcontraction} for contraction.

\section{Entanglement}\label{ENentangle}

The SVD was not the only choice for tensor decomposition. There are indeed several matrix decomposition methods for rectangular matrices, for example the $QR$ decomposition \cite{schollwock2005density,press1989numerical}. As there are many choices, we will demonstrate the importance of the SVD method in the context of tensor networks. The explanation is related to information theory, which focuses on how two objects communicate with each other~\cite{ash1965information}. In the case of a tensor network, the idea is the same: we are interested in how two tensors interact. An exhaustive discussion of this theory is not necessary to understand the basics of tensor networks, but it should still be mentioned that it plays an underlying role in algorithm development.

The key to understanding SVDs is to understand what the $\D$ matrix represents. In a physical system, the $\M{}$ matrix of the previous section is the wave function $\psi$.  Note that if we want to form the density matrix from $\psi$, $\densmat{}=\psi\psi^\dagger$, then the eigenvalues of $\densmat{}$, $\rho_i$, will be the same regardless of the basis.  Expressed in terms of the basis functions to the left (right), stored in $\U{}$ ($\V$), of the site that $\psi$ occupies, the density matrix is $\densmat{}=\U{}\bigOmega \U{}^\dagger$ where $\bigOmega$ is a diagonal matrix containing all $\rho_i$.  Solving for $\psi$ from $\densmat{}(=\U\sqrt{\bigOmega}\V^\dagger\V\sqrt{\bigOmega}\U^\dagger$) , we find $\psi=\U{}\D\V^\dagger$ where $\bigOmega=\D^2$.   So, the SVD gives us all necessary components of a density matrix. 

An interesting connection is the entanglement entropy, also called the von Neumann entropy \cite{reif2009fundamentals},
\begin{equation}\label{ENentropy}
S=-\sum_i\rho_i\log_2\rho_i.
\end{equation}

\begin{figure}[t]
  \includegraphics[height=0.3\columnwidth]{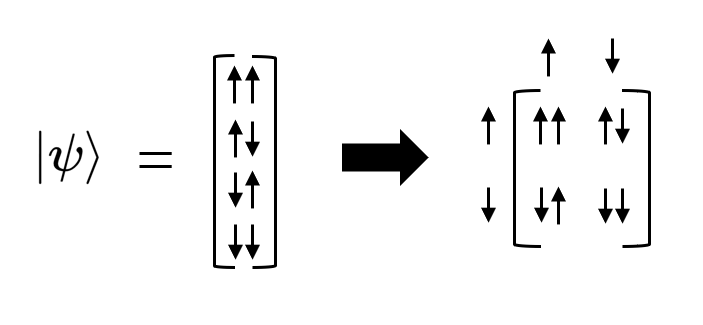}
  \caption{Vector describing a state of two spin particles a half. It can be reshaped as a matrix. Each index in the matrix represents the spin of each of the sites.}
  \label{ENvectmat}
\end{figure}

Some specific cases of quantum states can be analyzed in terms of entanglement. We present here two examples illustrating two extremes of entanglement. A first example is the state;
\begin{equation}
|\psi\rangle=(|\uparrow\uparrow\rangle+|\downarrow\uparrow\rangle)/\sqrt2=\left(\frac{|\uparrow\rangle+|\downarrow\rangle}{\sqrt2}\right)\otimes|\uparrow\rangle
\end{equation}
which has the SVD;
\begin{equation}
\U{}\D\V^\dagger=\left(\begin{array}{cc}
0 & 1\\
1 & 0
\end{array}\right)\left(\begin{array}{cc}
1 & 0\\
0 & 0
\end{array}\right)\left(\begin{array}{cc}
1 & 1\\
1 & -1
\end{array}\right)\Big/\sqrt2
\end{equation}
where $|\uparrow\rangle=(1\;0)^T$, $|\downarrow\rangle=(0\;1)^T$ and $|\psi\rangle$ have been reshaped as in \ENreffig{ENvectmat}. The $T$ symbol represents the transpose. The entanglement between these two spins is $S=0$ (we consider only non-zero values of $\D$) since we have only one eigenvalue in \ENrefeq{ENentropy}. These spins are called {\it un-entangled}.

\begin{figure}[b]
  \includegraphics[width=\columnwidth]{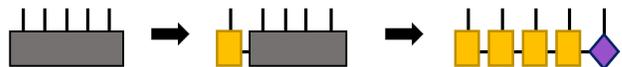}
\caption{A multi-site vector can be reshaped and divided by SVD several times so that each site is represented by a tensor. The last tensor is the only one that can be non-unitary, since it contains all the information in the $\D$ matrix of the subsequent SVD.}
  \label{ENMPSFIG}
\end{figure}

As a second example, let's consider the state;
\begin{equation}
|\psi\rangle=(|\uparrow\uparrow\rangle+|\downarrow\downarrow\rangle)/\sqrt2.
\end{equation}
This state is {\it maximally entangled} since the eigenvalues of $\D$ are both $1/\sqrt2$, and the entropy is maximal when the values of $\D$ are equal. Note that the singlet state $|\psi\rangle=(|\uparrow\downarrow\rangle-|\downarrow\uparrow\rangle)/\sqrt2$ is also maximally entangled with $S=1$.

In a typical physical system, the structure of $\D$ will give an entanglement somewhere between these two examples. An additional point is mentioned in Appendix~\ref{ENsec:correlation}, the $\D$ structure can be broadly classified into two instances for a physical problem. 

\section{Matrix Product States}\label{ENsec:MPS}

\begin{figure}[t]
  \includegraphics[width=0.9\columnwidth]{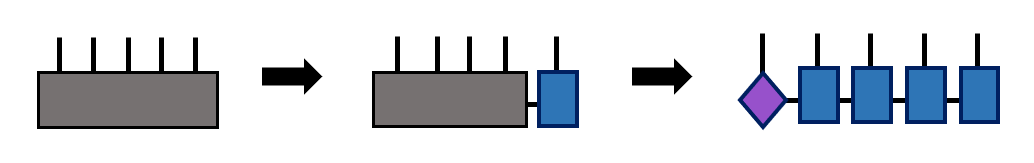}
  \caption{The MPS can also be built from right to left.}
  \label{ENleftMPS}
\end{figure}

We are ready to derive a tensor network from a quantum system from the tools in \ENrefsec{ENbasicops}. To do this, we will start from a complete wave function and then apply a series of reshapes and SVDs to obtain the site-by-site representation of the wave function, illustrated in \ENreffig{ENMPSFIG}. It is also possible to go from right to left as in \ENreffig{ENleftMPS}. Both results are called equivalent representations of a Matrix Product State (MPS). During this procedure, the only important step is to always reshape the physical index from the sites as in \ENreffig{ENMPSFIG}. Note that we start from a pre-solved wave function for demonstration purposes only. In general, the wave function will be derived from a Hamiltonian that is broken down into a tensor network representation.

Even though the operations can be determined from \ENreffig{ENMPSFIG}, it is worth going through each step in detail and we do so next.  While showing the steps for deriving the MPS, it is natural to ask what the newly introduced horizontal index means.  To fully understand the result, we demonstrate the meaning we go through the steps to generate the MPS twice in the following.  The first time is to explicitly show how each step is reshaped and where the physical indices are placed in the reshape.  Then, we go through each step again but now keeping track of the basis functions on each index. This will allow us to fully understand the meaning of a horizontal index that is introduced.  On each diagram, we will show the basis functions, even though this information is not usually provided.  The index--also called a link index--represents the basis functions of all other sites in the system.

\subsection{Separate a wave function into a matrix product state}

\begin{figure}[t]
  \includegraphics[width=0.7\columnwidth]{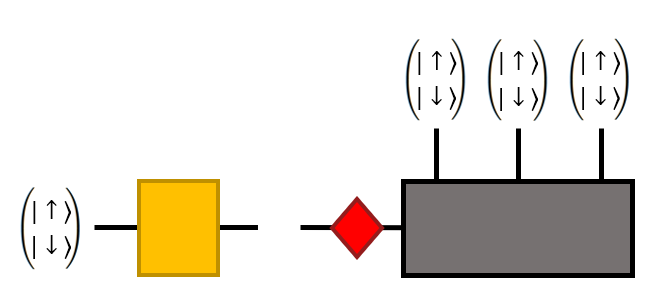}
  \caption{It is possible to perform a SVD on the reshaped wave function so that the tensor generating $\U{}$ represents locally our first site. The $\D$ matrix describing the weight of all vector combinations can be contracted to the rest of the wave function.}
  \label{ENMPSsecond}
\end{figure}

To demonstrate, we start with a wave function from an exact diagonalization. The complete wave function can be reshaped to isolate the degrees of freedom of one site from the rest, as shown in \ENreffig{ENMPSfirst}. To perform this task, a program is available in Appendix~\ref{ENMPScode} as an example. The vertical indices of a wave function are called physical indices, since they express the degrees of physical freedom of each site ({\it i.e.}, for a spin system the states are $|\uparrow\rangle,|\downarrow\rangle$ while for the Hubbard model it would be $|0\rangle,|\uparrow\rangle,|\downarrow\rangle,|\uparrow\downarrow\rangle$). A SVD can be taken as represented in \ENreffig{ENMPSsecond}. It is important to put an index on the left and the rest on the right: this is what allows the separation of the physical indices from the first site. The $\U{}$ matrix, resulting from the SVD, can be reshaped to a tensor representing the first site of the MPS. An additional index appeared during the procedure, it connects the left and right tensors. This type of index, drawn horizontally on the figure, is called a link index.

\begin{figure}[b]
  \includegraphics[width=0.9\columnwidth]{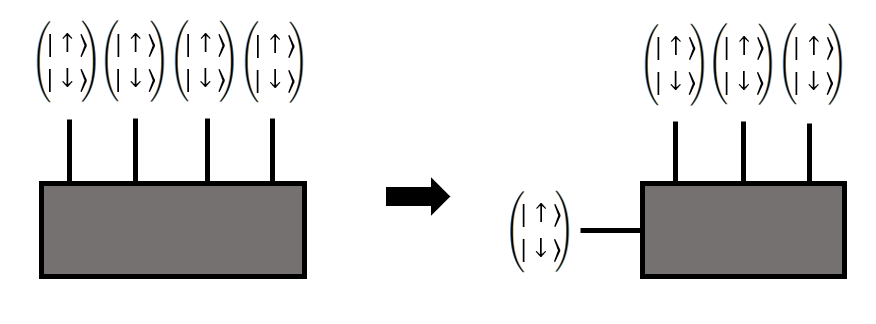}
  \caption{Reshaping an index in preparation for the first site to be separated.}
  \label{ENMPSfirst}
\end{figure}

\begin{figure}[b]
  \includegraphics[width=0.8\columnwidth]{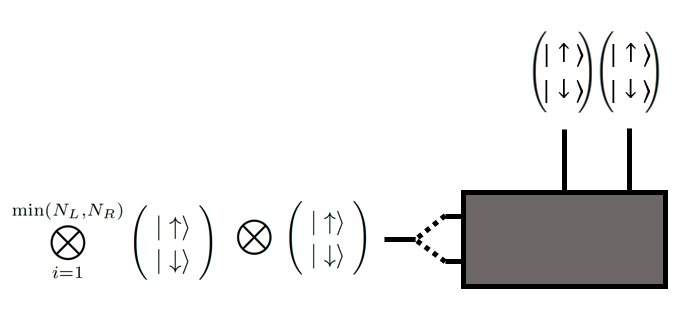}
  \caption{The rest of the wave function must be reshaped so that the following physical index is combined with the link index.}
  \label{ENrestreshape}
\end{figure}

\begin{figure}[t]
  \includegraphics[width=1\columnwidth]{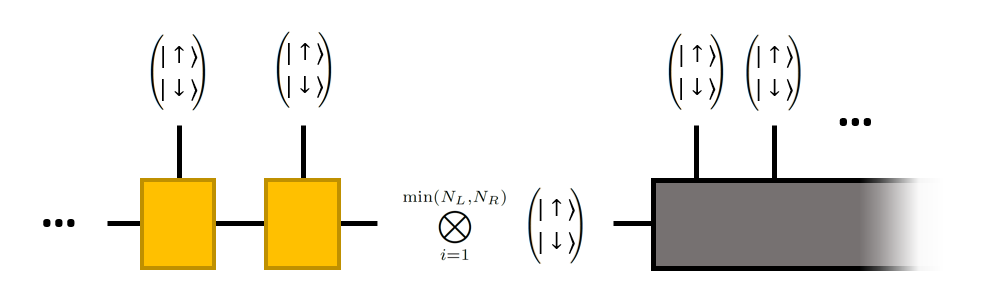}
  \caption{In the center of an MPS, the basis functions are the Kronecker products of all the basic states on the left or right (whichever is lower).}
  \label{ENMPSfourth}
\end{figure}

The same idea can then be applied to the remaining sites of the wave function. The remaining tensor containing the other degrees of freedom must be reshaped so that the next physical index is combined with the previous link index. This is what is shown in \ENreffig{ENrestreshape}. The $\U{}$ matrix can once again be reshaped to have a tensor of the required shape in \ENreffig{ENMPSfourth}. The procedure can be repeated until each site is represented by an individual tensor. Now, we have obtained a local representation of each site in our system as well as our first example of a wave function expressed as a tensor network, the MPS.

The price associated with separating a wave function is the addition of a new index, the same one that appears in \ENrefsec{ENsec:svd} when performing the SVD (see \ENreffig{ENSVD}). This new index contains information for each site that is not the current site. The following section aims to understand exactly how this link index appears and its meaning.

\subsection{Meaning of the horizontal indices}

Throughout the diagrams in the previous section, we have taken care to express the basis functions of each index, even if they are not normally displayed.

To understand the meaning of the horizontal index, we must understand what each value of the index represents, and we do so in terms of the basis.  First, we start by examining how basis functions are joined and separated on the physical indices before discussing the link indices.  As was already shown, the physical index's basis functions are $|\uparrow\rangle,|\downarrow\rangle$.  If we want to write the basis for two physics indices we use a Kronecker product ($\otimes$, see Appendix~\ref{ENkronecker}). This operation takes all the states to the right of the operator and attaches them to the states on the left. The two-site wave function can be written
\begin{equation}\label{ENjoinbasis}
\left(\begin{array}{c}
|\uparrow\rangle\\
|\downarrow\rangle
\end{array}\right)\otimes\left(\begin{array}{c}
|\uparrow\rangle\\
|\downarrow\rangle
\end{array}\right)\equiv\left(\begin{array}{c}
|\uparrow\uparrow\rangle\\
|\uparrow\downarrow\rangle\\
|\downarrow\uparrow\rangle\\
|\downarrow\downarrow\rangle
\end{array}\right),
\end{equation}
where the first and second vector represent the possible states of the first and second site respectively.

For several sites, we can write
\begin{equation}
\left(\begin{array}{ccccc}
|\ldots\uparrow\uparrow\uparrow\uparrow\uparrow\uparrow\uparrow\rangle\\
|\ldots\uparrow\uparrow\uparrow\uparrow\uparrow\uparrow\downarrow\rangle\\
|\ldots\uparrow\uparrow\uparrow\uparrow\uparrow\downarrow\uparrow\rangle\\
\vdots
\end{array}\right)\equiv\bigotimes_{i=1}^N\left(\begin{array}{c}
|\uparrow\rangle\\
|\downarrow\rangle
\end{array}\right).
\end{equation}

When the vector is reshaped, the basis functions are also separated according to this grouping. For example, in our case with two sites, \ENreffig{ENvectmat} shows a reshaped two-site wave function.

If we take a wavefunction and apply a SVD, the new index can be understood in terms of the basis of the physical indices.  Consider the SVD of the two-site wavefunction from \ENrefeq{ENjoinbasis} reshaped as in  \ENreffig{ENvectmat}. We will only consider the basis functions involved in the following and not how they are weighted.  Examining the $\hat U$ ($\hat V$), one column (row) is a linear combination of the outer indices basis functions, weighted by the values in $\hat U$ ($\hat V$).  Each column (row) represents another value of the horizontal index and a different linear combination of the basis functions.

The largest dimension of the horizontal index is the minimum between the two sides, $\mathrm{min}(N_L,N_R)$ based on the definition of the SVD that is used here.  We could have chosen a different definition, but this is the standard choice here for both computational speed and the fact that the physical information should be independent of the basis set, so the size of the new index can be represented in either the left or the right.  To write out what the basis functions of the horizontal index appear as, we can write
\begin{equation}
\bigotimes_{i=1}^{\mathrm{min}(N_L,N_R)}\left(\begin{array}{c}
|\uparrow\rangle\\
|\downarrow\rangle
\end{array}\right).
\end{equation}
This is the basis of the horizontal index.  This expression means that the link indices are linear combinations of all other basis functions on all other physical indices.

\subsection{Truncation}

When decomposing a complete wave function using successive SVDs (see \ENrefsec{ENsec:MPS}), the size of the link index between each site can be determined by knowing the dimensions of the matrix on which the SVD is applied. For a spin-$1/2$ chain, the physical index is always size 2. The first link index, linking the first site to the rest of the wave function, will be size two, because the matrix to which the SVD is applied is $2\times 2^{N-1}$. The size of the $\D$ matrix then corresponds to the minimum of the two dimensions. Following the same logic, the next link index will be of size 4, then 8, 16, 32, etc. At some point, the link indices will shrink, limited by the size of the second index of the matrix to be decomposed, to a size of 2 at the opposite end of the MPS.

However, this exponential growth in the link index precisely brings us back to the problem of the exact diagonalization of the wave function in \ENrefsec{ENwhytensors}, which justified the need for tensor networks. Fortunately, it is sometimes possible to truncate the link index without affecting the accuracy of the state too much by using the SVD introduced in the \ENrefsec{ENsec:svd}.

\begin{figure}[t]
  \includegraphics[width=\columnwidth]{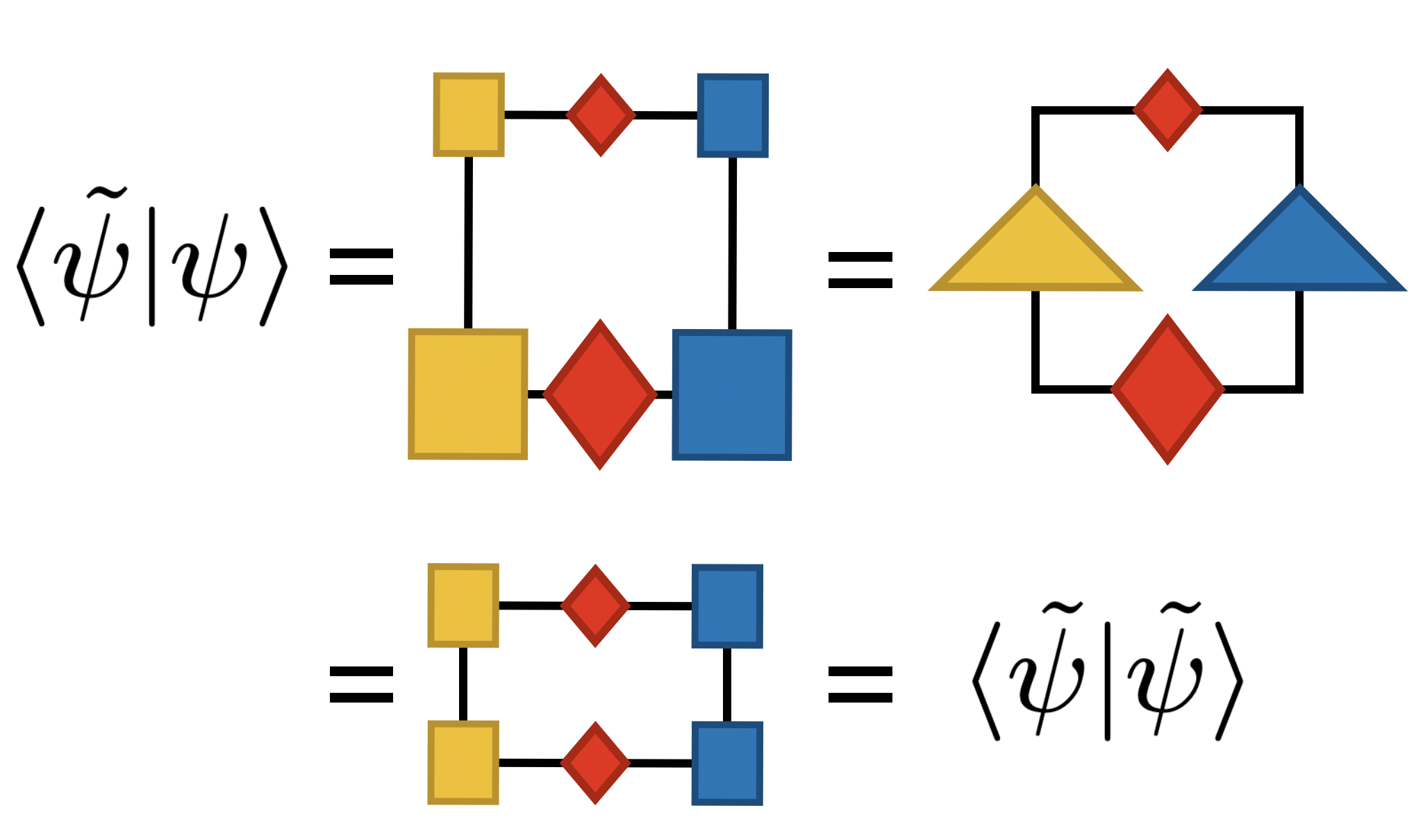}
\caption{Equality $\langle\tilde{\psi}|\psi\rangle=\langle\tilde{\psi}|\tilde{\psi}\rangle$ (\ENrefeq{ENtruncerr}) is justified here. The truncated wave function by the SVD is represented by smaller tensors at the top of the diagram. $\tilde \V^\dagger \V$ and $\tilde \U{}^\dagger \U{}$ then contract into isometries that will only truncated the $\D$ values retained in the truncated wave function. This contraction is therefore equivalent to $\langle\tilde{\psi}|\tilde{\psi}\rangle$.}
  \label{ENerrorjust}
\end{figure}

The $\D$ matrix is typically arranged by ordering the singular values in descending order on the diagonal from the upper left corner of the matrix. One may wonder how much the accuracy of the system is affected by truncating lower values. A good way to evaluate this loss of accuracy is to evaluate the resulting approximate wave function (noted with a tilde), $|\tilde\psi\rangle$. The difference between the truncated (or approximated) wave function and the full wave function is;
\begin{equation}\label{ENtruncerr}
\left\||\psi\rangle-|\tilde{\psi}\rangle\right\|^2=1+\delta-2\delta=1-\delta
\end{equation}
where $\delta=\langle\tilde{\psi}|\psi\rangle=\langle\tilde{\psi}|\tilde{\psi}\rangle = \sum_{i=1}^{m}\lambda_i^2$ is called the truncation error.  The equality ($\langle\tilde{\psi}|\psi\rangle=\langle\tilde{\psi}|\tilde{\psi}\rangle$) is observed when the SVD was applied to the complete wave function and its truncated version, as shown in \ENreffig{ENerrorjust}. We notice then that $\tilde \V^\dagger \V$ and $\tilde \U{}^\dagger \U{}$ contract to isometries which will truncate the $\D$ values not present in $|\tilde\psi\rangle$.

In order to show the impact of this procedure on the observables of the problem, it should be recalled that the energy, $E$, is related to the density matrix according to $\mathrm{Tr}(\densmat{}\Ham{})\equiv E$ \cite{reif2009fundamentals}. By truncating only small elements of $\densmat{}$, the overall energy of the system will change very little. Fortunately, for several cases of interest, a large part of the singular values are small (see \ENrefsec{ENmpslimits} or Appendix~\ref{ENsec:correlation} for more information). The size of the $\D$ matrix can therefore be greatly reduced for many physical systems.

Truncation eliminates the basis states associated with a low weight in the density matrix. This is a fundamental concept in tensor networks. 

\subsection{Measurements and Gauges}\label{ENgauge}

Measurements related to local operators are perfectly adapted to a MPS. Remember that the SVD produces two isometric tensors ($\U{}$ and $\V^\dagger$) as well as the $\D$ matrix, the only object that associates any weight to the decomposition. The tensors on the left of the orthogonality center ($\U{}$) are called left-normalized while those on the right ($\V^\dagger$) are right-normalized. If two left-normalized tensors are contracted, the result is equivalent to the identity (see \ENreffig{ENuniso}). The equivalent is also true for the right-normalized tensors. The general property that all non-unitary information of an MPS is contained in a single tensor is called a gauge~\cite{vidal2003efficient}. In the case of a local measure, not all of the network's tensors need to be contracted explicitly. A practical application of this property is shown here. 

\begin{figure}[t]
  \includegraphics[width=0.8\columnwidth]{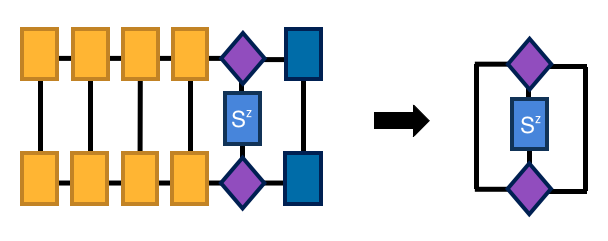}
\caption{Local measurements can be made much easier by using the MPS. Here, the operator is equivalent to a local spin measurement on a single site. The application of this operator involves only three tensors.}
  \label{ENMPOFIG}
\end{figure}

An example of a local measurement of the operator $\Spin{}^z$ with $\hbar=1$,
\begin{equation}\label{ENSzDef}
\Spin{}^z=\frac12\left(\begin{array}{cc}
1 & 0\\
0 & -1
\end{array}\right),
\end{equation}
on a single site is shown in \ENreffig{ENMPOFIG}. Applying the same measure to the full wave function, we should have written $\Spin_5^z=\I\otimes \I\otimes \I\otimes \I\otimes \Spin{}^z\otimes \I$, which is a much larger matrix than the one in \ENrefeq{ENSzDef}.

\begin{figure}[b]
  \includegraphics[width=0.8\columnwidth]{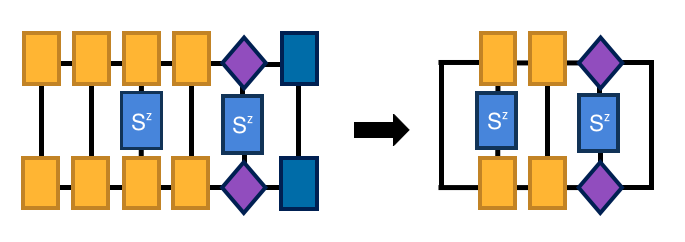}
\caption{For a measurement applyied to two sites, the problem is reduced to the tensors of the sites of interest and the tensors located between them. This is a consequence of the orthogonality breakage when contracting from the left.}
  \label{ENMPO2site}
\end{figure}

For a measurement applied to two or more sites, as shown in \ENreffig{ENMPO2site} ($\langle\psi|\Spin_3^z\Spin_5^z|\psi\rangle$), orthogonality is not maintained for all sites. By contracting the normalized tensors on the left, the appearance of an operator breaks the orthogonality condition. The tensors between the different operators do not therefore contract the identity. Since the orthogonality center has been placed on the same site as the other operator, the rest of the tensors, all normalized on the right, contract to the identity.

\begin{figure}[t]
  \includegraphics[width=\columnwidth]{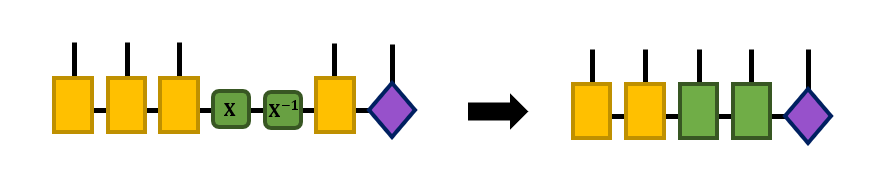}
\caption{The MPS is not unique, as it always represents a complete wave function after the introduction of a matrix and its inverse between two tensors.}
  \label{ENnounique}
\end{figure}

There is no single MPS representing a wave function. Indeed, between two link indices, an identity can be introduced as a term $\X\X^{-1}$. By contracting $\X$ with the MPS tensor to its left and $\X^{-1}$ with the tensor to its right, both MPS tensors are affected. As this operation is equivalent to the application of an identity, the MPS remains equivalent to the original state, as shown in the \ENreffig{ENnounique}. As a result, two MPSs may have a different structure while achieving the same results.

\subsection{{Matrix Product Operators}}\label{ENMPO}

The MPS is a local representation, in terms of tensors, of a wave function. To have a complete representation of a quantum mechanics problem in terms of local tensor networks, we need an equivalent construction for operators: the matrix product operator (MPO).

\begin{figure}[b]
  \includegraphics[width=0.45\columnwidth]{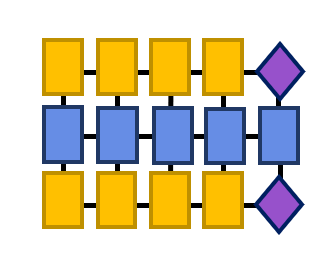}
\caption{This diagram is the average value of a MPO operator in MPS state, $\langle\psi|\Ham{}|\psi\rangle$.}
  \label{ENUnitMPO}
\end{figure}

\begin{figure}[t]
  \includegraphics[width=0.9\columnwidth]{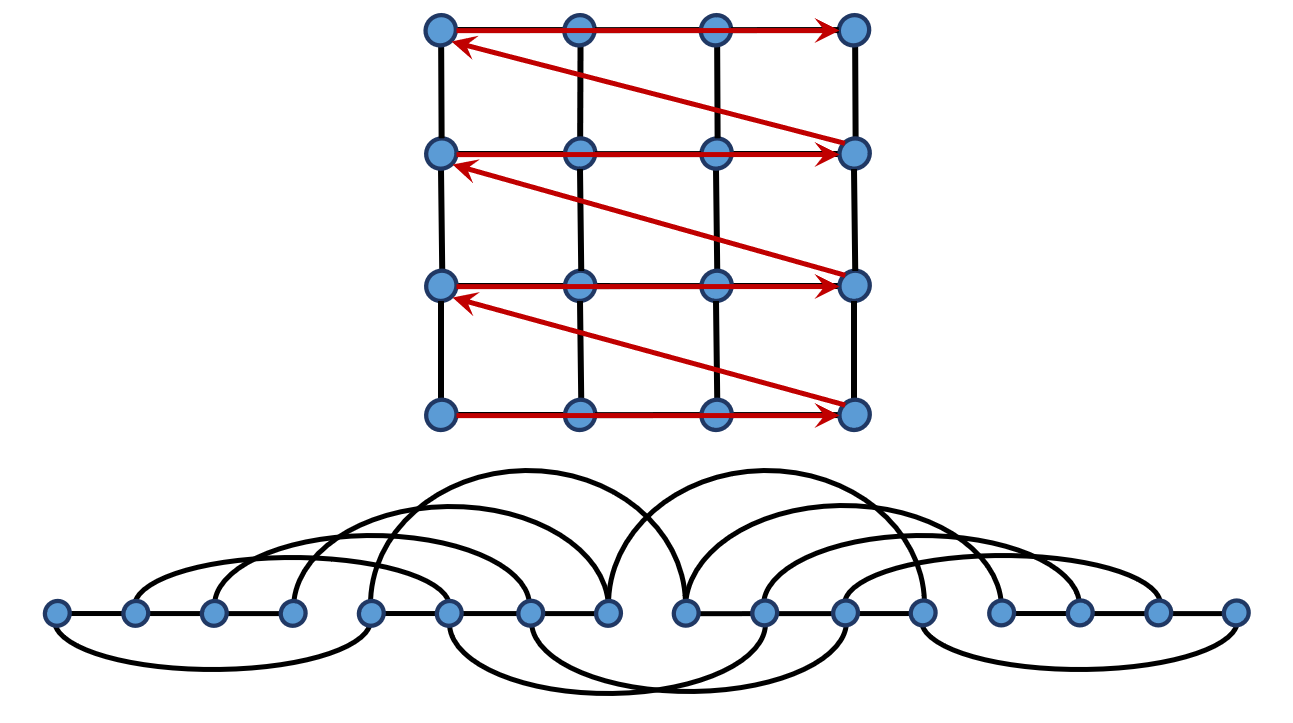}
\caption{A two-dimensional network can be expressed in one-dimension. The path used may vary depending on the situation.}
  \label{ENsnake}
\end{figure}

For example, the Hamiltonian of the Ising model can be constructed with the following MPO;
\begin{equation}\label{ENbulkMPO}
\Ham{}_{a,a'}^\mathrm{(Ising)}=\left(\begin{array}{ccc}
\I & \0 & \0\\
\Spin{}^z & \0 & \0\\
\0 & -J\Spin{}^z & \I\\
\end{array}\right).
\end{equation}
As each of the elements of this ``matrix" are themselves operators, it is in fact a tensor of rank 4 (see \ENreffig{ENUnitMPO}). The rows and columns of the matrix signify the horizontal indices of the MPO, and the individual operators have a physical index to connect to the wavefunction and a second to connect to the dual of the wavefunction. When a chain of these operator matrices is contracted along the link indices to form the Hamiltonian of a system, it is located in the lower left corner of the final matrix. So, for a system with open boundary conditions, the ends of the chain are two vectors: $(0, 0, 1)$ on the left and $(1,0,0)^T$ on the right.  Determining the form of the MPO follows a few simple rules, but there are also methods to automatically determine the form of the operators \cite{hubig2017generic}.

Like the MPS, the representation as a MPO is not unique. Trivially, rows and zero columns could be added to the matrices without changing the operator represented. 

To add terms that couple more distant sites, it is necessary to increase the size of the operator matrix. For example, the Ising model with coupling to the first and second neighbour,
\begin{equation}
\Ham{}=\sum_i \left(-J_1 \Spin{}^z_{i}\Spin{}^z_{i+1}-J_2 \Spin{}^z_i\Spin{}^z_{i+2}\right),
\end{equation}
requires the MPO
\begin{equation}\label{ENLRI}
\Ham{}_{a,a'}^\mathrm{(Ising)}=\left(\begin{array}{cccc}
\I & \0 & \0 & \0\\
\Spin{}^z & \0 & \0 & \0\\
\0 & \I & \0 & \0\\
\0 & -J_1 \Spin{}^z & -J_2 \Spin{}^z & \I\\
\end{array}\right).
\end{equation}
Typically, for each term acting at a given site in the system, an additional column and row must be added. To build a MPO for a two-dimensional network, a path that crosses all the sites of the system once must be chosen. This path defines an effective problem in one dimension; an example is illustrated in \ENreffig{ENsnake}. The actual bond dimension for a MPO depends on the width of the system, and therefore the size of the MPO is proportional to the width of the system. This limits the applicability of MPOs and MPSs to the system in a dimension larger than one.

Some long-distance couplings can be written with small MPOs. For example, an exponential coupling of the distance between two sites
\begin{equation}
\Ham{}=\sum_{i\neq j}e^{-\frac{|i-j|}\xi}\Spin{}^z_{i}\Spin{}^z_{j},
\end{equation}
 couples all sites in a system to each other. The MPO generating this Hamiltonian is written as follows
\begin{equation}
\Ham{}_{a,a'}=\left(\begin{array}{cccc}
\I & \0 & \0\\
\Spin{}^z & \kappa\I & \0\\
\0 & \Spin{}^z & \I
\end{array}\right)
\end{equation}
with $\kappa\equiv\exp(-1/\xi)$. Writing the MPO of this system using the method described for \ENrefeq{ENLRI} would have required matrices of size proportional to the number of sites in the system. Finding the smallest MPO expression for the Hamiltonian of a system is critical to the performance of MPS algorithms.

\subsection{Limitations of the Matrix Product State}\label{ENmpslimits}

What magnitude should a wave function have to describe a given system? How much information can we truncate and still retain the good properties of the system we want to solve? The answer is problem dependent.  However, there are a few simple common properties that will apply to physical systems of interest.

Roughly, there are two cases that will differentiate the performance of our tensor network ansatz and the maximum truncation necessary to get a good answer. A good way to think about these two cases is in the thermodynamic limit (during the calculations, we use a finite system, but it is approximating the limit). In the first case, there will be a difference between the two lowest energies (see Appendix~\ref{ENsec:correlation}).  In the other case, there is a small difference between the eigenvalues and they form a continuum. The MPS is well suited for states with a gap between these states. Moreover, if the interaction is local ({\it i.e.}, the interaction extends only to a few sites in a one-dimensional path), then the MPS can be truncated without sacrificing too much precision~\cite{verstraete2006matrix}.  In a simple example, such as the 10-site Heisenberg model, 50 states can be kept on the link index and we will still get a good accuracy compared to the exact diagonalization. If we have long-range interactions that span multiple sites, then the link index should keep more information.  Note that on a finite size system, there is always at least a small gap and in many cases, the MPS can approximate a system of interest and will only require a larger bond dimension if the system is not ideally suited for the MPS \cite{schollwock2005density}.

We can always unfold a higher dimensional system into a one-dimensional system. It works with a varied effectiveness, as in \ENreffig{ENsnake}.  The increase in size is the result of the increase in interaction length, which is introduced in \ENrefsec{ENMPO}. The MPO's growth with dimension refers to the area law of entanglement and takes its name from the study of black holes~\cite{eisert2010colloquium}. The area law of entanglement says that the amount of information passed between two domains of a system can be expressed at the boundary between those regions.  {The area law is correct for states that are at the extremes of the eigenvalue spectrum ({\it i.e.}, the ground state).  For the states in the center of the spectrum, there is a volume law that has a different behavior for the entanglement.} 

For a one-dimensional problem, the boundary between the right and left side of the system is zero-dimensional. The boundary between the right and left sides of a two-dimensional problem is one-dimensional. In general, the area between the two sides is $d-1$ dimensions for a problem with $d$ dimensions. So, increasing the size of the problem will increase the amount of information from one side of the system to the other. 

\subsection{Beyond the Matrix Product State}

The MPS is only one ansatz which can represent a wave function. In this section, we present two additional forms here.

\begin{figure}[t]
  \includegraphics[width=0.9\columnwidth]{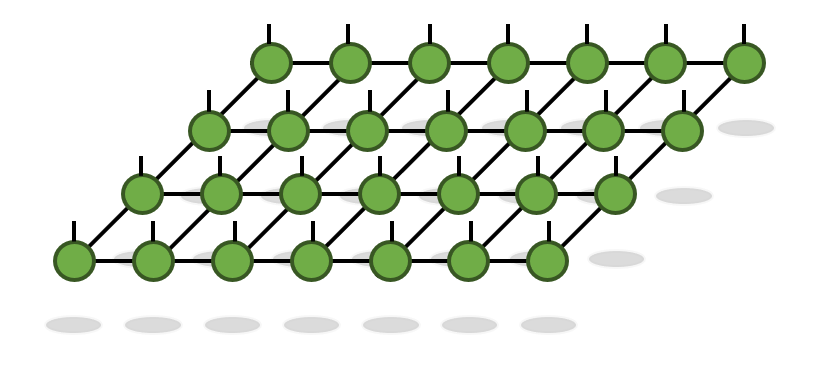}
\caption{The PEPS is a two-dimensional representation of the wave function. This sometimes allows a better representation of two-dimensional systems than the MPS.}
  \label{ENPEPS}
\end{figure}

All our discussions so far have been for systems where the right and left sides are easily identifiable ({\it i.e.}, one-dimensional system), and MPS works better when correlations decrease exponentially as in a gapped system with local interactions. For a two-dimensional system, it may be advantageous to use the two-dimensional version of the MPS called the projected entanglement pair state (PEPS)~\cite{verstraete2004renormalization,verstraete2008matrix}.  PEPS (\ENreffig{ENPEPS}) uses rank 5 tensors (with a physical index and 4 link indices) as members of the network for a square lattice (and a different lattice would require a different rank; for example, a honeycomb lattice would have tensors of rank 4). The size of the link index may be smaller than in the case of calculating a two-dimensional MPS by choosing this ansatz, but the rank of the tensor is higher. 

Completely contracting the tensor network in the PEPS is extremely difficult~\cite{valiant1979complexity}.  There is no known algorithm to contract an arbitrary tensor network effectively, however there are algorithms that try to do this in the best way~\cite{pfeifer2014faster,kourtis2019fast}.

\begin{figure}[b]
  \includegraphics[width=0.8\columnwidth]{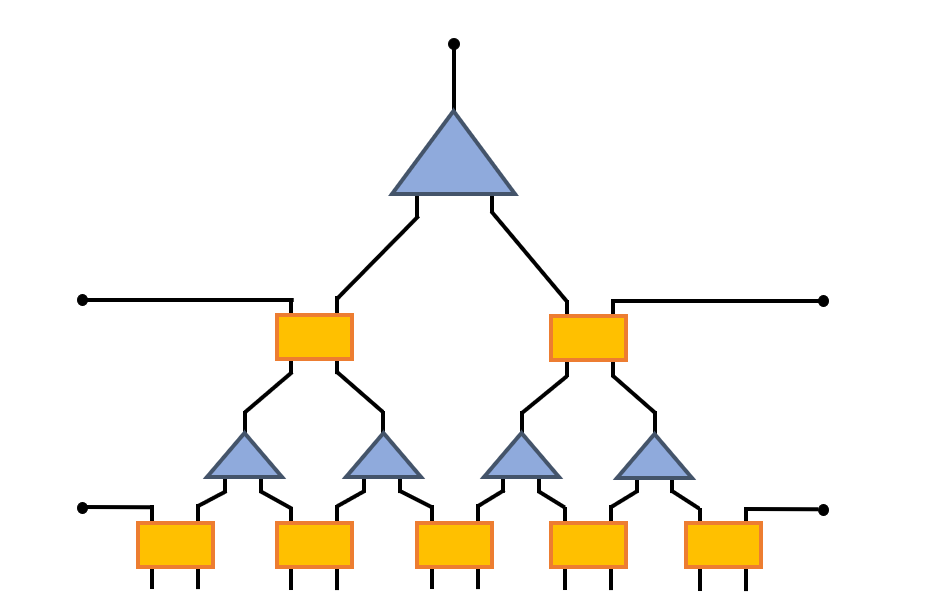}
\caption{Representation of a MERA. }
  \label{ENMERA}
\end{figure}

\begin{figure}[t]
  \includegraphics[width=0.9\columnwidth]{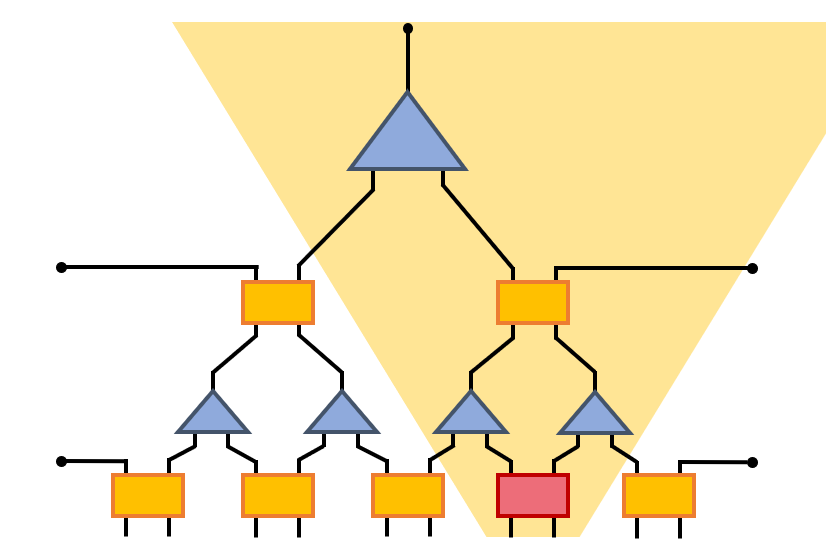}
\caption{A change applied to a MERA tensor will only affect those tensors with which it shares an index. So, all the tensors in the yellow cone will be affected.}
  \label{ENlightcone}
\end{figure}

\begin{figure}[b]
  \includegraphics[width=0.8\columnwidth]{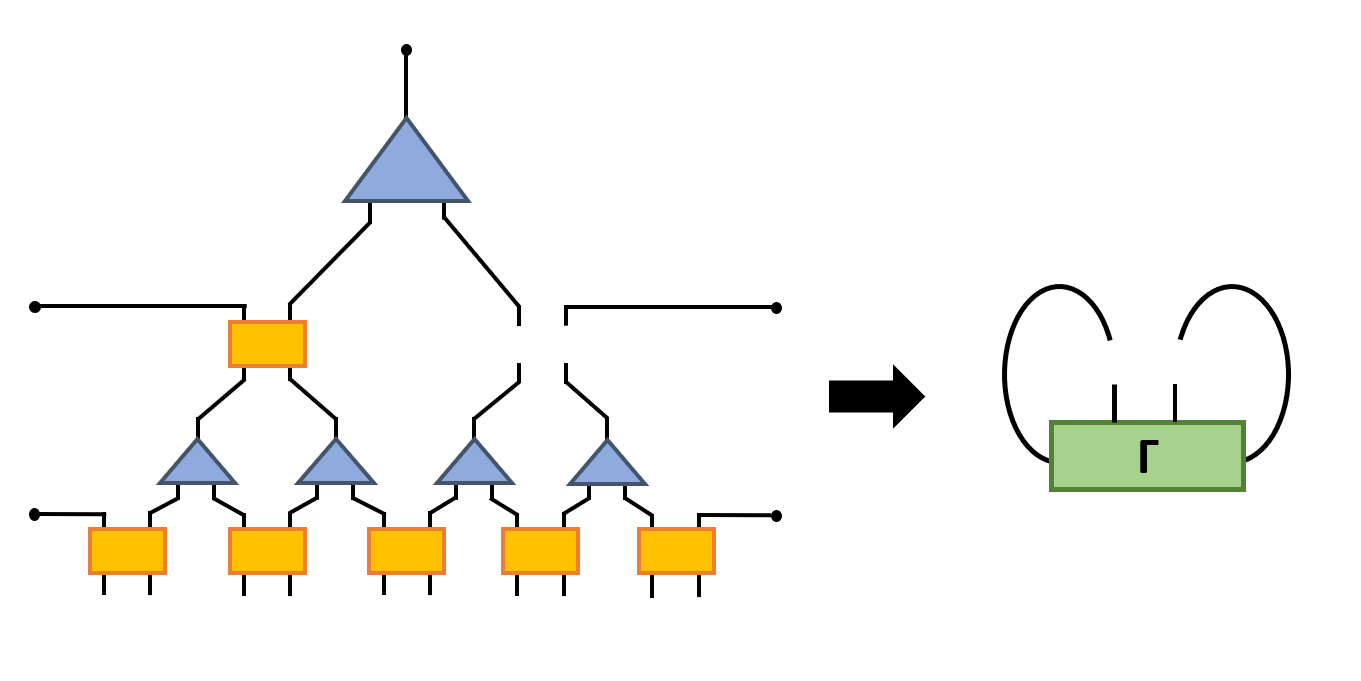}
\caption{To solve for a MERA tensor, we create an environment made of the contraction of all the other tensors in the network. The resulting tensor will be the one that minimizes the trace (see text).}
  \label{ENMERAsolve}
\end{figure}

To introduce the second ansatz, it should be remembered that an MPS can also manage correlations without a gap, but this requires keeping more singular values of the SVD. We also have the possibility to use another tensor network called the multi-scale entanglement renormalization ansatz (MERA) which is specifically designed to represent gapped and gapless systems \cite{vidal2007entanglement,vidal2008class}. A MERA is presented in \ENreffig{ENMERA}. To see why MERA is better for gapless systems, we will refer to the Appendix~\ref{ENsec:correlation}.

A property of the MERA is that it has a causal cone, only the tensors inside this cone will be affected when another tensor is altered. For example, changing the tensor in \ENreffig{ENlightcone} on the last line will not affect the sites to the left or right. All other tensors higher in the network whose indices are connected to them, will be affected by the change. Since only a few tensors are altered by the inclusion of an operator, this property can be seen as similar to the gauge for MPS, but it is a different property.

Note that there is also a two-dimensional version of the MERA~\cite{evenbly2010frustrated}. MERAs have also been used to simulate conformal field theories that may be useful in the study of holography, for example~\cite{swingle2012entanglement}.

A common method for determining unitary tensors or isometric tensors in a MERA is to maximize the operator's trace when applied to the network. For example, we can remove one of the tensors from the network (see \ENreffig{ENMERAsolve}) and contract all the other indices. This produces a tensor $\bigGamma$. To generate the correct unitary tensor ($\W$), the condition $\max\mathrm{Tr}\left(\W\bigGamma\right)$ must be met. To determine how to maximize this expression exactly, note that $\bigGamma$ can be decomposed with a SVD, $\bigGamma=\U{}\D\V^\dagger$. If we choose $\W=\V\U{}^\dagger$, then under cyclic property of the trace, the condition is reduced to the trace on $\D$. This then relates to maximizing the entanglement of the system.

More information about these tensor networks can be found in Ref.~\onlinecite{evenbly2009algorithms,evenbly2016entanglement}.

\section{{Algorithms}}\label{ENalgorithms}

To solve a quantum mechanics problem, the tensors of a network are transformed using the four basic operations of the \ENrefsec{ENbasicops}. The different tensor network algorithms are different uses of the basic operations.

\subsection{Time evolution}
There are two flavors of time evolution algorithms. One uses the system evolution operator $e^{-i\Ham{}t}$, with $t$ the time interval, to simulate the dynamics of a state, the other uses the imaginary time evolution operator $e^{-\beta\Ham{}}$ with $\beta$ the imaginary time to obtain the fundamental state. By repeated application of $e^{-\beta\Ham}$ with a finite value of $\beta$ on any state, the ground state can be approached.

To apply these exponential operators to an MPS, we must carry out a factorization that allows us to act locally on the MPS. Unfortunately, there is no exact identity to accomplish this for matrix exponentials. In practice, we use the identity of Trotter-Suzuki \cite{suzuki1985decomposition} to perform the decomposition in an approximate way:
\begin{equation}\label{ENBCH}
e^{-i(\Ham{}_A+\Ham{}_B)t}\approx e^{-i\Ham{}_At}e^{-i\Ham{}_Bt}.
\end{equation}
This expression makes an error in the order of $t^2[\Ham{}_A,\Ham{}_B]$. The approximation is therefore good for sufficiently short times. A solution for a time $Nt$ can be obtained by applying the expression $N$ times on a state. 

Time evolving block decimation (TEBD) is a method to simulate the time evolution of a MPS \cite{vidal2003efficient,vidal2004efficient}. The algorithm does this:
\begin{enumerate}
  \item All exponential terms acting on the orthogonality center are applied to the orthogonality center and the relevant neighbors. Following this step, the center and the neighbors just acted on are combined into a single tensor of higher rank.
  
   \includegraphics[width=0.9\columnwidth]{bakerINFIG1.PNG}  
  \item A singular value decomposition is performed to extract the tensor from the site that was the center, and moves the orthogonality center to the first neighboring site.
  \item Step 1 is repeated until the end of the MPS is reached, making sure to apply all the terms of the evolution operator only once.
\end{enumerate}
  \includegraphics[width=\columnwidth]{bakerINFIG2.PNG}
Once the end of the MPS is reached, one time step is performed. To take another step, we start again in the opposite direction.

\subsection{Tensor Renormalization Group }

Tensor networks can also be used to solve classical spin statistical physics problems. These systems are doubly interesting because of the correspondence principle \cite{bohr1920serienspektra} which tells us that a quantum system in $d$ dimensions has a classical analog in $d+1$ dimensions.

Kadanoff introduced a renormalization method for spin systems in 1966 \cite{kadanoff1966scaling}. The principle of this method is to group spins into domains. These domains then form the sites of a network at a coarser level. By performing this repeated renormalization operations on a system, we eventually reach a fixed point that reveals important characteristics of the system. Tensor renormalization algorithms have the same property of creating a sequence of networks with increasingly large steps, and the state of the domains is filtered with a decomposition to singular values.

\paragraph{Setting up the problem.--}
The starting point for tensor network algorithms for conventional systems is the partition function $\mathcal{Z}$ in the form of a local tensor network at the finest scale:
  \includegraphics[width=\columnwidth]{bakerINFIG3.PNG}

For example, the contribution of a group of four sites forming a plaquette to the partition function of the Ising model on a square lattice is:
\begin{equation}
\mathcal{Z}_{ijk\ell}=e^{-\beta\Ham{}_{ijk\ell}}=\exp\Big(\beta J(\sigma_i\sigma_j+\sigma_j\sigma_k+\sigma_k\sigma_\ell+\sigma_\ell\sigma_i)\Big)
\end{equation}
where $\sigma_i=\pm1$ is the spin at the $i$ site. Each of the elements of this tensor corresponds to a different configuration of the spins. The partition function is then given by the sum of the configurations of each of these local tensors: it is the contraction of a network of tensors. To contract this network exactly is only possible for very small systems. For others, it is necessary to use the ideas of the renormalization group to extract the most relevant information at each scale.

\paragraph{Algorithm.--}
The tensor renormalization group (TRG) \cite{levin2007tensor} is an example of applying the ideas of the renormalization group to the tensor network. The local tensors are contracted together, forming a larger network, and a decomposition with singular values is used to identify the most relevant configurations and reject the others. These steps are then repeated.

The first step is to perform a SVD sequence to separate the initial tensor:

  \includegraphics[width=\columnwidth]{bakerINFIG4.PNG}

The $\D$ matrix of each of the links is factorized $\sqrt{\D}\cdot\sqrt{\D}$ and only one tensor is contracted with the tensors $\U{}$ and $\V^\dagger$ respectively.

\includegraphics[width=\columnwidth]{bakerINFIG5.PNG}

After decomposing a site, the tensors are contracted to form a larger network.

\includegraphics[width=\columnwidth]{bakerINFIG6.PNG}

The tensors are contracted with the previous pattern because the system is translationally invariant. Below, we see in more detail the tensors of the network, and that their decompositions lead to the contraction pattern shown above:

\includegraphics[width=0.95\columnwidth]{bakerINFIG7.PNG}

The site at the topmost scale consists of a plaquette of four tensors. At each step of renormalization, the number of tensors present in the network is divided by four. This algorithm can be used for finite or infinite systems, and for other problems than Ising spins on a square network.
Other algorithms, such as the tensor network renormalization (TNR) method \cite{evenbly2015tensor,evenbly2015tensorB,evenbly2017algorithms}, are able to solve these classic two-dimensional systems. The TNR is a refinement of the TRG that solves for the partition function with greater accuracy.

\section{Other applications of tensor networks}

We have paid particular attention to methods for which it is necessary to consider the entanglement and interactions between sites. Although, there are several other applications of these methods.

For example, in quantum computing, it is possible to represent the logic gates acting on the qubits using diagrams~\cite{nielsen2002quantum,bravyi2014efficient,darmawan2017tensor,tremblay2018depth}. These gates can generate entanglement between qubits. It should be noted, however, that it is not common practice to renormalize by truncating certain values in the density matrix. The operations presented in \ENrefsec{ENsec:svd} are not necessary since all the logic gates are unitary.

For problems where it is not necessary to consider entanglement to perform renormalization, it is sometimes possible to use the graphical notation presented. Among other tasks, it is possible to use tensor networks in machine learning (neural networks) \cite{novikov2016exponential,stoudenmire2016supervised,evenbly2019number} or in image compression \cite{evenbly2018representation}. The graphical representation is also useful for quickly illustrating a problem. It should be noted that for the methods we have presented in this review, it is important to consider the entanglement between sites, whereas this is not always the case in other contexts.

\section{Conclusion}

Tensor network methods are applicable in many situations and converge quickly for problems. Understanding how to use these tools effectively determines the properties of a quantum or classical system. A graphical representation is used which allows the tensor networks to be applied to other areas.

We conducted an overview of the basic operations of tensor networks: reshaping, index permutation, contraction, and decomposition. In addition, we have shown the link between the singular value decomposition and the entanglement between two parts of a network. To model a wave function, we introduced the matrix product state. The tensor network representation of a Hamiltonian, the matrix product operator, can be used with this representation of wave function. Finally, we presented algorithms, including time evolution and the tensor renormalization group, that use these tools to find the ground state of a quantum system or solve for classical systems.

\section{{Acknowledgement}}

S.D. gratefully acknowledges the support of the Institut Transdisciplinaire d'Information Quantique (INTRIQ). T.E.B.~thanks the support of the Postdoctoral Fellowship from Institut quantique and support from INTRIQ.  This research was undertaken in part thanks to funding from the Canada First Research Excellence Fund (CFREF).

The authors gratefully acknowledge the feedback from Sarah Labb\'e, Sara Turcotte, Agustin Di Paolo, Sarah E.~Grefe, Martin Schnee, Maxime Dion, Colin Trout, Jessica Lemieux, Alexandre Pr\'emont-Foley, Prosper Reulet, Yunlong Lian, Catherine Leroux, Thomas Gobeil, David Poulin, David Sénéchal, Andreas Bill, Yves B\'erub\'e-Lauzi\`ere, Yves Grosdidier, and Glen Evenbly.

\begin{appendix}

\section{More details on the difference between a MPS and a MERA}\label{ENsec:correlation}

Expressing the quantum ground state with entanglement can be more generally reformulated in terms of correlations. The concept of a correlation is very important in physics. Since physics is the study of the consequences of the interaction between different bodies, studying correlations (or what happens to other particles when one is disturbed) is essentially the study of physics itself. Correlations can occur in several different contexts, but we will extend this concept here to illustrate the differences between the MPS and the MERA in a different way.

\begin{figure}[b]
  \includegraphics[width=\columnwidth]{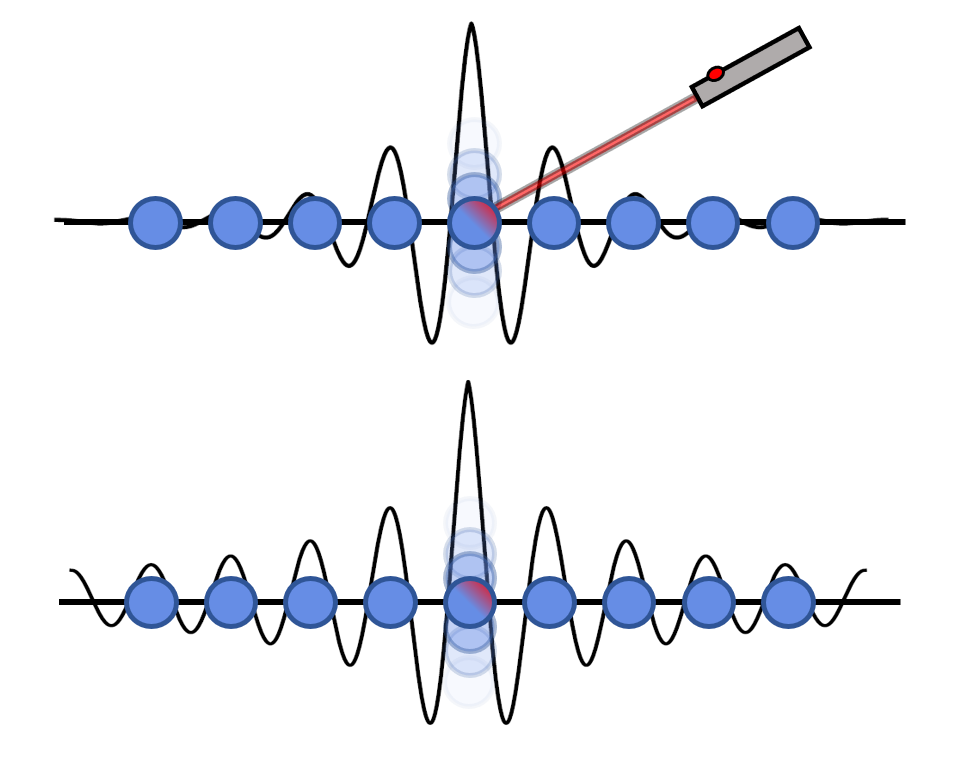}
\caption{Representation of the correlations after a disturbance for a gap (top) and without gap (bottom) system. Gapless correlations extend beyond the exponential decay in gapped systems.}
  \label{ENperturbe}
\end{figure}

When a system is disturbed, the disturbance can affect other remote sites (see \ENreffig{ENperturbe}). The magnitude, in general, depends on distance and there are only two types of correlations that can be present in a system, and there is a mathematical proof of this~\cite{hastings2004locality}. If the system has a difference between the first and second eigenvalue, then the correlations will decrease exponentially.  If the eigenvalues have no difference between the first two states (or if the eigenvalues are really very close to each other), then the correlations decrease like a power law. In summary, a correlation $\mathcal{C}$ can be one of the following forms:
\begin{equation}\label{ENcorrelations}\tag{A1}
\mathcal{C}\sim\begin{cases}
\exp(-x/\xi)&\mathrm{with\;gap}\\
(x/\xi)^\gamma&\mathrm{gapless.}\\
\end{cases}
\end{equation}

\begin{figure}[t]
  \includegraphics[width=\columnwidth]{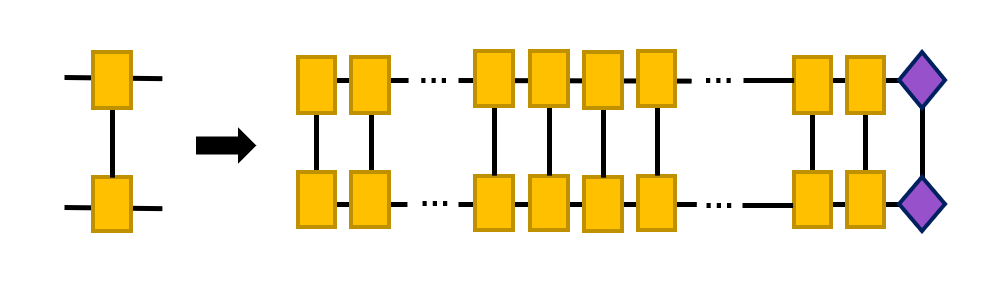}
\caption{A series of transfer matrices.}
  \label{ENtransfermatrix}
\end{figure}

By forming the transfer matrix of an MPS with its Hermitian conjugate as in \ENreffig{ENtransfermatrix}, we can contract a chain of these operators, assuming a sufficiently long chain. It is well known that multiplying matrices in succession can reveal the smallest eigenvalue. For example, each transfer matrix $T$ can be decomposed by a decomposition into eigenvalues $\U{}\bigLambda \U{}^\dagger$. A product of these can give 
\begin{equation}\label{ENtransfers}\tag{A2}
\prod_{i=1}^N T_i=\left(\U{}\bigLambda\U{}^\dagger\right)\left(\U{}\bigLambda \U{}^\dagger\right)\left(\U{}\bigLambda\U{}^\dagger\right)\ldots=\U{}\bigLambda^N\U{}^\dagger.
\end{equation}
The $\bigLambda$ matrix is diagonal and contains the eigenvalues with
\begin{equation}\label{ENlambdaN}\tag{A3}
\bigLambda^N=\left(\begin{array}{cccc}
\varepsilon_1^N&0&0&\cdots\\
0&\varepsilon_2^N&0&\cdots\\
0&0&\varepsilon_3^N&\ddots\\
\vdots&\vdots&\ddots&\ddots
\end{array}\right).
\end{equation}
If there is a difference between the eigenvalues, the largest eigenvalue, in magnitude, ($\varepsilon_1$) will appear much larger than all the others when it is raised to the power $N$.  Therefore, the only value in \ENrefeq{ENlambdaN} that is not negligible is $\varepsilon_1$.

The series of transfer matrices defined in \ENrefeq{ENtransfers} can carry information from one perturbed site to another that is $N$ sites further away. It is then reasonable to ask how the two sites are correlated. Since we know that the structure of the correlation is given as in \ENrefeq{ENcorrelations}, we see that a correlation can be rewritten
\begin{equation}\tag{A4}
\exp\left(-\frac{N}{{\ln(\varepsilon_1)^{-1}}}\right)
\end{equation}
where we can clearly identify the coherence length $\xi^{-1}=\ln(\varepsilon_1)$.  This shows how the smallest eigenvalue is characterized by an MPS. A system without a gap would give a sum of exponential terms that would become a power law.

If we form a transfer matrix for a MERA, we can do it at any level. If we choose the lowest level, then we get the same result as for a MPS. However, the equivalent transfer matrix formed at the first level of the MERA extends over 4 sites rather than 2, the next level extends over 8 sites, etc. In this way, the MERA allows correlations to be exposed on different scales. Thus, with this ansatz, we can encode correlations with a much greater range, including that of a gapless system that MERA can manage very well.

\section{Kronecker product and direct sum}
\label{ENkronecker}

The Kronecker product between two tensors, $\A \otimes \B$, is defined as the multiplication of all tensor elements $\B$ with all tensor elements $\A$. For example,
\begin{align}
\A\otimes \B&=\left(\begin{array}{cc}
a_{11}&a_{12}\\
a_{21}&a_{22}
\end{array}\right)\otimes\left(\begin{array}{cc}
b_{11}&b_{12}\\
b_{21}&b_{22}
\end{array}\right)\tag{B1}\\
&=\left(\begin{array}{cc}
a_{11}\B&a_{12}\B\\
a_{21}\B&a_{22}\B
\end{array}\right)\tag{B2}\\
&=\left(\begin{array}{cccc}
a_{11}b_{11} & a_{11}b_{12} & a_{12}b_{11} & a_{12}b_{12}\\
a_{11}b_{21} & a_{11}b_{22} & a_{12}b_{21} & a_{12}b_{22}\\
a_{21}b_{11} & a_{21}b_{12} & a_{22}b_{11} & a_{22}b_{12}\\
a_{21}b_{21} & a_{21}b_{22} & a_{22}b_{21} & a_{22}b_{22}\\
\end{array}\right)\tag{B3}
\end{align}

Kronecker's product can be considered as the fifth operation on tensors. However, the latter is rarely used in algorithms and has not been included in the list in \ENrefsec{ENbasicops}. A similar reasoning can be used for other operations such as the direct sum $\A \oplus \B$, defined for two matrices $\A$ and $\B$ as
\begin{equation}\tag{B4}
  \A\oplus \B=\left(\begin{array}{cc}
  \A&\0\\
  \0&\B
  \end{array}\right).
\end{equation}

\section{Sample code for constructing a Matrix Product State}\label{ENMPScode}

Here, we provide a simple code to obtain any wave function as an MPS. This code is written in the Julia programming language.

\begin{widetext}
\begin{lstlisting}
#############################################################
#
#  Converting wavefunctions into the matrix product state
#
#############################################################
# Made by T.E. Baker, S. Desrosiers, M. Tremblay, and M.P. Thompson (2019)
# See accompanying license with this program
# This code is native to the julia programming language (v1.1.1)
#

using LinearAlgebra

# creating random normalized state vector of size N
physInd = 2 #size of the physical index
vect = rand(ComplexF64, physInd^N, 1) #we initialize a random state
vect /= norm(vect) #this normalizes the wavefunction

# reshaping into 2 * 2^N tensor
function makeMPS(vect,physInd,N)
  mps = []

  #reshaped vector isolating the first index of size 2
  M = reshape(vect, physInd, physInd^(N-1))

  Lindsize = 1 #current size of the left index

  # MPS building loop
  for i=1:N-1
    U, D, V = svd(M) # applying SVD.
    temp = reshape(U,Lindsize,physInd,size(D,1))
    push!(mps, temp)
    Lindsize = size(D,1)
    DV = Diagonal(D)*V'
    if i == N-1
      temp = reshape(DV,Lindsize[1],physInd,1)
      push!(mps,temp)
    else
      Rsize = cld(size(M,2),physInd) #integer division, round up
      M = reshape(DV,size(D,1)*physInd,Rsize)
    end
  end
  return mps
end

mps = makeMPS(vect,physInd,N)
\end{lstlisting}
\end{widetext}

\end{appendix}